%% file: 2LepStopPaper2013.tex
\pdfoutput=1 
\documentclass[11pt]{article}
\usepackage{jheppub}
\usepackage{rotating}
\usepackage{atlasphysics} 
\graphicspath{{/}}
\usepackage{multirow}
\usepackage{subfigure}
\usepackage{xspace}

\def\etmiss{\ensuremath{E_{\mathrm{T}}^{\mathrm{miss}}}\xspace}

\newcommand{\tone}   {\tilde{t}_{1}}
\newcommand{\neut}  {\tilde{\chi}^{0}_{1}}
\newcommand{\chipm}  {\tilde{\chi}^{\pm}_{1}}
\def\etmiss{\ensuremath{E_{\mathrm{T}}^{\mathrm{miss}}}\xspace}
\def\EtMiss{\ensuremath{E_{\mathrm{T}}^{\mathrm{miss}}}\xspace}

\renewcommand{\ttbar} {\ensuremath{t\bar{t}}\xspace}

\def\MET{\ensuremath{E_{\mathrm{T}}^{\mathrm{miss}}}\ }

\newcommand{\mttwo}{\ensuremath{m_{\mathrm{T2}}}\xspace}
\newcommand{\ttbarW}{\ensuremath{t\bar{t}W}\xspace}
\newcommand{\ttbarZ}{\ensuremath{t\bar{t}Z}\xspace}

\usepackage{preprintcover}  % load package

\PreprintCoverPaperTitle{Search for direct top-squark pair production in final states with two leptons in \boldmath{pp} collisions at $\sqrt{s}=8$~TeV with the ATLAS detector}  % paper title
\PreprintIdNumber{CERN-PH-EP-2014-014}  % CERN preprint number
\PreprintCoverAbstract{A search is presented for direct top-squark pair production in final states with two leptons (electrons or muons) of opposite charge using 20.3~fb$^{-1}$ of $pp$ collision data at $\sqrt{s}=8$~TeV, collected by the ATLAS experiment at the Large Hadron Collider in 2012. No excess over the Standard Model expectation is found. The results are interpreted under the separate assumptions (i) that the top squark decays to a $b$-quark in addition to an on-shell chargino whose decay occurs via a real or virtual $W$ boson, or (ii) that the top squark decays to a $t$-quark and the lightest neutralino. A top squark with a mass between 150~GeV and 445~GeV decaying to a $b$-quark and an on-shell chargino is excluded at 95\% confidence level for a top squark mass equal to the chargino mass plus 10~GeV, in the case of a 1~GeV lightest neutralino. Top squarks with masses between 215 (90)~GeV and 530 (170)~GeV decaying to an on-shell (off-shell) $t$-quark and a neutralino are excluded at 95\% confidence level for a  1~GeV neutralino. }  % paper abstract
\PreprintJournalName{JHEP}  % journal name

\title{Search for direct top-squark pair production in final states with two leptons in \boldmath{pp} collisions at $\sqrt{s}=8$~TeV with the ATLAS detector}
 
\author{The ATLAS Collaboration}
\abstract{
A search is presented for direct top-squark pair production in final states with two leptons (electrons or muons) of opposite charge using 20.3~fb$^{-1}$ of $pp$ collision data at $\sqrt{s}=8$~TeV, collected by the ATLAS experiment at the Large Hadron Collider in 2012. No excess over the Standard Model expectation is found. The results are interpreted under the separate assumptions (i) that the top squark decays to a $b$-quark in addition to an on-shell chargino whose decay occurs via a real or virtual $W$ boson, or (ii) that the top squark decays to a $t$-quark and the lightest neutralino. A top squark with a mass between 150~GeV and 445~GeV decaying to a $b$-quark and an on-shell chargino is excluded at 95\% confidence level for a top squark mass equal to the chargino mass plus 10~GeV, in the case of a 1~GeV lightest neutralino. Top squarks with masses between 215 (90)~GeV and 530 (170)~GeV decaying to an on-shell (off-shell) $t$-quark and a neutralino are excluded at 95\% confidence level for a  1~GeV neutralino. 
}

\begin{document}

\maketitle

%%%%%%%%%%%%%%%%%%%%%%%%%%%%%%%%%%%%%%%%%%%%%%%%%%%%%%%%%%%%%%%%%%%%%%%%%%%%%%
\section{Introduction}

\label{sec:introduction}
Supersymmetry (SUSY)~\cite{Miyazawa:1966,Ramond:1971gb,Golfand:1971iw,Neveu:1971rx,Neveu:1971iv,Gervais:1971ji,Volkov:1973ix,Wess:1973kz,Wess:1974tw} is an extension to the Standard Model (SM) 
which introduces supersymmetric partners of the known fermions and bosons.
For each known boson or fermion, SUSY introduces a particle with identical quantum numbers except for a difference of half a unit of spin ($S$). The introduction of gauge-invariant and renormalisable interactions into SUSY models can violate the conservation of baryon number ($B$) and lepton number ($L$), resulting in a proton lifetime shorter than current experimental limits~\cite{PhysRevD.86.012006}.
This is usually solved by assuming that the multiplicative quantum number $R$-parity ($R$), defined as $R = (-1)^{3(B-L)+2S}$, is conserved.
In the framework of a generic $R$-parity-conserving minimal supersymmetric extension of the SM (MSSM)~\cite{Fayet:1976et,Fayet:1977yc,Farrar:1978xj,Fayet:1979sa,Dimopoulos:1981zb}, SUSY particles are produced in pairs where the lightest supersymmetric particle (LSP) is stable, and is a candidate for dark matter.
In a large variety of models, the LSP is the lightest neutralino ($\tilde{\chi}^{0}_{1}$).
The scalar partners of right-handed and left-handed quarks (squarks), $\tilde q_{R}$ and $\tilde q_{L}$,  
mix to form two mass eigenstates, $\tilde q_1$ and $\tilde q_2$, with $\tilde q_1$ defined to be the lighter 
one. 
In the case of the supersymmetric partner of the top quark (top squark, $\tilde{t}$), large mixing effects can lead to one top-squark mass eigenstate, $\tilde{t}_1$, that is significantly lighter than the other squarks. 
Consideration of naturalness and its impact on the SUSY particle spectrum, suggests that top squarks cannot be too heavy, to keep the Higgs boson mass close to the electroweak scale~\cite{Barbieri:1987fn,Carlos:1993yy}. Thus $\tilde{t}_1$ could be pair-produced with relatively large cross-sections at the Large Hadron Collider (LHC). 

The top squark can decay into a variety of final states, depending, amongst other factors, on the hierarchy of the mass eigenstates formed from the linear superposition of the SUSY partners of the Higgs boson and electroweak gauge bosons. In this paper the relevant mass eigenstates are the lightest chargino ($\tilde{\chi}^\pm_1$) and the $\tilde{\chi}^{0}_{1}$.
Two possible sets of SUSY mass spectra are considered, assuming that the mixing of the neutralino gauge eigenstates is such that the $\neut$ is mostly the supersymmetric partner of the SM boson B (before electroweak symmetry breaking) and taking into account previous experimental constraints from the LEP experiments~\cite{lepsusy} that $m(\chipm) >103.5$~GeV.

\begin{figure}[!h]
\begin{center}
\includegraphics[width=0.7\textwidth]{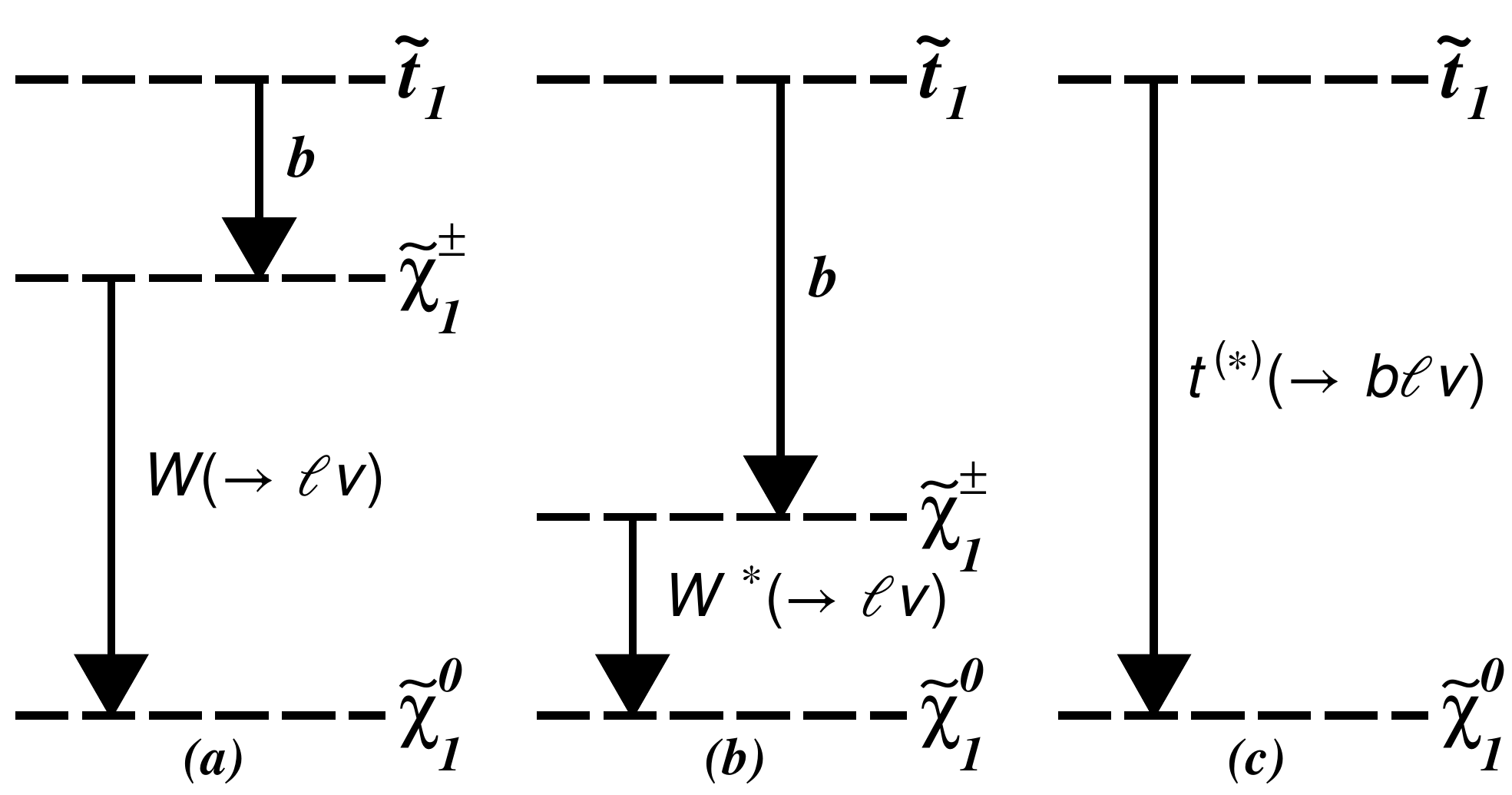}
\label{hierarchy}
\caption{Schematic diagrams of mass hierarchy for the $\tilde{t}_1\rightarrow b+\tilde{\chi}^\pm_1$ decay mode ((a) larger than the $W$ mass $(\tilde{\chi}^\pm_1,\tilde{\chi}^0_1)$ mass splitting and (b) smaller than the $W$ mass $(\tilde{\chi}^\pm_1,\tilde{\chi}^0_1)$ mass splitting), and (c) the $\tilde{t}_1\rightarrow t\tilde{\chi}^0_1$ decay mode.}
\end{center}
\end{figure}

In both sets of spectra (figure~\ref{hierarchy}) the $\tilde{t}_1$ is the only coloured particle contributing to the production processes. In the first scenario the $\tilde{t}_1$, assumed to be $\tilde{t}_L$, decays via $\tilde{t}_1\rightarrow b+\tilde{\chi}^\pm_1$, where $m(\tilde{t_1})-m(\tilde{\chi}_1^\pm)>m(b)$, and the $\tilde{\chi}^\pm_1$ (assumed to be mostly the supersymmetric partner of the SM $W$ boson before electroweak symmetry breaking) subsequently decays into the lightest neutralino (assumed to be the LSP) and a real (figure~\ref{hierarchy} (a)) or virtual (figure~\ref{hierarchy} (b)) $W$ boson. In the second scenario (figure~\ref{hierarchy} (c)), the $\tilde{t}_1$, assumed to be 70\% $\tilde t_R$, decays via $\tilde{t}_1\rightarrow t+\tilde{\chi}^0_1$. Both on-shell, kinematically allowed for $m(\tilde{t}_1)>m(t)+m(\tilde{\chi}^0_1)$, and off-shell (resulting in a three-body decay to $bW\tilde{\chi}^0_1$) top quarks are considered.

In all scenarios the top squarks are pair-produced and, since only the leptonic decay mode of the $W^{(*)}$ is considered, the events are characterised by the presence of two isolated leptons ($e$, $\mu$)\footnote{Electrons and muons from $\tau$ decays are included.} with opposite charge, and two $b$-quarks. Significant missing transverse momentum $\mathbf p^\mathrm{miss}_\mathrm{T}$, whose magnitude is referred to as $E^\mathrm{miss}_\mathrm{T}$, is also expected from the neutrinos and neutralinos in the final states.

In this paper, three different analysis strategies are used to search for $\tilde{t}_1$ pair production, with a variety of signal regions defined for each. Two of the analyses target the $\tilde{t}_1\rightarrow b+\tilde{\chi}^\pm_1$ decay mode and the three-body $\tilde{t}_1\rightarrow bW\tilde{\chi}^0_1$ decay via an off-shell top-quark, whilst one targets the $\tilde{t}_1\rightarrow t+\tilde{\chi}^0_1$ to an on-shell top-quark decay mode. 

The kinematics of the $\tilde{t}_1\rightarrow b+\tilde{\chi}^\pm_1$ decay mode depend upon the mass hierarchy of the $\tilde{t}_1$, $\tilde{\chi}^\pm_1$ and $\tilde{\chi}^0_1$ particles (figure~\ref{hierarchy}~(a) and~\ref{hierarchy}~(b)). In order to be sensitive to all the possible mass splittings, two complementary cut-based analysis strategies are designed: one to target large $\tilde{\chi}^\pm_1-\tilde{\chi}^0_1$ mass splittings (larger than the $W$ bosons mass), and one to target small $\tilde{\chi}^\pm_1-\tilde{\chi}^0_1$ mass splittings (smaller than the $W$ bosons mass); the first one provides the sensitivity to the $\tilde{t}_1$ three-body decay. 

These signatures have both very small cross-section and low branching ratios (BRs) (of top-quark pairs to dileptonic final states). A multivariate approach is used to target the on-shell top $\tilde{t}_1\rightarrow t+\tilde{\chi}^0_1$ decay mode (figure~\ref{hierarchy}~(c)), to enhance sensitivity beyond what can be achieved with cut-and-count techniques. 

Previous ATLAS analyses using data at $\sqrt{s}=7$~TeV and 8~TeV have placed exclusions limits at 95\% confidence level (CL) on both the $\tilde{t}_1\rightarrow b+\tilde{\chi}^\pm_1$~\cite{Aad:2012yr,Aad:2012tx,Aad:2013ija} and $\tilde{t}_1\rightarrow t+\tilde{\chi}^0_1$~\cite{:2012si,:2012ar,Aad:2012uu} decay modes. This search is an update of the 7 TeV analysis targeting the two-lepton final state~\cite{Aad:2012uu} with a larger dataset, including additional selections sensitive to various signal models and exploiting a multivariate analysis technique. Limits on top squarks direct production have also been placed by the CMS~\cite{Chatrchyan:2012uea,Chatrchyan:2012wa,Chatrchyan:2013lya,Chatrchyan:2013xna}, CDF~\cite{Aaltonen:2009sf} and D0~\cite{Abazov:2008kz} collaborations.

\section{The ATLAS detector}

ATLAS is a multi-purpose particle physics experiment~\cite{Aad:2008zzm} at the LHC.
The detector layout\footnote{ATLAS uses a right-handed coordinate system with its origin at the nominal interaction point (IP) in the centre of the detector and the $z$-axis coinciding with the axis of the beam pipe. The $x$-axis points from the IP to the centre of the LHC ring, and the $y$-axis points upwards. Cylindrical coordinates
$(r,\phi)$ are used in the transverse plane, $\phi$ being the azimuthal angle around the beam pipe. The pseudorapidity is defined in terms of the polar angle $\theta$ as $\eta = -\ln \tan(\theta/2)$.} consists of inner tracking devices surrounded by a superconducting solenoid, electromagnetic and hadronic calorimeters and a muon spectrometer. 
The inner tracking detector (ID) covers $| \eta | < 2.5$ and consists of a silicon pixel detector, a semicondictor microstrip detector, and a transition 
radiation tracker. The ID is surrounded by a thin superconducting solenoid providing a 2T axial magnetic field and it 
provides precision tracking of charged particles and vertex reconstruction. The calorimeter system covers the pseudorapidity range $| \eta | < 4.9$.
In the region  $| \eta | < 3.2$, high-granularity liquid-argon electromagnetic sampling calorimeters are used. A steel/scintillator-tile calorimeter provides energy measurements for hadrons within  $| \eta | < 1.7$. The end-cap and forward regions, which cover the range $1.5 < | \eta | < 4.9$, are instrumented with liquid-argon calorimeters for both electromagnetic and hadronic particles.
The muon spectrometer surrounds the calorimeters and consists of three large superconducting air-core toroid magnets, each with eight coils, 
a system of precision tracking chambers ($|\eta| < 2.7$) and fast trigger chambers ($| \eta | < 2.4$). 

\section{Monte Carlo simulations and data samples}

\label{sec:MC}
Monte Carlo (MC) simulated event samples are used to model the signal and
to describe all the backgrounds which produce events with two prompt leptons from $W$, $Z$ or $H$ decays. 
All MC samples utilised in the analysis are produced using the ATLAS Underlying Event Tune 2B~\cite{ATLAS-CONF-2011-009} and are processed through the ATLAS detector simulation~\cite{Aad:2010ah} based on GEANT4~\cite{Agostinelli:2002hh} or passed through a fast simulation using a parameterisation of the performance of the ATLAS electromagnetic and hadronic calorimeters~\cite{ATLAS-PUB-atlfast}. 
Additional $pp$ interactions in the same (in-time) and nearby (out-of-time) bunch crossings (pile-up) are included in the simulation.

Processes involving supersymmetric particles are generated using {\tt HERWIG++2.5.2}~\cite{Bahr:2008pv} ($\tilde{t}_1\rightarrow t+\tilde{\chi}^0_1$) and {\tt MADGRAPH-5.1.4.8}~\cite{Alwall:2011uj} ($\tilde{t}_1\rightarrow b+\tilde{\chi}^\pm_1$) interfaced to~{\tt PYTHIA-6.426}~\cite{Sjostrand:2006za} (with the PDF set CTEQ6L1~\cite{Nadolsky:2008zw}). Different initial-state (ISR) and final-state radiation (FSR) and $\alpha_s$ parameter values are used to generate additional samples in order to evaluate the effect of their systematic uncertainties.
Signal cross-sections are calculated at next-to-leading order (NLO) in $\alpha_s$, including the resummation of soft gluon emission at next-to-leading-logarithm accuracy (NLO+NLL) \cite{Beenakker:1997ut,Beenakker:2010nq,Beenakker:2011fu}, as described in ref.~\cite{Kramer:2012kl}.

Top-quark pair and $Wt$ production are simulated with {\tt MC@NLO-4.06}~\cite{Frixione:2002ik,Frixione:2005vw}, interfaced with {\tt HERWIG-6.520}~\cite{Corcella:2000bw} for the fragmentation and the hadronisation processes, and using {\tt JIMMY-4.31}~\cite{Butterworth:1996zw} for the underlying event description. 
In addition, {\tt ACERMC-3.8}~\cite{Kersevan:2004yg} samples and {\tt POWHEG-1.0}~\cite{Frixione:2007vw} samples, interfaced to both {\tt PYTHIA-6.426} and {\tt HERWIG-6.520}, are used to estimate the event generator, fragmentation and hadronisation systematic uncertainties. Samples of \ttbarZ and \ttbarW production (referred to as $t\bar{t}V$) are generated with
{\tt MADGRAPH-5.1.4.8} interfaced to {\tt PYTHIA-6.426}.
Samples of $Z/\gamma^\star$ produced in association with jets
are generated with {\tt SHERPA-1.4.1}~\cite{Gleisberg:2008ta}, while
{\tt ALPGEN-2.14}~\cite{Mangano:2002ea} samples  are used for evaluation of systematic uncertainties.
Diboson samples ($WW$, $WZ$, $ZZ$)
are generated with {\tt POWHEG-1.0}.
Additional samples generated with {\tt SHERPA-1.4.1} are used to estimate the systematic arising from choice of event generator. Higgs boson production, including all decay modes,\footnote{An SM-like 125~GeV Higgs boson, with the same BR as in the SM, is assumed.} is simulated with {\tt PYTHIA-8.165}~\cite{Sjostrand:2007gs}.
Samples generated with {\tt MC@NLO-4.06}, {\tt POWHEG-1.0} and {\tt SHERPA-1.4.1} are produced using the parton distribution function (PDF) set CT10~\cite{Pumplin:2002vw}. All other samples are generated using the PDF set CTEQ6L1.

The background predictions are normalised to the theoretical cross-sections, including
higher-order QCD corrections where available, or are normalised to data in dedicated control regions (CRs).
The inclusive cross-section for $Z/\gamma^*+$jets is calculated with DYNNLO~\cite{dynnlo}
with the MSTW 2008 NNLO PDF set~\cite{Martin:2009iq}.
The $t\bar{t}$ cross-section for $pp$ collisions at a centre-of-mass energy of $\sqrt{s} = 8$~TeV is $\sigma_{t\bar{t}}= 253^{+13}_{-15}$~pb for a top-quark mass of $172.5 \gev$. It has been calculated at next-to-next-to-leading order (NNLO) in QCD including resummation of next-to-next-to-leading-logarithmic (NNLL) soft gluon terms with {\sc top++2.0}~\cite{Cacciari:2011hy,Baernreuther:2012ws,Czakon:2012zr,Czakon:2012pz,Czakon:2013goa,Czakon:2011xx}. The uncertainties due to the choice of PDF set and $\alpha_s$ were calculated using the PDF4LHC prescription~\cite{Botje:2011sn} with the MSTW2008 NNLO~\cite{Martin:2009iq,Martin:2009bu}, CT10 NNLO~\cite{Lai:2010vv,Gao:2013xoa} and NNPDF2.3 5f FFN~\cite{Ball:2012cx} PDF sets, and were added in quadrature to the uncertainty due to the choice of renormalisation and factorisation scale.
The approximate NNLO+NNLL cross-section is used for the normalisation of the $Wt$~\cite{Kidonakis:2010ux} sample.
The cross-sections calculated at NLO are used for the diboson~\cite{Binoth:2006mf}, \ttbarW and \ttbarZ~\cite{Garzelli:2012bn} samples.

The data sample used was recorded between March and December 2012 with the LHC operating at a $pp$ centre-of-mass energy of $\sqrt{s}=8$~TeV. Data were collected based on the decision of a three-level trigger system.
The events accepted passed either a single-electron, a single-muon, a double-electron, a double-muon, or an electron--muon trigger. The trigger efficiencies are approximately 99\%, 96\% and 91\% for the events passing the full $ee$, $e\mu$ and $\mu\mu$ selections described below, respectively.
After beam, detector and data-quality requirements, data corresponding to a total integrated luminosity of $20.3 \,\ifb $ were analysed~\cite{lumi}.

\section{Physics object selection}
\label{sec:obj}

Multiple vertex candidates from the proton--proton interaction are reconstructed using the tracks in the inner detector. The vertex with the highest scalar sum of the transverse momentum squared, $\Sigma p^2_\mathrm{T}$, of the associated tracks is defined as the primary vertex.

Jets are reconstructed from three-dimensional energy clusters~\cite{atlas-clustering} in the calorimeter
using the anti-$k_t$ jet clustering algorithm~\cite{Cacciari:2005hq,Cacciari:2008gp}
with a radius parameter of 0.4. The cluster energy is corrected using calibration factors based on MC simulation and validated with extensive test-beam and collision-data studies~\cite{ATLAS-LARG-2009-001}, in order to take into
account effects such as non-compensation and inhomogeneities, the presence of dead material
and out-of-cluster energy deposits. Corrections for converting to the jet energy scale
and for in-time and out-of-time pile-up are
also applied, as described in~Ref.~\cite{ATLAS-CONF-2013-004}.
Jet candidates with transverse momentum ($p_\mathrm{T}$) greater than 20~GeV, $|\eta| < 2.5$ and a ``jet vertex fraction" larger than 0.5 for those with $p_\mathrm{T}<50$~GeV, are selected as jets in the analysis. The jet vertex fraction quantifies the fraction of the total jet momentum of the event that originates
from the reconstructed primary vertex. This requirement rejects jets originating from additional proton--proton interactions.
Events containing jets that are likely to have arisen from detector noise or cosmic rays are also removed using the procedures described in ref.~\cite{Aad:2011he}.

A neural-network-based algorithm is used to identify which of the selected jet candidates contain a $b$-hadron decay ($b$-jets). The inputs to this algorithm are the impact parameter of inner detector tracks, secondary vertex reconstruction and the topology of $b$- and $c$-hadron decays inside a jet~\cite{btagalgo}. The efficiency for tagging $b$-jets in an MC sample of $t\bar{t}$ events using this algorithm is 70\% with rejection factors of 137 and 5 against light quarks and $c$-quarks, respectively. To compensate for differences between the $b$-tagging efficiencies and mis-tag rates in data and MC simulation, correction factors derived using $t\bar{t}$ events are applied to the jets in the simulation as described in ref.~\cite{btagalgoeff}.

Electron candidates are required to have $\pT > 10 \GeV$, $|\eta| < 2.47$ and
to satisfy ``medium" electromagnetic shower shape and track selection
quality criteria~\cite{Aad:2011mk}. These are defined as preselected electrons. Signal electrons
are then required to satisfy ``tight'' quality
criteria~\cite{Aad:2011mk}.                                               
They are also required to be isolated within the tracking volume:
the scalar sum,
$\Sigma \pT$, of the $\pT$
of inner detector tracks with $\pT > 1 \GeV$, not including the electron track,
within a cone of radius
$\Delta R = \sqrt{(\Delta\eta)^2+(\Delta\phi)^2} = 0.2$ around the electron candidate must be less
than 10\% of the electron \pT, where $\Delta\eta$ and $\Delta\phi$ are the separations in $\eta$ and $\phi$.

Muon candidates are reconstructed either from muon segments matched to inner detector tracks, or from combined tracks in the inner detector and muon spectrometer~\cite{ATLAS-CONF-2011-063}.
They are required to have $p_\mathrm{T}>10$~GeV  and $|\eta|<2.4$. Their longitudinal and transverse
impact parameters must be within 1~mm and 0.2~mm of the primary vertex, respectively.
Such preselected candidates are then
required to have $\Sigma \pT < 1.8 \GeV$, where $\Sigma \pT$ is defined in analogy to the electron
case. Event-level weights are applied to MC events to correct for differing lepton reconstruction and identification efficiencies between the simulation and those measured in data.

Ambiguities exist in the reconstruction of electrons and jets as they use the same calorimeter energy clusters as input: thus any jet whose axis lies within $\Delta R = 0.2$ of a preselected electron is discarded. Moreover, preselected electrons or muons within $\Delta R = 0.4$ of any remaining jets are rejected to discard leptons from the decay of a $b$- or $c$-hadron.

$E^\mathrm{miss}_\mathrm{T}$ is defined as the magnitude of the two-vector $\mathbf p^\mathrm{miss}_\mathrm{T}$ obtained
from the negative vector sum of the transverse momenta of all reconstructed electrons, jets and muons, and calorimeter energy clusters not associated with any objects.
Clusters associated with electrons with $p_\mathrm{T} >$ 10 GeV, and those associated with jets with $p_\mathrm{T} >$ 20 GeV make use of the electron and jet calibrations of these respective objects. For jets the calibration includes the pile-up correction described above whilst the jet vertex fraction requirement is not applied. Clusters of calorimeter cells with $|\eta |< 2.5$ not associated with these objects are calibrated using both calorimeter and tracker information~\cite{Aad:2012re}.

\section{Event selection}
\label{sec:selection}

\subsection{Preselection and event variables}
\label{sec:preselections}
A common set of preselection requirements, and some discriminating variables are shared by the three analysis strategies. The following event-level variables are defined, and their use in the various analyses is detailed in sections~\ref{sec:selection-lep},~\ref{sec:selection-had} and~\ref{sec:selection-mva}:

\begin{itemize}
\item[-] $m_{\ell\ell}$: the invariant mass of the two oppositely charged leptons.

\item[-] $m_\textup{T2}$ and $m^{b-\textup{jet}}_\textup{T2}$: lepton-based and jet-based stransverse mass. The stransverse mass ~\cite{Lester:1999tx,Barr:2003rg} is a kinematic variable that can be used
to measure the masses of pair-produced semi-invisibly decaying heavy particles. This quantity is defined as
\begin{equation*} 
\mttwo
( \mathbf p_{\mathrm{T,1}}, \mathbf p_{\mathrm{T,2}}, \mathbf
q_{\mathrm{T}}) = \min_{\mathbf q_{\mathrm{T,1}} + \mathbf q_{\mathrm{T,2}} = 
\mathbf q_{\mathrm{T}} }
\left\{ \max [\; 
    m_{\mathrm{T}}( \mathbf p_{\mathrm{T,1}}, \mathbf q_{\mathrm{T,1}} ),  
    m_{\mathrm{T}}( \mathbf p_{\mathrm{T,2}}, \mathbf q_{\mathrm{T,2}} 
    ) 
\;] 
\right\} ,
\label{mt2eq}
\end{equation*}
\noindent where $m_{\mathrm T}$ indicates the transverse mass,\footnote{The transverse mass is defined by the equation $m_\textup{T}=\sqrt{2|\mathbf p_{\mathrm{T,1}}||\mathbf p_{\mathrm{T,2}}|(1-\mathrm{cos}(\Delta\phi))}$, where $\Delta\phi$ is the angle between the particles with transverse momenta $\mathbf p_{\mathrm{T,1}}$ and
$\mathbf p_{\mathrm{T,2}}$ in the plane perpendicular to the beam axis.} $\mathbf p_{\mathrm{T,1}}$ and
$\mathbf p_{\mathrm{T,2}}$ are the transverse momentum vectors of two particles (assumed to be massless), and $\mathbf
q_{\mathrm{T,1}}$ and $\mathbf q_{\mathrm{T,2}}$ are vectors and $ \mathbf q_{\mathrm{T}} = \mathbf q_{\mathrm{T,1}} + \mathbf q_{\mathrm{T,2}}$.
The minimisation is performed over all the possible decompositions 
of  $\mathbf q_{\mathrm{T}}$. For $t\bar{t}$ or $WW$ decays, if the transverse momenta of the two leptons in each event are taken as  $\mathbf p_{\mathrm{T,1}}$ and
$\mathbf p_{\mathrm{T,2}}$, and $E^\mathrm{miss}_\mathrm{T}$ as $\mathbf q_\mathrm{T}$, $m_\textup{T2}(\ell,\ell,E^\mathrm{miss}_\mathrm{T})$ is bounded sharply from above by the mass of the $W$ boson~\cite{Cho:2007dh,Burns:2008va}. In the $\tilde{t}_1\rightarrow b+\tilde{\chi}^\pm_1$ decay mode the upper bound is strongly correlated with the mass difference between the chargino and the lightest neutralino. If the transverse momenta of the two reconstructed
$b$-quarks in the event are taken as $\mathbf p_{\mathrm{T,1}}$ and
$\mathbf p_{\mathrm{T,2}}$, and the lepton transverse momenta are added vectorially to the missing transverse momentum in the event to form $\mathbf q_\mathrm{T}$, the resulting $m_\textup{T2}(b,b,\ell+\ell+E^\mathrm{miss}_\mathrm{T})$ has a very different kinematic limit: for top-quark pair production it is approximately bound by the mass of the top quark, whilst for top-squark decays the bound is strongly correlated to the mass difference between the top squark and the chargino. In this paper, $m_\textup{T2}(\ell,\ell,E^\mathrm{miss}_\mathrm{T})$ is referred to simply as $m_\textup{T2}$, whilst $m_\textup{T2}(b,b,\ell+\ell+E^\mathrm{miss}_\mathrm{T})$ is  referred to as $m^{b-\textup{jet}}_\textup{T2}$. The mass of the $\mathbf
q_{\mathrm{T}}$ is always set to zero in the calculation of these stransverse variables.

\item[-] $\Delta\phi_j$: the azimuthal angular distance between
	the $\mathbf p^{\mathrm{miss}}_{\mathrm{T}}$ vector and the direction of the closest jet.

\item[-] $\Delta\phi_{\ell}$: the azimuthal angular distance between
	the $\mathbf p^{\mathrm{miss}}_{\mathrm{T}}$ vector and the direction of the highest-$p_\mathrm{T}$ lepton.
	
\item[-] $\Delta\phi_\mathrm{b}$ and $\mathbf p^{\ell\ell}_{\mathrm{Tb}}$: the azimuthal angular distance between the 
	$\mathbf p^{\mathrm{miss}}_{\mathrm{T}}$ vector
	and the $\mathbf p^{\ell\ell}_{\mathrm{Tb}} = \mathbf p^{\mathrm{miss}}_{\mathrm{T}} +
	                      \mathbf p^{\ell_1}_{\mathrm{T}}+
	                      \mathbf p^{\ell_2}_{\mathrm{T}}$
	vector\footnote{Note that the b in $\mathbf p^{\ell\ell}_{\mathrm{Tb}}$ (and consequently $\Delta\phi_\textup{b}$) does not bear any relation to $b$-jet. In Ref.~\cite{Polesello:2009rn} it was so named to indicate that it represents the transverse momentum of {boosted} objects.}. The $\mathbf p^{\ell\ell}_{\mathrm{Tb}}$ variable, with magnitude $p^{\ell\ell}_{\mathrm{Tb}}$,
	is the opposite of the vector sum of all the transverse hadronic activity in the event.
	\item[-] $m_\textup{eff}$: the scalar sum of the $E^\mathrm{miss}_\mathrm{T}$, the transverse momenta of the two leptons and that of the two jets with the largest $p_\mathrm{T}$ in each event.
\item[-] $\Delta\phi_{\ell\ell}$ ($\Delta\theta_{\ell\ell}$): the azimuthal (polar) angular distance between the two leptons.

\item[-] $\Delta\phi_{j\ell}$: the azimuthal angular distance between the highest-$p_\mathrm{T}$ jet and lepton.

\end{itemize}
The three different analyses are referred to in this paper as the ``leptonic $m_\textup{T2}$'', ``hadronic $m_\textup{T2}$'' and ``multivariate analysis (MVA)'', respectively. The first two are so named as they use, in the first case, $m_\textup{T2}$, and in the second case, $m^{b-\textup{jet}}_\textup{T2}$, as the key discriminating variable.
The $m_\textup{T2}$ selection is used to ensure orthogonality between these two analyses, allowing for their results to be combined.
The third uses an MVA technique and targets the on-shell top $\tilde{t}_1\rightarrow t+\tilde{\chi}^0_1$ decay. 

In all cases, events are required to have exactly two oppositely charged signal leptons (electrons, muons or one of each). At least one of these electrons or muons must have $p_\mathrm{T}>25 \GeV$, in order for the event to be triggered with high efficiency, and $m_{\ell\ell}>20 \GeV$ (regardless of the flavours of the leptons in the pair), in order to remove leptons from low mass resonances.\footnote{The $m_{\ell\ell}$ requirement also resolves overlap ambiguities between electron and muon candidates by implicitly removing events with close-by electrons and muons.}
If the event contains a third preselected electron or muon, the event is rejected. This has a negligible impact on signal acceptance, whilst simplifying the estimate of the fake and non-prompt lepton background (defined in section~\ref{sec:fakes}) and reducing diboson backgrounds.

All three analyses consider events with both different-flavour (DF) and same-flavour (SF) lepton pairs. These two event populations are separately used to train the MVA decision\footnote{MVA uses events which are known to belong to signal or background to determine the mapping function from which it is possible to subsequently classify any given event into one of these two categories. This ``learning" phase is usually called ``training".} and are explicitly separated when defining the signal regions (SRs). The decay $\tilde{t}_1\rightarrow b+\tilde{\chi}^\pm_1$ is symmetric in flavour and the $Z/\gamma^*$ background is small, hence the populations are therefore not separated in the hadronic and leptonic $m_\textup{T2}$ analyses.
All three analyses exploit the differences between the DF and SF populations when evaluating and validating background estimates.

\label{sec:selection-common}

\subsection[Leptonic $m_\textup{T2}$ selection]{Leptonic \boldmath$m_\textup{T2}$ selection}
\label{sec:selection-lep}

After applying the preselection described in section~\ref{sec:selection-common}, events with SF leptons are required to have the invariant mass of the lepton pairs outside the 71--111~GeV range. This is done in order to reduce the number of background events containing two leptons produced by the decay of a $Z$ boson.
Two additional selections are applied to reduce the number of background events with high $\mttwo$ 
arising from events with large \etmiss  due to mismeasured jets: $\Delta \phi_b<1.5$ and $\Delta \phi_j>1$. 
After these selections the background is dominated by \ttbar events for DF
lepton pairs and $Z/\gamma^\star$+jets for SF lepton pairs. 
The \mttwo distribution for $Z/\gamma^\star$+jets is, however, steeply falling and by requiring  $\mttwo > 40 \GeV$ 
the \ttbar becomes the dominant background in the SF sample as well.

The leptonic $m_\textup{T2}$ selection has been optimised to target models with $\Delta m(\chipm, \ninoone) > m(W)$ (figure~\ref{hierarchy}~(a)). The jet \pT\ spectrum is exploited in order to provide sensitivity to models with varying jet multiplicity.
Four non-exclusive SRs are defined, with different selections on $\mttwo$ and 
on the transverse momentum of the two leading jets, as reported in table~\ref{tab:SRL}. 
The SRs L90 and L120 require $\mttwo > 90 \GeV$ and $\mttwo > 120 \GeV$, respectively, with no additional requirement on jets.
They provide sensitivity to scenarios with a small $\Delta m(\tone, \chipm)$ (almost degenerate top squark and chargino), where the production of high-$\pT$ jets is not expected. 
The SR L100 has a tight jet selection, requiring at least two jets with $\pT > 100 \GeV$ and $\pT > 50 \GeV$, respectively, and $m_\textup{T2}>100$~GeV.
This SR provides sensitivity to scenarios with both large $\Delta m(\tone, \chipm)$ and $\Delta m(\chipm, \ninoone)$, where large means bigger than the $W$ boson mass.
SR L110 has a looser selection on jets, requiring two jets with $\pT > 20 \GeV$ each and $\mttwo > 110 \GeV$.
It provides sensitivity to scenarios with small to moderate (up to around the $W$ boson mass) values of $\Delta m(\tone, \chipm)$ resulting in moderate jet 
activity.

\begin{table}[!h]
\caption{\label{tab:SRL} Signal regions used in the leptonic $m_\textup{T2}$ analysis. The last two rows give the relative sizes of the mass splittings that the SRs are sensitive to: small (almost degenerate), moderate (up to around the $W$ boson mass) or large (bigger than the $W$ boson mass). }

\begin{center}
\begin{tabular}{l|cccc}
\hline
SR        & L90 & L100 & L110 & L120 \\
\hline
\hline
leading lepton $\pT$ [GeV]                               & \multicolumn{4}{c}{$> 25 $} \\
\hline
$\Delta\phi_j$ [rad] &  \multicolumn{4}{c}{$> 1.0$} \\
$\Delta\phi_\mathrm{b}$ [rad] &  \multicolumn{4}{c}{$< 1.5$} \\ 
\hline
 $\mttwo$ [GeV]    & $> 90 $ & $> 100 $ & $> 110 $ & $> 120 $\\ 
\hline
Leading jet $\pT$ [GeV] & - & $> 100 $ & $> 20 $ &  - \\
Second jet $\pT$ [GeV] & - & $>  50 $  & $> 20 $ & - \\
\hline
$\Delta m(\tilde{t}_1,\tilde{\chi}^\pm_1)$ & small & large & moderate & small \\
$\Delta m(\tilde{\chi}^\pm_1,\tilde{\chi}^0_1)$ & moderate & large & moderate & large \\
\hline
\end{tabular}
\end{center}
\end{table}

\subsection[Hadronic $m_\textup{T2}$ selection]{Hadronic \boldmath$m_\textup{T2}$ selection}
\label{sec:selection-had}

In contrast to the leptonic $m_\textup{T2}$ selection, the hadronic $m_\textup{T2}$ selection is designed to be sensitive to the models with chargino--neutralino mass differences smaller than the $W$ mass (figure~\ref{hierarchy}~(b)). In addition to the preselection described in section~\ref{sec:selection-common}, events in the SR (indicated as H160) are required to satisfy the requirements given in table~\ref{tab:SRH}. The requirement of two $b$-jets favours signal over background; the targeted signal events have in general higher-$p_\mathrm{T}$ $b$-jets as a result of a large $\Delta m(\tone, \chipm)$ (figure~\ref{hierarchy}~(b)).
The $t\bar{t}$\hspace{1mm} background is then further reduced by the $m^{b-\textup{jet}}_\textup{T2}$ requirement, which preferentially selects signal models with large $\Delta m(\tilde{t}_1,\tilde{\chi}^\pm_1)$ over the SM background.
The requirement on leading lepton $p_\mathrm{T}$ has little impact on the signal, but reduces the remaining $Z/\gamma^*+$jets background to a negligible level.

\begin{table}[!h]

\caption{Signal region used in the hadronic $m_\textup{T2}$ analysis. The last two rows give the relative sizes of the mass splittings that the SR is sensitive to\label{tab:SRH}: small (almost degenerate), moderate (up to around the $W$ boson mass) or large (bigger than the $W$ boson mass).}
\begin{center}
\begin{tabular}{c|c}
\hline
SR & H160 \\
 \hline
 \hline
$b$-jets & $=2$ \\
Leading lepton $p_\mathrm{T}$ [GeV]  & $<60$  \\
$m_\textup{T2}$ [GeV] & $<90$ \\
$m^{b-\textup{jet}}_\textup{T2}$ [GeV] & $>160$ \\
\hline
$\Delta m(\tilde{t}_1,\tilde{\chi}^\pm_1)$ & large \\
$\Delta m(\tilde{\chi}^\pm_1,\tilde{\chi}^0_1)$ & small \\
\hline
\end{tabular}
\end{center}
\end{table}

\subsection{Multivariate analysis}
\label{sec:selection-mva}

In this analysis, $\tilde{t}_1\rightarrow t+\tilde{\chi}^0_1$ signal events are separated from SM backgrounds using an MVA technique based on boosted decision trees (BDT) that uses a gradient-boosting algorithm (BDTG) \cite{BDTG}.
In addition to the preselection described in section \ref{sec:selection-common}, events are required to have at least two jets, a leading jet with $p_\mathrm{T}>50$~GeV and $m_\textup{eff}>300$~GeV. The selected events are first divided into four (non-exclusive) categories, with the requirements in each category designed to target different $\tilde{t}_1$ and $\ninoone$ masses:

\begin{itemize}
\item[-] { (C1)}
 \MET$> 50$ GeV: provides good sensitivity for $m(\tilde{t}_1)$ in the range 200--500 GeV and for low neutralino masses;
\item[-] { (C2)}
\MET$> 80$ GeV: provides good sensitivity along the ${m}({\tilde{t}_1})={m}({t})+{m}({\tilde{\chi}^{0}_{1}})$ boundary;
\item[-] { (C3)}
\MET$> 50$ GeV and leading lepton \pT\ $>50$ GeV: provides good sensitivity for $m(\tilde{t}_1)$ in the range 400--500 GeV, and $m(\tilde{t}_1)>$~500~GeV for high neutralino masses;
\item[-] { (C4)}
 \MET$> 50$ GeV and leading lepton \pT\ $>80$ GeV: provides good sensitivity for $m(\tilde{t}_1)> 500$ GeV.
\end{itemize}
Events are then further divided into those containing an SF lepton pair and those containing a DF lepton pair.  Categories (C1), (C2) and (C4) are considered for DF events, and categories (C1) and (C3) for SF events.

A BDTG discriminant is employed to further optimise the five subcategories (three for DF, two for SF) described above.
The following variables are given as input to the BDTG: $E^\mathrm{miss}_\mathrm{T}$, $m_{\ell\ell}$, $m_\textup{T2}$, $\Delta\phi_{\ell\ell}$, $\Delta\theta_{\ell\ell}$, $\Delta\phi_{l}$ and $\Delta\phi_{j\ell}$.
These variables are well modelled by the simulation and are effective in discriminating $t+\tilde{\chi}_1^0$ signal from SM background; the distributions in data and MC simulation for the four ``best ranked'' (their correlation with the BDTG ranges from $\sim80\%$ to $\sim95\%$) input variables for the SF and DF channels after C1 cuts are shown in figures~\ref{MVA_variablesA} and~\ref{MVA_variablesB}, respectively. In each of the sub-figures, the uncertainty band represents the total uncertainty, from all statistical and systematic uncertainty sources (section~\ref{sec:systematics}). 
The correlation coefficient between each pair of variables is found to be in good agreement (within 1--2$\%$)  between data and MC.

\begin{figure}[hb]
\begin{center}
\includegraphics[width=0.5\textwidth]{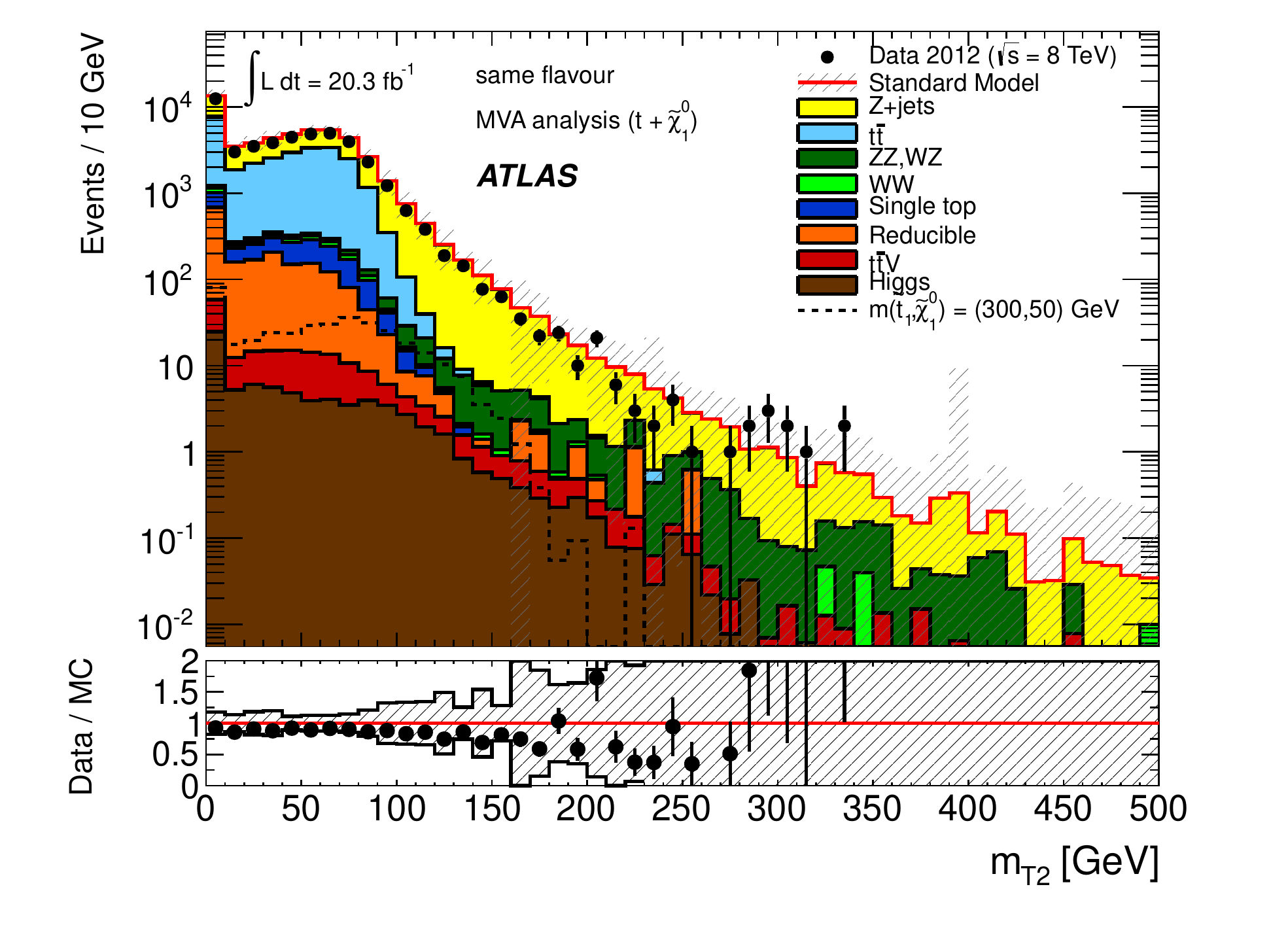}\includegraphics[width=0.5\textwidth]{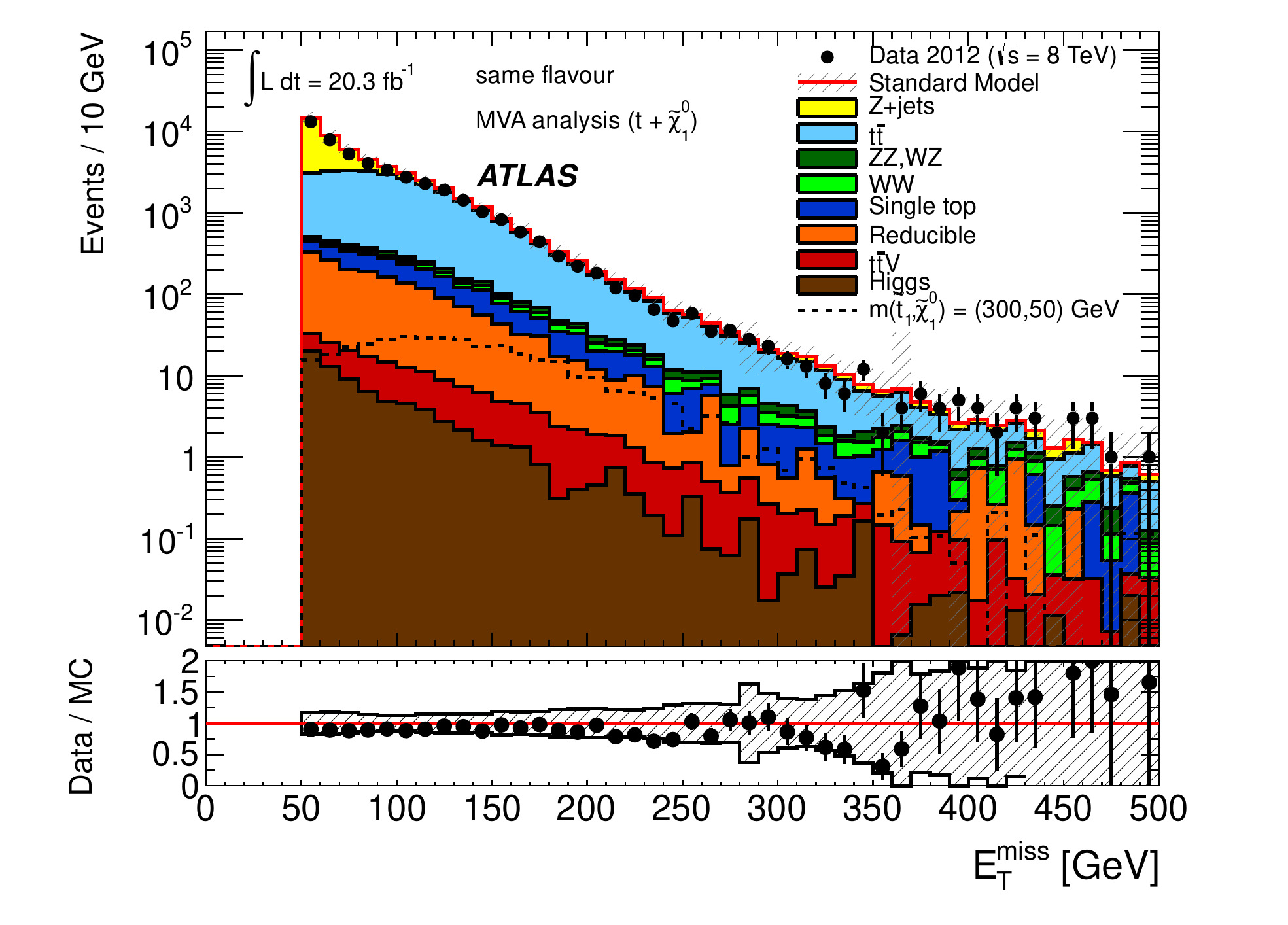}
\includegraphics[width=0.5\textwidth]{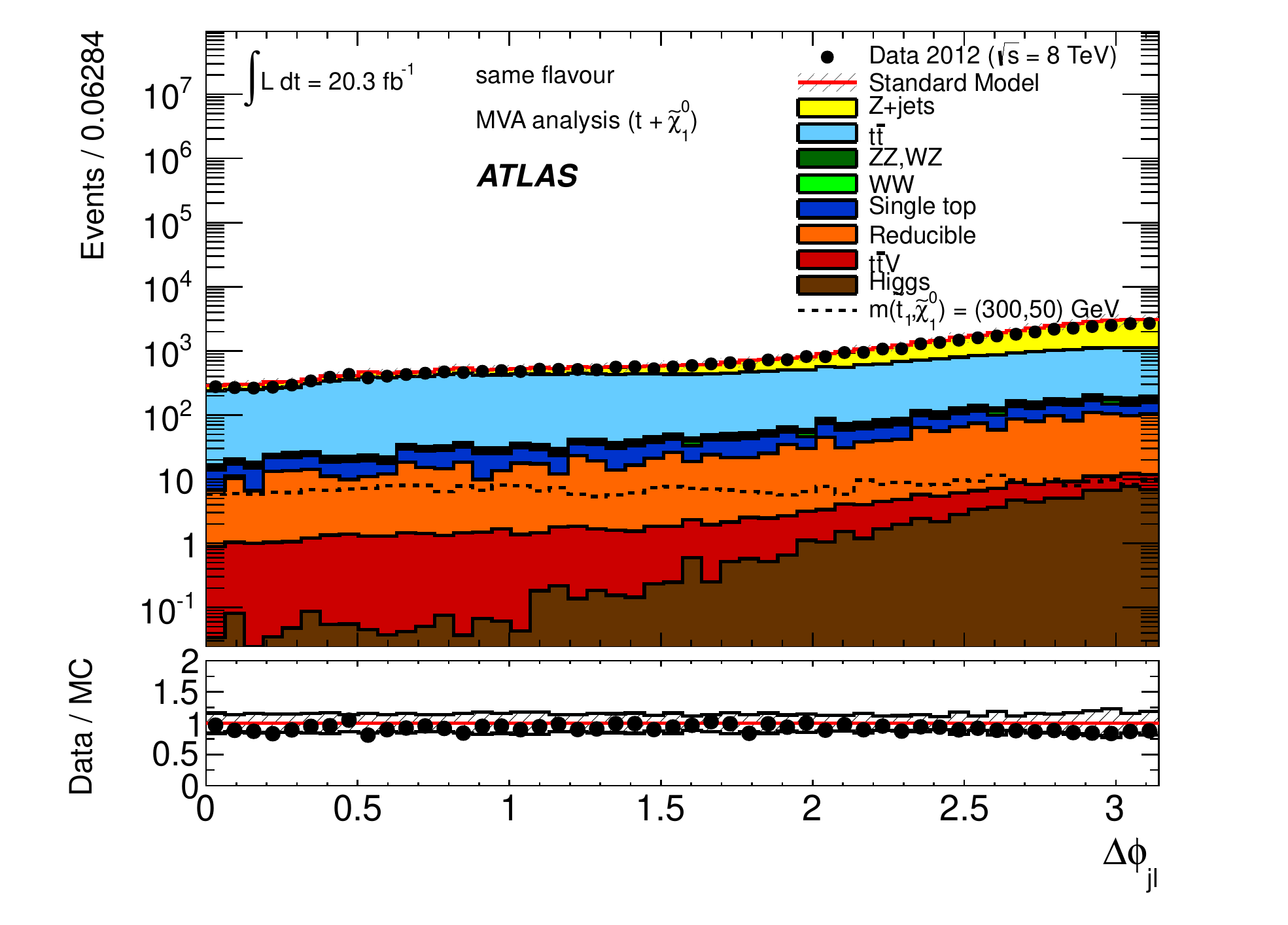}\includegraphics[width=0.5\textwidth]{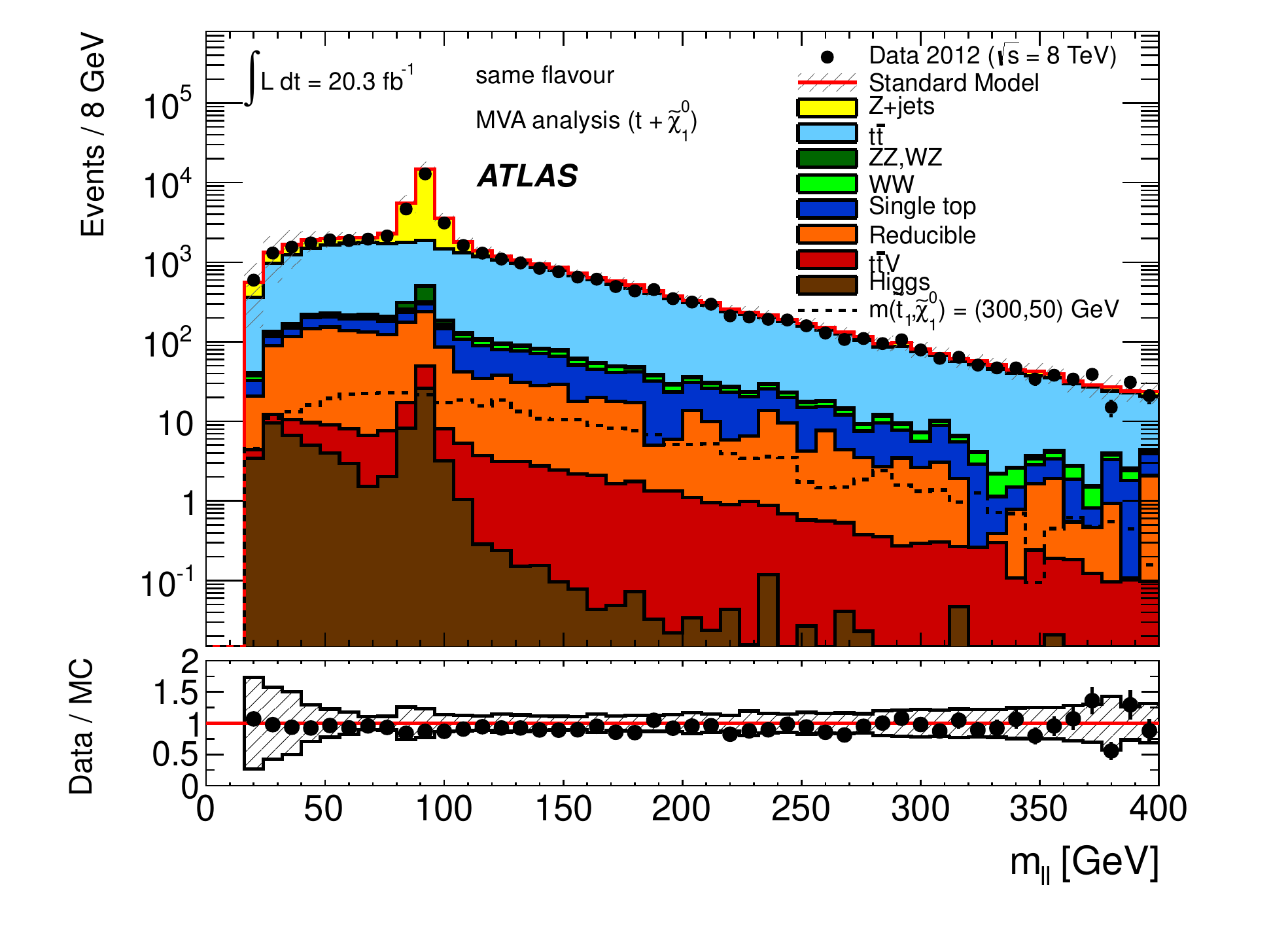}
\end{center}
\caption{\label{MVA_variablesA} The four best ranked input variables for the MVA analysis. SF channel: $m_\textup{T2}$, $E^\mathrm{miss}_\mathrm{T}$,  $\Delta\phi_{j\ell}$ and $m_{\ell\ell}$ after C1 cuts (\MET$> 50$ GeV). The contributions from all SM backgrounds are shown as a histogram stack; the bands represent the total uncertainty from statistical and systematic sources. The components labelled ``Reducible'' correspond to the fake and non-prompt lepton backgrounds and are estimated from data as described in section~\ref{sec:fakes}; the other backgrounds are estimated from MC simulation.}
\end{figure}

\begin{figure}[ht]
\begin{center}
\includegraphics[width=0.5\textwidth]{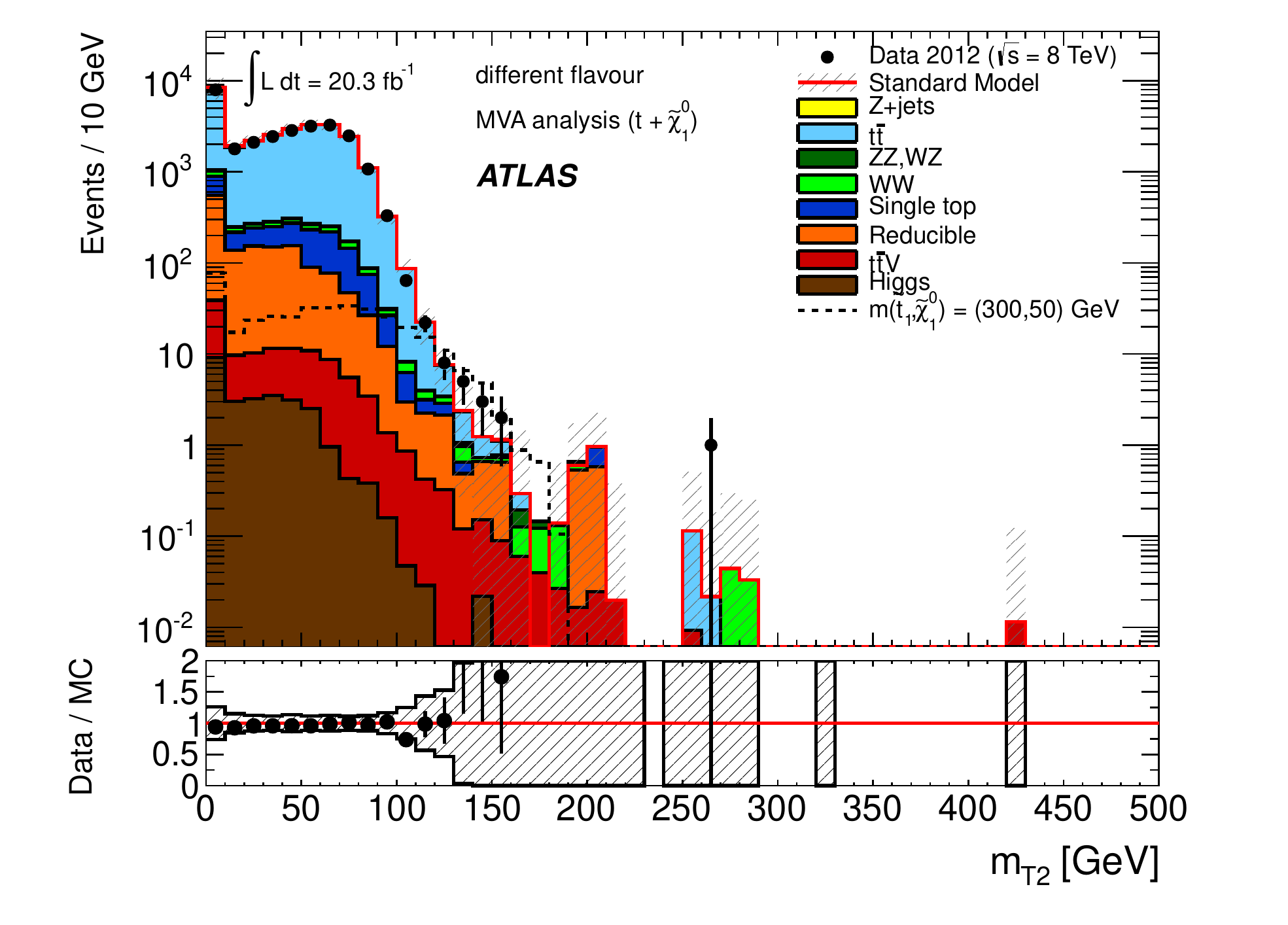}\includegraphics[width=0.5\textwidth]{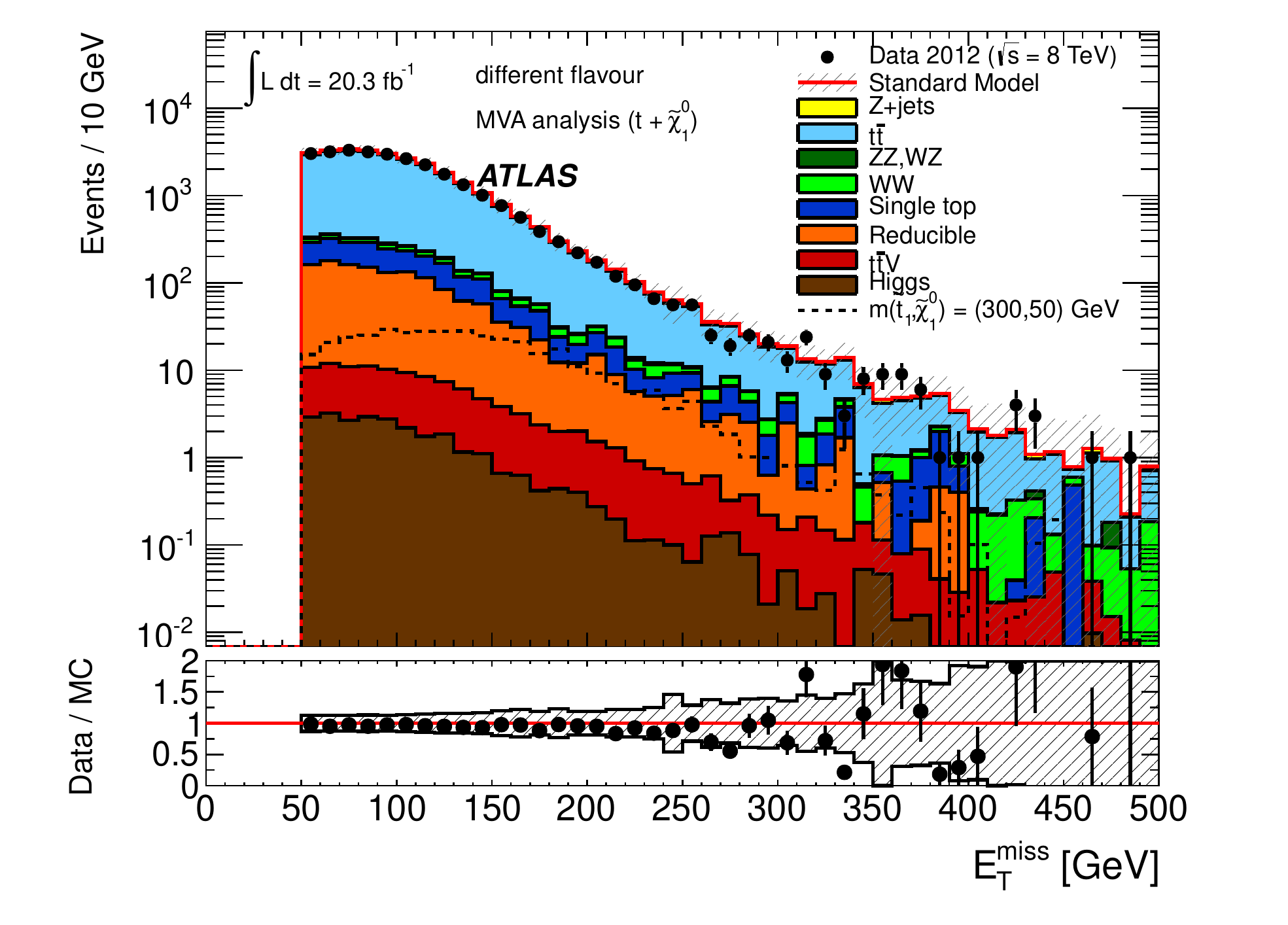}
\includegraphics[width=0.5\textwidth]{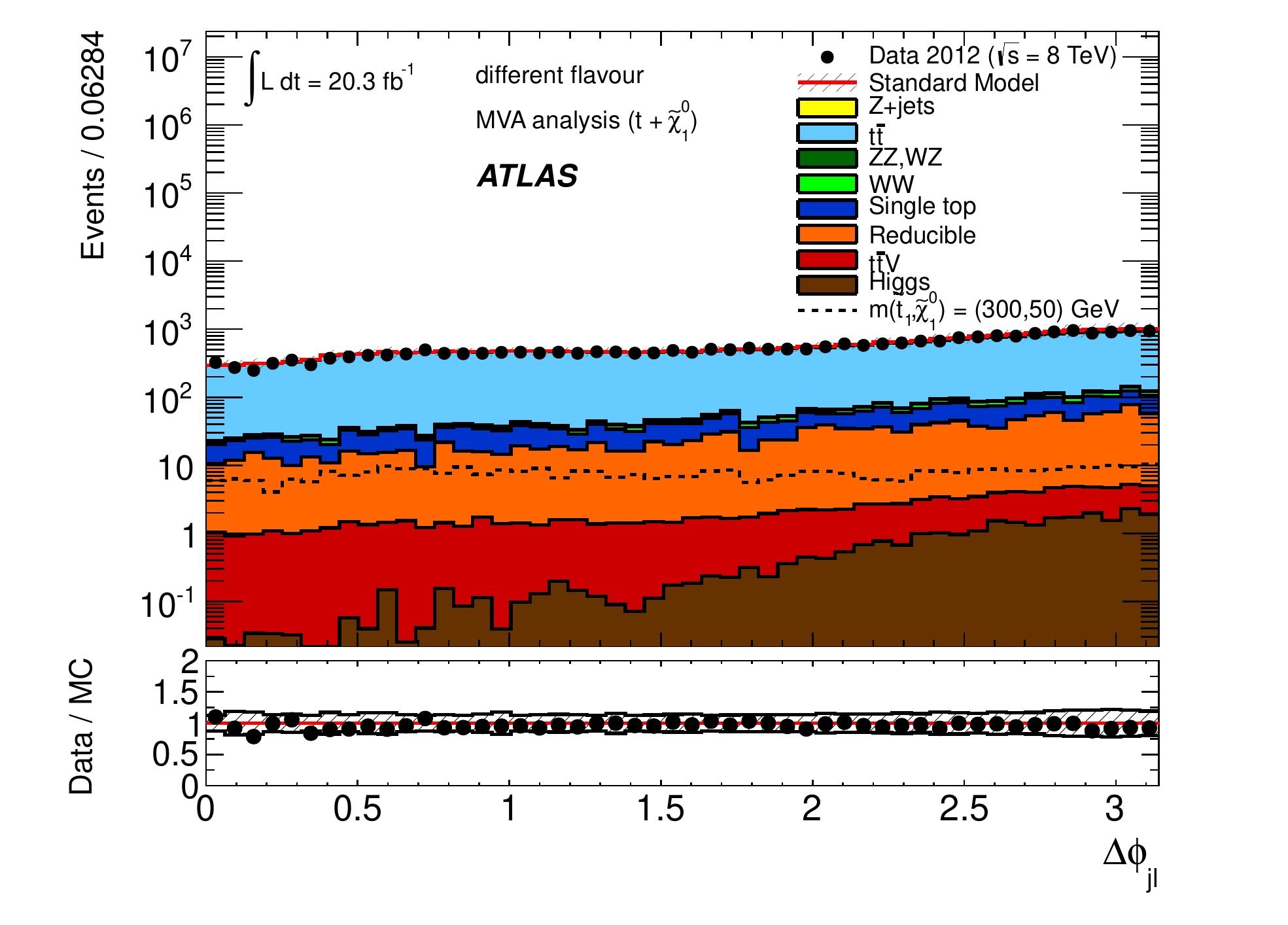}\includegraphics[width=0.5\textwidth]{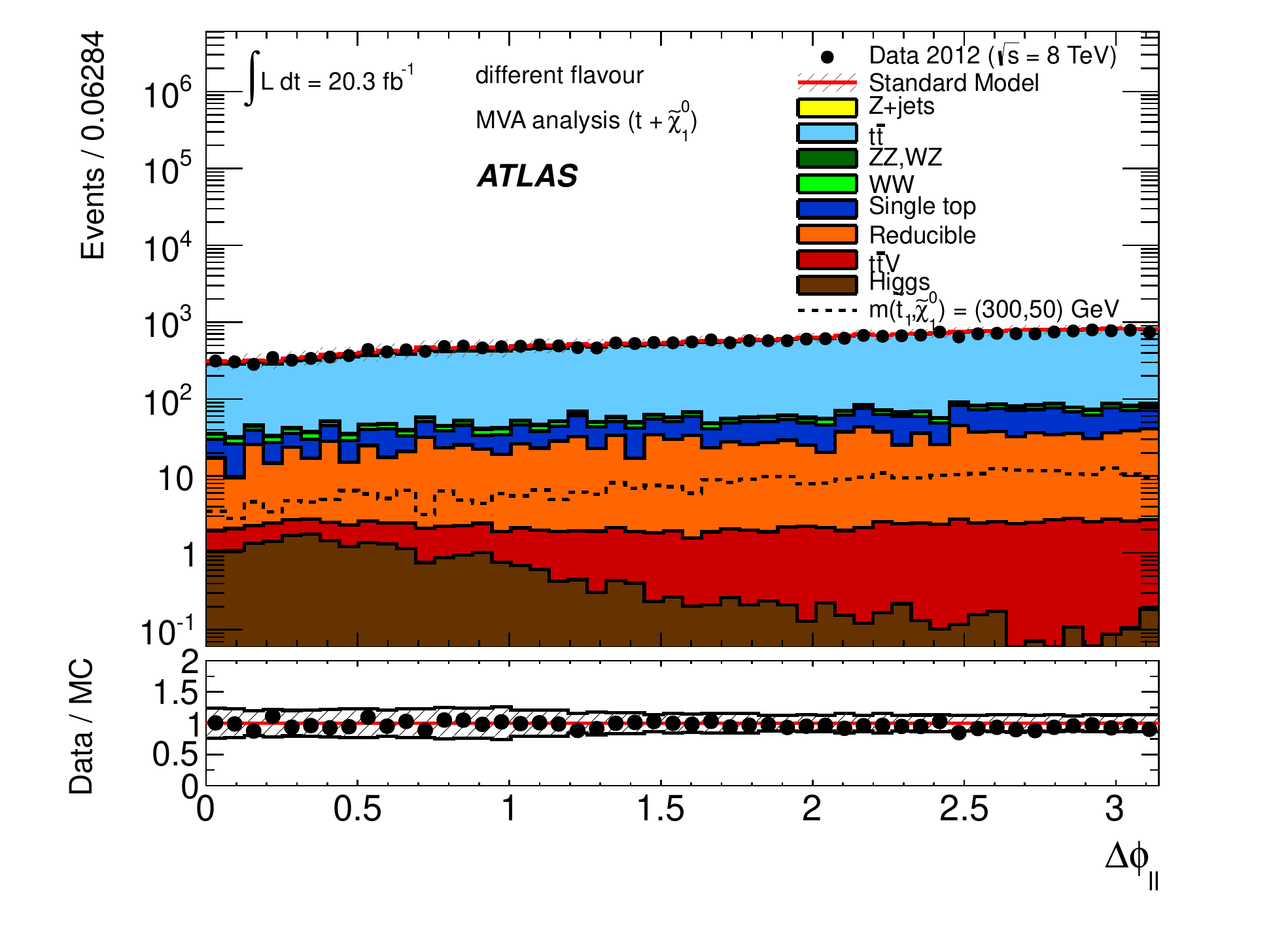}
\end{center}
\caption{\label{MVA_variablesB} The four best ranked input variables for the MVA analysis. DF channel: $m_\textup{T2}$, $E^\mathrm{miss}_\mathrm{T}$, $\Delta\phi_{j\ell}$ and $\Delta\phi_{\ell\ell}$ after C1 cuts. The contributions from all SM backgrounds are shown as a histogram stack; the bands represent the total uncertainty from statistical and systematic sources. The components labelled ``Reducible'' correspond to the fake and non-prompt lepton backgrounds and are estimated from data as described in section~\ref{sec:fakes}; the other backgrounds are estimated from MC simulation.}
\end{figure}

Several BDTGs are trained using the simulated SM background against one or more representative signal samples, chosen appropriately for each of the five subcategories. The BDTG training parameters are chosen to best discriminate signal events from the background, without being overtrained (MC sub-samples, which are statistically independent to the training sample, are used to check that the results are reproducible). The resulting discriminants are bound between $-1$ and $1$. The value of the cut on each of these discriminants is chosen to maximise sensitivity to the signal points considered, with the possible values of the BDTG threshold scanned in steps of 0.01. A total of nine BDTGs (five for DF events, four for SF events) and BDTG requirements are defined, setting the SRs. They are summarised in table~\ref{tab:sigreg}.

\begin{table} [htb!]
\caption{Signal regions for the MVA analysis. The first column gives the name of each SR, where DF and SF indicate different and same flavours, respectively. The second column gives the signal sample used to train the BDTG. The third column lists the selection requirements applied in addition to the BDTG requirement given in the fourth column and the common SR requirements: $\geq2$~jets, leading jet $p_\mathrm{T}>50$~GeV, $m_\textup{eff}>300$~GeV. }
\label{tab:sigreg}
\begin{center}
\begin{tabular}{l|ccc}
\hline
 SR       & Training Sample [GeV] & Category & BDTG range\\
 & ($m(\tilde{t}_1),m(\tilde{\chi}^0_1)$)  & & \\
\hline
\hline
$\textup{M1}^\textup{DF}$ & (225,0)   & C1 ($E^\mathrm{miss}_\mathrm{T}>50$~GeV)   &  $>-0.13$\\

$\textup{M2}^\textup{DF}$ & (250,25)   & C1 ($E^\mathrm{miss}_\mathrm{T}>50$~GeV)   &  $>-0.18$\\

$\textup{M3}^\textup{DF}$ & (300,50)  & C1 ($E^\mathrm{miss}_\mathrm{T}>50$~GeV)   &  $>0.19$\\

$\textup{M4}^\textup{DF}$ & (350,170)   & C2 ($E^\mathrm{miss}_\mathrm{T}>80$~GeV)  &  $>-0.65$\\

$\textup{M5}^\textup{DF}$ & (550,0)  & C4 ($E^\mathrm{miss}_\mathrm{T}>50$~GeV,&  $>-0.33$\\
& &  leading lepton $p_\mathrm{T}>80$~GeV)   & \\
\hline
$\textup{M1}^\textup{SF}$ & (225,25)   & C1 ($E^\mathrm{miss}_\mathrm{T}>50$~GeV)   &  $>-0.66$\\

$\textup{M2}^\textup{SF}$ & (300,50)   & C1 ($E^\mathrm{miss}_\mathrm{T}>50$~GeV)   &  $>-0.11$\\

$\textup{M3}^\textup{SF}$ & (300,100)   & C1 ($E^\mathrm{miss}_\mathrm{T}>50$~GeV)   &  $>-0.77$\\

$\textup{M4}^\textup{SF}$ & (500,250)   & C3 ($E^\mathrm{miss}_\mathrm{T}>50$~GeV,  &  $>-0.76$\\
 & &  leading lepton $p_\mathrm{T}>50$~GeV)& \\
\hline
\end{tabular} 
\end{center}
\end{table}

\clearpage

\section{Standard Model background determination}
\label{sec:backgrounds}

All backgrounds containing prompt leptons from $W$, $Z$ or $H$ decay are estimated directly from MC simulation. The dominant backgrounds (top-quark pair production for all analyses, and diboson and $Wt$ single-top production for the leptonic $m_\textup{T2}$ and hadronic $m_\textup{T2}$ analyses respectively) are normalised to data in dedicated CRs, and then extrapolated to the SRs using the MC simulation (with a likelihood fit as described in section~\ref{sec:background:fit}). Whilst it is not a dominant background, $Z/\gamma^*+$jets is also normalised in a dedicated CR in the hadronic $m_\textup{T2}$ analysis. All other such contributions are normalised to their theoretical cross-sections.

The backgrounds due to non-prompt leptons (from heavy-flavour decays or photon conversions) or jets misidentified as leptons are estimated using a data-driven technique. Events with these types of lepton are referred to as ``fake and non-prompt'' lepton events. The estimation procedure is common to all three analyses and is described in section~\ref{sec:fakes}.

\subsection{Background fit}
\label{sec:background:fit}

The observed numbers of events in the CRs are used to derive SM background estimates in each SR via a profile likelihood fit~\cite{2011EPJC...71.1554C}. This procedure takes into account the correlations across the CRs due to common systematic uncertainties and the cross-contamination in each CR from other SM processes. 
The fit takes as input, for each SR:
\begin{enumerate}
\item The number of events observed in each CR and the corresponding number of events predicted in each by the MC simulation for each (non-fake, prompt) background source.
\item The number of events predicted by the MC simulation for each (non-fake, prompt) background source.
\item The number of fake and non-prompt lepton events in each region (CRs and SR) obtained with the data-driven technique (see section~\ref{sec:fakes}).
\end{enumerate}
Each uncertainty source, as detailed in section~\ref{sec:systematics}, is treated as a nuisance parameter in the fit, constrained with a Gaussian function taking into account the correlations between sample estimates. The likelihood function is the product of Poisson probability functions describing the observed and expected number of events in the control regions and the Gaussian constraints on the nuisance parameters. For each analysis, and each SR, the free parameters of the fit are the overall normalisations of the CR-constrained backgrounds: $t\bar{t}$, $WW$ and  $(WZ,ZZ)$ for the leptonic $m_\textup{T2}$ analysis; $t\bar{t},Wt$ and $Z/\gamma^*+$jets for the hadronic $m_\textup{T2}$ analysis and $t\bar{t}$ for the MVA analysis. The contributions from all other non-constrained prompt-lepton processes are set to the MC expectation, but are allowed to vary within their respective uncertainties. The contribution from fake and non-prompt lepton events is also set to its estimated yield and allowed to vary within its uncertainty. The fitting procedure maximises this likelihood by adjusting the free parameters; the fit constrains only the background normalisations, while the systematic uncertainties are left unchanged (i.e. the nuisance parameters always have a central value very close to zero with an error close to one). Background fit results are cross-checked in validation regions (VRs) located between, and orthogonal to, the control and signal regions.
Sections~\ref{sec:backgrounds-lep} to~\ref{sec:backgrounds-mva} describe the CR defined for each analysis and, in addition, any VRs defined to cross-check the background fit results.

\subsection{Fake and non-prompt lepton background estimation}
\label{sec:fakes}

The fake and non-prompt lepton background arises from semi-leptonic $t\bar{t}$, $s$-channel and $t$-channel single-top, $W+$jets and light- and heavy-flavour jet production. 
The main contributing source in a given region depends on the topology of the events: low-$m_\textup{T2}$ regions are expected to be dominated by the multijet background, while regions with moderate/high $m_\textup{T2}$ are expected to be dominated by the $W+$jets and \ttbar production.
The fake and non-prompt lepton background rate is estimated
for each analysis from data using a matrix method estimation, similar to that described in refs.~\cite{top1,top2}.
In order to use the matrix method, two types of lepton identification criteria are defined: ÒtightÓ, corresponding to the full set of identification criteria described above, and ÒlooseÓ, corresponding to preselected electrons and muons. 
The number of events containing fake leptons in each region is obtained by acting on a vector of observed (loose, tight) counts with a $4\times 4$ matrix with terms containing probabilities ($f$ and $r$) that relate real--real, real--fake, fake--real and fake--fake lepton event counts to tight--tight, tight--loose, loose--tight and loose--loose counts.

The two probabilities used in the prediction are defined as follows: $r$ is the probability for real leptons satisfying the loose selection criteria to also pass the tight selection and $f$ is the equivalent probability for fake and non-prompt leptons.
The probability $r$ is measured using a $Z\rightarrow\ell\ell (\ell=e,\mu)$ sample, while the probability $f$ is measured from two background-enriched control samples.
The first of these requires exactly one lepton with $p_\mathrm{T} > 25$~GeV, at least one jet, $E^\mathrm{miss}_\mathrm{T} < 25$~GeV, and an angular distance $\Delta R < 0.5$ between the leading jet and the lepton, in order to enhance the contribution from the multijet background. The probability is parameterised as a function of the lepton $\eta$ and $p_\mathrm{T}$ and the number of jets.
For leptons with $p_\mathrm{T} < 25$~GeV, in order to avoid trigger biases, a second control sample which selects events containing a same-charge DF lepton pair is used. 
The probability $f$ is parameterised as a function of lepton $p_\mathrm{T}$ and $\eta$, the number of jets, $m_\textup{eff}$ and $m_\textup{T2}$. The last two variables help to isolate the contributions expected to dominate from multijet, $W+$jets or \ttbar productions.
In both control samples, the probability is parameterised by the number of $b$-jets when a $b$-jet is explicitly required in the event selection (i.e. in the hadronic $m_\textup{T2}$), in order to enhance the contribution from heavy-flavour jet production.

Many sources of systematic uncertainty are considered when evaluating this background. Like the probabilities themselves, the systematic uncertainties are also parameterised as a function of the lepton and event variables discussed above. The parameterised uncertainties are in general dominated by differences in the measurement of the fake lepton probabilities obtained when using the two control regions above. The limited number of events in the CR used to measure the probabilities are also considered as a source of systematic uncertainty. The overall systematic uncertainty ranges between 10$\%$ and 50$\%$ across the various regions (control, validation and signal). Ultimately, in SRs with very low predicted event yields the overall uncertainty on the fake and non-prompt lepton background yield is dominated by the statistical uncertainty arising from the limited number of data events in the SRs, which reaches 60--80$\%$ in the less populated SRs. In these regions, however, the contributions from fake and non-prompt lepton events are small or negligible.

The predictions obtained using this method are validated in events with same-charge lepton pairs. As an example, figure~\ref{fig:lepmt2:fakes} shows the distribution of $m_\textup{eff}$ and $m_\textup{T2}$ in events with a same-charge lepton pair after the preselection described in section~\ref{sec:preselections}, prior to any additional selection.

\begin{figure}[ht]
\begin{center}
\includegraphics[width=0.5\textwidth]{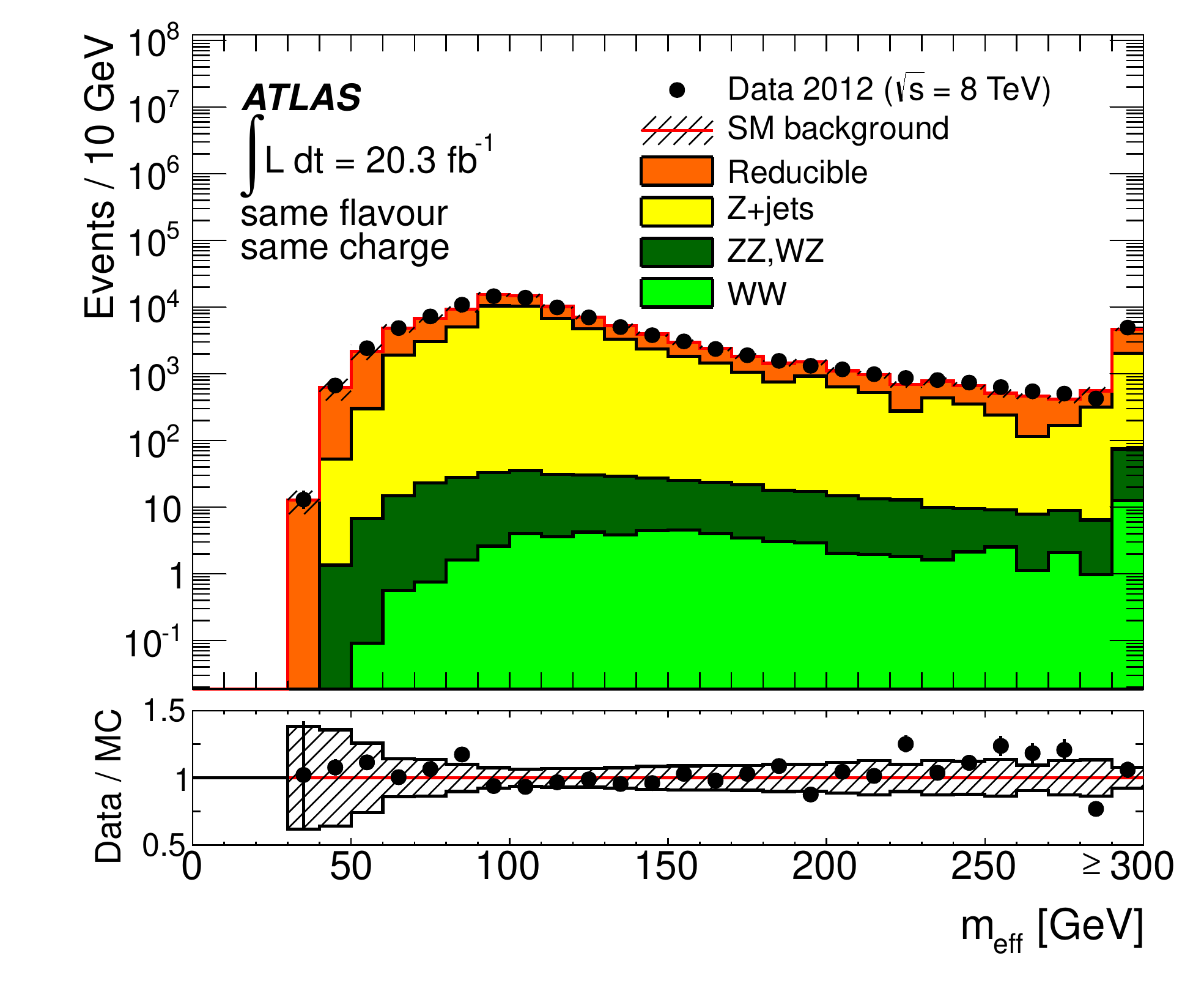}\includegraphics[width=0.5\textwidth]{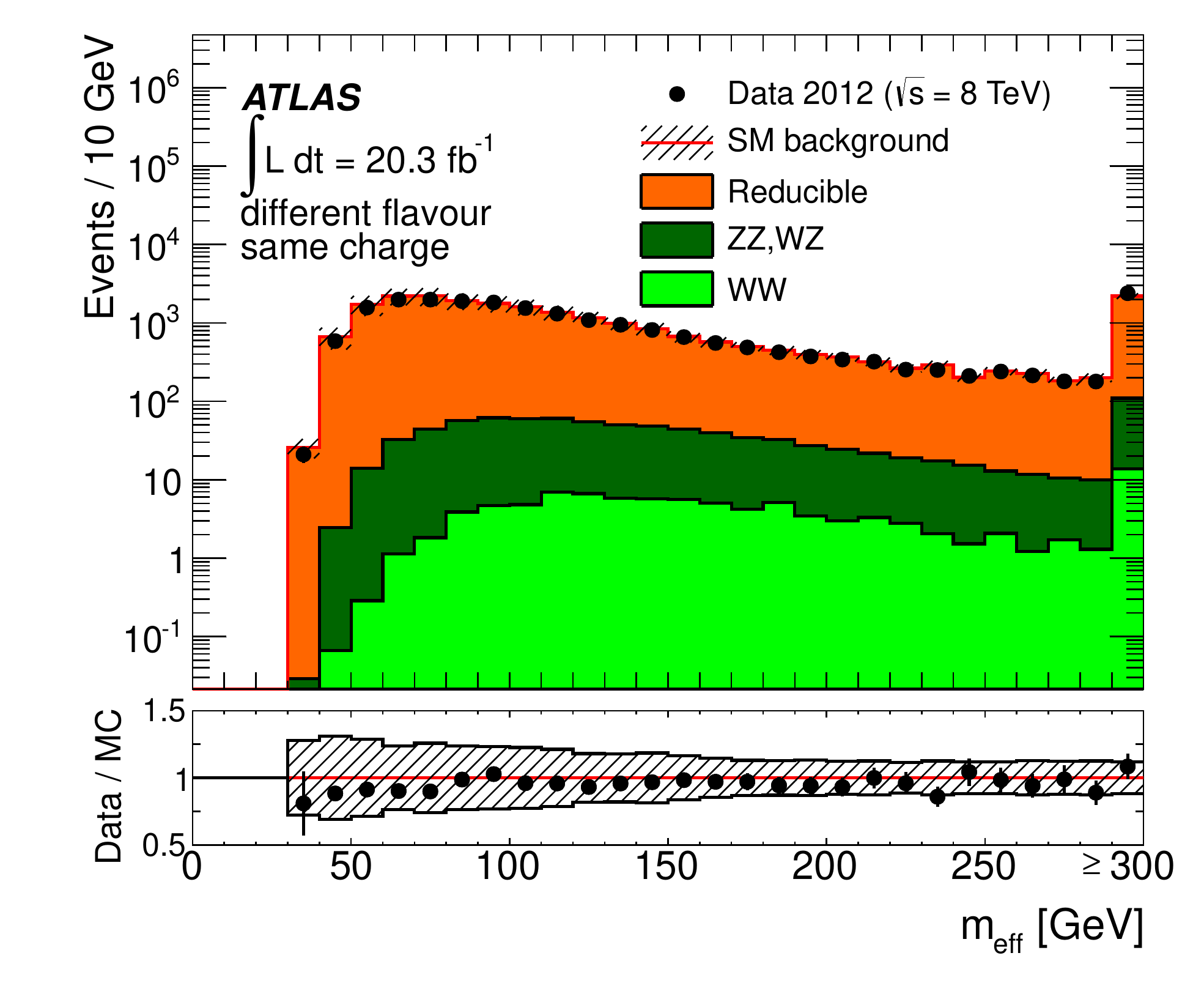}
\includegraphics[width=0.5\textwidth]{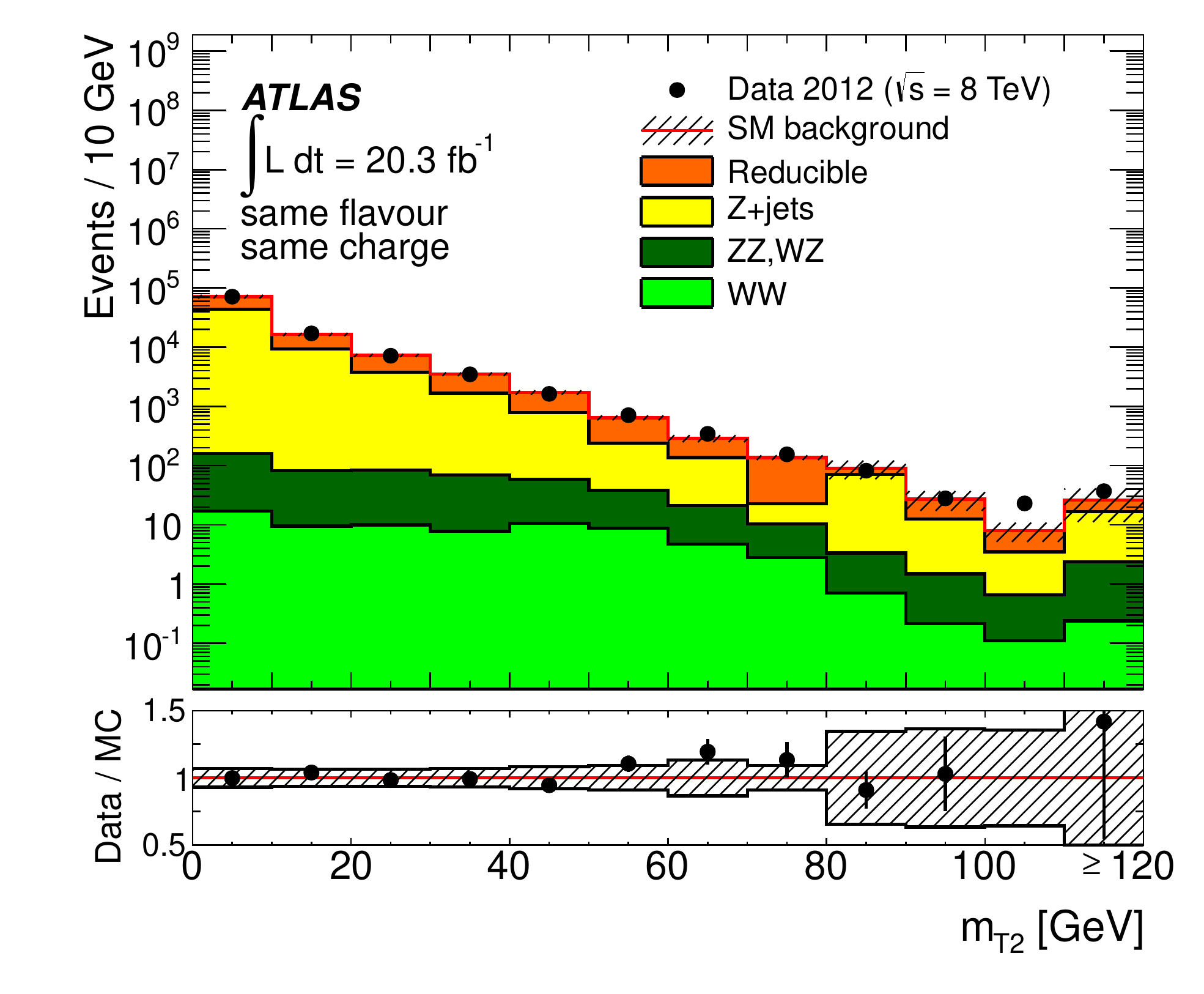}\includegraphics[width=0.5\textwidth]{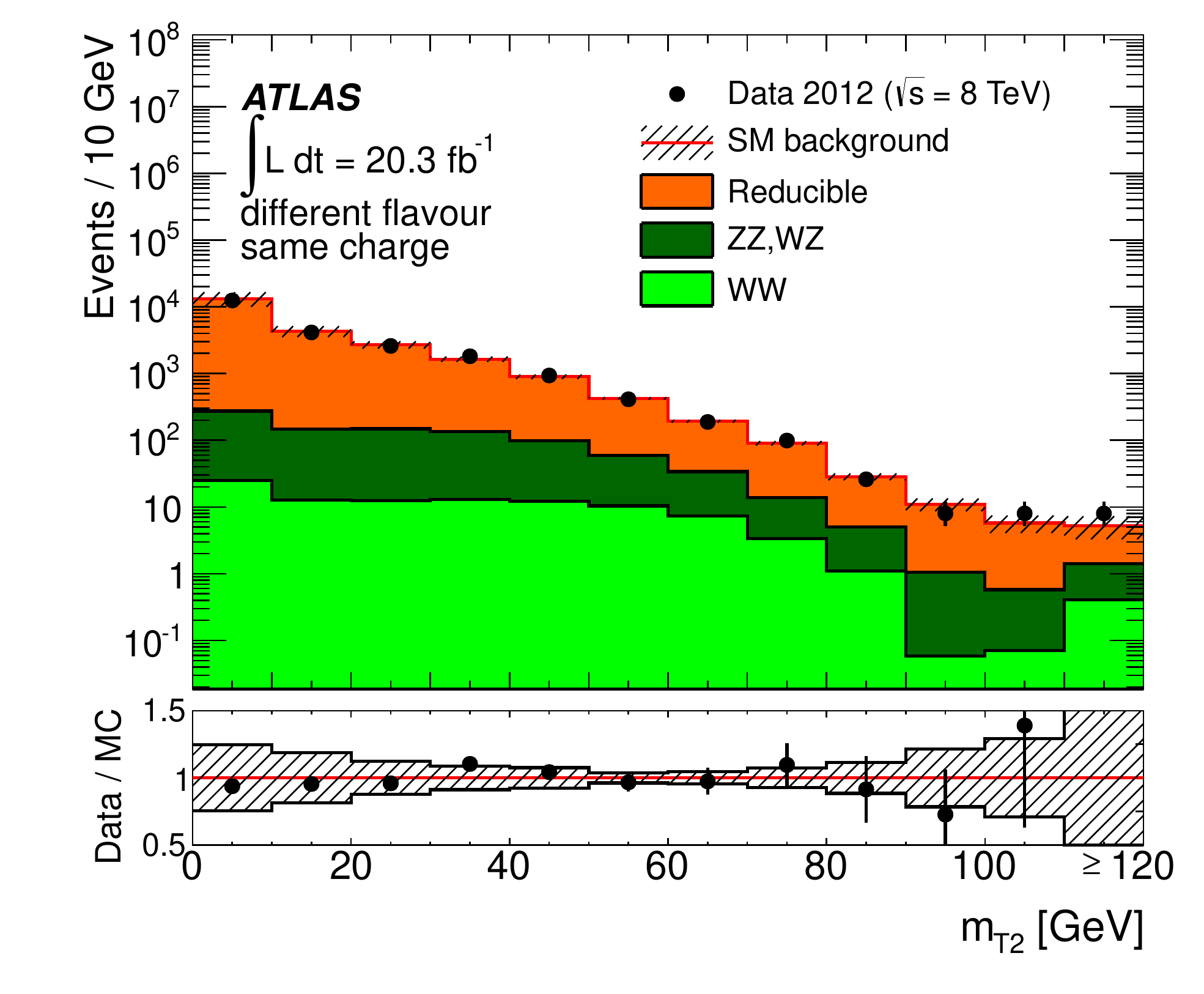}
\end{center}
\caption{\label{fig:lepmt2:fakes} Distributions of $m_\textup{eff}$ (top) and $m_\textup{T2}$ (bottom), for SF (left) and DF (right) same-charge lepton pairs, after the preselection requirements described in section~\ref{sec:preselections}. The components labelled ``Reducible'' correspond to the fake and non-prompt lepton backgrounds and are estimated from data as described
 in section~\ref{sec:fakes}.
The other SM backgrounds processes which are expected to contribute events with two real leptons are shown and are estimated from MC simulation. The reconstructed leptons are required to match with a generator-level lepton in order to avoid any double counting of the total fake and non-prompt lepton contribution. The bands represent the total uncertainty. }
\end{figure}

\subsection[Leptonic $m_\textup{T2}$ analysis]{Leptonic \boldmath$m_\textup{T2}$ analysis}
\label{sec:backgrounds-lep}

The dominant SM background contributions in the SRs are $t\bar{t}$ and $WW$ decays. Other diboson processes also expected to contribute significantly are: $WZ$ in its 3-lepton decay mode and $ZZ$ decaying to two leptons and two neutrinos. 
A single dedicated CR is defined for each of these backgrounds (CRX$_\textup{L}$, where X$=$T,W,Z for the $t\bar{t}$, $WW$ and other diboson productions respectively).
Predictions in all SRs make use of the three common CRs.
This choice was optimised considering the background purity and the available sample size.

The validity of the combined background estimate is tested using a set of four validation regions (VR$^\textup{X}_\textup{L}$, where X describes the specific selection under validation). The definitions of the CRs and VRs are given in table~\ref{tab:lepmT2:CVR}.
The validity of the \ttbar\ background prediction for different jet selections is checked in VR$^\textup{100}_\textup{L}$ and VR$^\textup{110}_\textup{L}$.

Additional SM processes yielding two isolated leptons and large \EtMiss (Higgs, $Wt$, $Z/\gamma^*\rightarrow\ell\ell$+jets and $t \bar tV $) and providing a sub-dominant contribution to the SRs  are determined from MC simulation. The fake and non-prompt lepton background is a small contribution (less than 10\% of the total background). 
The composition before and after the likelihood fit is given in table~\ref{tab:lepMT2:CRcomp} for the CRs and table~\ref{tab:lepMT2:VRcomp} for the VRs. In these (and all subsequent) composition tables the quoted uncertainty includes all the sources of statistical and systematic uncertainty considered (see section~\ref{sec:systematics}.). The purity of the CRs is improved by exploiting flavour information and selecting either DF or SF pairs depending on the process being considered. The normalisation factors derived are, however, applied to all the events in a given process (both DF and SF). Checks were performed to demonstrate that the normalisation factors are not flavour-dependent. Good agreement is found between data and the SM prediction before and after the fit, leading to normalisation factors compatible with unity. The normalisations of the $t\bar{t}$, $WW$ and $WZ,ZZ$ backgrounds as obtained from the fit are $0.91~\pm~0.07$, $1.27\pm0.24$ and $0.85\pm0.16$ respectively.

The number of expected signal events in the CRs was investigated for each signal model considered.
The signal contamination in CRT$_\textup{L}$ and CRW$_\textup{L}$ is negligible, with the exception of signal models with top squark masses close to the top-quark mass. In this case, the signal contamination can be as high as $20\%$ in CRT$_\textup{L}$ and up to $100\%$ in CRW$_\textup{L}$. The signal contamination in CRZ$_\textup{L}$ is typically less than $10\%$, with a few exceptions; for signal models with top-squark masses below $250$~GeV, the contamination is closer to $30\%$, and for signal models with small $\Delta m(\tone, \chipm)$ the signal contamination is as high as 100\%.
The same CRs can be kept also for these signal models, despite the high signal contamination, since the expected yields in the SRs would be large enough for these signal models to be excluded even in the hypothesis of null expected background.
The signal contamination in the VRs can be up to $\sim100\%$ for signal models with top-quark-like kinematics and becomes negligible when considering models with increasing top-squark masses.

\begin{table}[!ht]
\caption{Definitions of the CRs and VRs in the leptonic $m_\textup{T2}$ analysis: CRT$_\textup{L}$ (used to constrain $t\bar{t}$), CRW$_\textup{L}$ (used to constrain $WW$), CRZ$_\textup{L}$ (used to constrain $WZ$ and $ZZ$), VR$^\textup{DF}_\textup{L}$ (validation region for DF), VR$^\textup{SF}_\textup{L}$ (validation region for SF), VR$^\textup{110}_\textup{L}$  (validation region for L110 jet requirements) and VR$^{100}_\textup{L}$ (validation region for L100 jet requirements).
 \label{tab:lepmT2:CVR}}
\begin{center}

\begin{tabular}{l|ccccccc}
\hline
Selection Variable & CRT$_\textup{L}$ & CRW$_\textup{L}$ & CRZ$_\textup{L}$ & VR$^\textup{DF}_\textup{L}$ & VR$^\textup{SF}_\textup{L}$ & VR$^{110}_\textup{L}$ & VR$^{100}_\textup{L}$ \\
 \hline
 \hline
Flavour &DF & DF & SF & DF &  SF & DF & DF\\
$m_{\ell\ell}$ [GeV] & - &  -  & 71--111 & - & $<71$ or $>111$ & - & - \\
$m_\textup{T2}$ [GeV] & 40--80 & -40-80 & $>90$ & 80--90 & 80--90 & 40--80 & 40--80  \\
$p_{\rm Tb}^{ll}$ [GeV] & $>30$ & $<15$ & - & - & - & $>30$ & $>30$\\
$\Delta\phi_j$ [rad] & $>1.0$ & $>1.0$ & $>1.0$ & $>1.0$ & $>1.0$ & $>1.0$ & $>1.0$ \\
$\Delta\phi_\mathrm{b}$ [rad] & $<1.5$ & $<1.5$ & $<1.5$& $<1.5$& $<1.5$& $<1.5$& $<1.5$\\
Leading jet $p_\mathrm{T}$ [GeV] & - & - & - &  - & - & $>20$ & $>100$ \\
Second leading jet $p_\mathrm{T}$ [GeV] & - & - & - & - & - &  $>20$ &  $>50$\\
\hline
\end{tabular}
\end{center}
\end{table}

\begin{table}[b]
\begin{center}
\caption{Background fit results for the three CRs in the leptonic $m_\textup{T2}$ analysis. The nominal expectations from MC simulation
are given for comparison for those backgrounds ($t\bar{t}$, $WW$, $WZ$ and $ZZ$) which are normalised to data. Combined statistical and systematic uncertainties are given. Events with fake or non-prompt leptons are estimated with the data-driven technique described in section~\ref{sec:fakes}. The observed events and the total (constrained) background are the same by construction. Entries marked - - indicate a negligible background contribution. Uncertainties on the predicted background event yields are quoted as symmetric except where the negative error reaches down to zero predicted events, in which case the negative error is truncated. }
\label{tab:lepMT2:CRcomp}
\setlength{\tabcolsep}{0.0pc}
{\small
\begin{tabular*}{\textwidth}{@{\extracolsep{\fill}}lrrr}
\noalign{\smallskip}\hline\noalign{\smallskip}
{\bf  Channel}           & CRT$_\textup{L}$            & CRW$_\textup{L}$            & CRZ$_\textup{L}$             \\[-0.05cm]
\noalign{\smallskip}\hline\noalign{\smallskip}
Observed events          & $12158$              & $913$              & $174$                    \\
\noalign{\smallskip}\hline\noalign{\smallskip}
Total (constrained) bkg events         & $12158 \pm 110$          & $913 \pm 30$          & $174 \pm 13$              \\
\noalign{\smallskip}\hline\noalign{\smallskip}
        Fit output, \ttbar events         & $8600 \pm 400$          & $136 \pm 24$          & $27 \pm 6$              \\
        Fit output, $WW$ events         & $1600 \pm 400$          & $630 \pm 50$          & $14 \pm 4$              \\
        Fit output, $WZ$, $ZZ$ events         &  $64 \pm 14$          & $14 \pm 5$          & $112 \pm 19$               \\
\noalign{\smallskip}\hline\noalign{\smallskip}
Total expected bkg events              & $12700 \pm 700$          & $800 \pm 90$          & $190 \pm 20$              \\
\noalign{\smallskip}\hline\noalign{\smallskip}
        Fit input, expected \ttbar events         & $9500 \pm 600$          & $150 \pm 25$          & $30 \pm 7$              \\
        Fit input, expected $WW$ events         & $1260\pm 110$          & $490 \pm 80$          & $10.7\pm 2.5$              \\
        Fit input, expected $WZ$, $ZZ$ events         &$76 \pm 12$          & $17 \pm 4$          & $132 \pm 11$               \\
        Expected $Z/ \gamma^{*} \rightarrow \ell \ell$ events         & $9_{-9}^{+11}$          & $1.5_{-1.5}^{+2.2}$          & $19 \pm 8$              \\
        Expected \ttbar$V$ events         & $10.8 \pm 3.4$          & $0.08 \pm 0.04$          & $0.64 \pm 0.21$              \\
        Expected $Wt$ events         & $1070 \pm 90$          & $35 \pm 7$          & $1.6\pm 1.1$              \\
        Expected Higgs boson events         & $67 \pm 21$          & $20 \pm 6$          & $0.08 \pm 0.04$              \\
        Expected events with fake and non-prompt leptons         & $740 \pm 90$          & $81 \pm 16$          & - -             \\
\noalign{\smallskip}\hline\noalign{\smallskip}
\end{tabular*}
%%%%
}
\end{center}
\end{table}

\begin{table}
\caption{Background fit results for the four VRs in the leptonic $m_\textup{T2}$ analysis. Combined statistical and systematic uncertainties are given. Events with fake or non-prompt leptons are estimated with the data-driven technique described in section~\ref{sec:fakes}. The observed events and the total (constrained) background are the same in the CRs by construction; this is not the case for the VRs, where the consistency between these event yields is the test of the background model. Entries marked - - indicate a negligible background contribution. Uncertainties on the predicted background event yields are quoted as symmetric except where the negative error reaches down to zero predicted events, in which case the negative error is truncated. }
\label{tab:lepMT2:VRcomp}
\begin{center}
\setlength{\tabcolsep}{0.0pc}
{\small
\begin{tabular*}{\textwidth}{@{\extracolsep{\fill}}lrrrr}
\noalign{\smallskip}\hline\noalign{\smallskip}
{\bf  Channel}           & VR$_\textup{L}^\textup{SF}$            & VR$_\textup{L}^\textup{DF}$           & VR$_\textup{L}^{110}$            & VR$_\textup{L}^{100}$              \\[-0.05cm]
\noalign{\smallskip}\hline\noalign{\smallskip}
Observed events          & $494$              & $622$              & $8162$              & $1370$                    \\
\noalign{\smallskip}\hline\noalign{\smallskip}
Total  bkg events         & $500 \pm 40$          & $620 \pm 50$          & $7800 \pm 400$          & $1390 \pm 110$              \\
\noalign{\smallskip}\hline\noalign{\smallskip}
        Fit output, \ttbar events         & $338 \pm 19$          & $430 \pm 29$          & $6800 \pm 400$          & $1230 \pm 110$              \\
        Fit output, $WW$ events         & $97 \pm 22$          & $121 \pm 27$          & $290 \pm 70$          & $38 \pm 15$              \\
        Fit output, $WZ$, $ZZ$ events         & $5.8 \pm 1.1$          & $2.2 \pm 1.4$          & $13.5 \pm 3.2$          & $1.5 \pm 1.2$ \\
        Expected $Z/ \gamma^{*} \rightarrow \ell \ell$ events         & $4_{-4}^{+5}$          & - -          & $3_{-3}^{+5}$          & $1_{-1}^{+1}$              \\
        Expected \ttbar$V$ events         & $0.48 \pm 0.18$          & $0.80 \pm 0.27$          & $10.1 \pm 3.1$          & $4.1 \pm 1.3$              \\
        Expected $Wt$ events         & $39 \pm 8$          & $60 \pm 10$          & $430 \pm 50$          & $62 \pm 8$              \\
        Expected Higgs boson events         & $0.39 \pm 0.16$          & $0.55 \pm 0.20$          & $14 \pm 4$          & $1.7 \pm 0.6$              \\
        Expected events with fake and non-prompt leptons         & $10.5 \pm 3.5$          & $13 \pm 4$          & $275 \pm 33$          & $45 \pm 7$              \\
%%     \\
\noalign{\smallskip}\hline\noalign{\smallskip}
\end{tabular*}
%%%%
}
\end{center}
\end{table}

\begin{figure}[p]
\begin{center}
\includegraphics[width=0.8\textwidth]{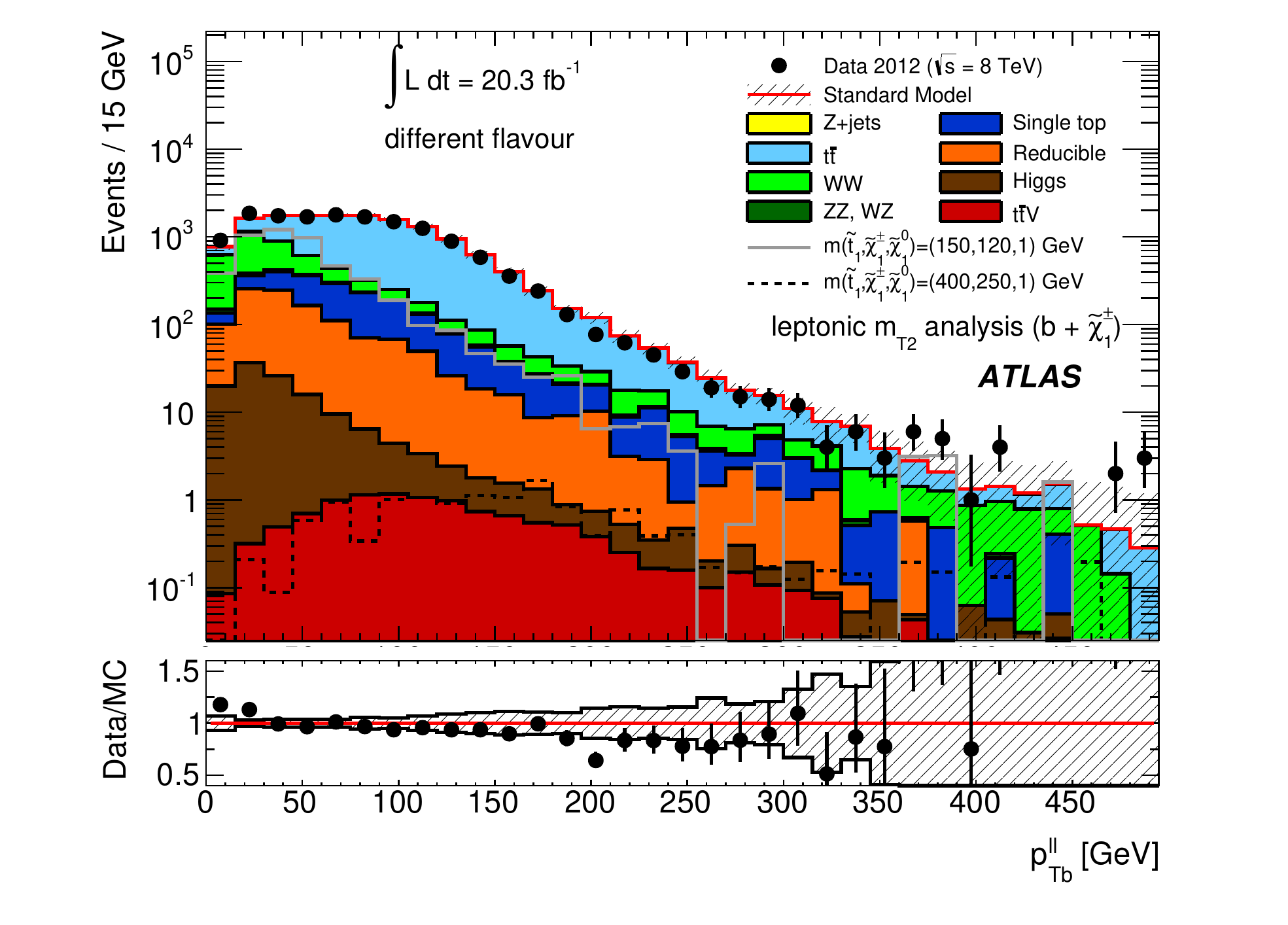}
\includegraphics[width=0.8\textwidth]{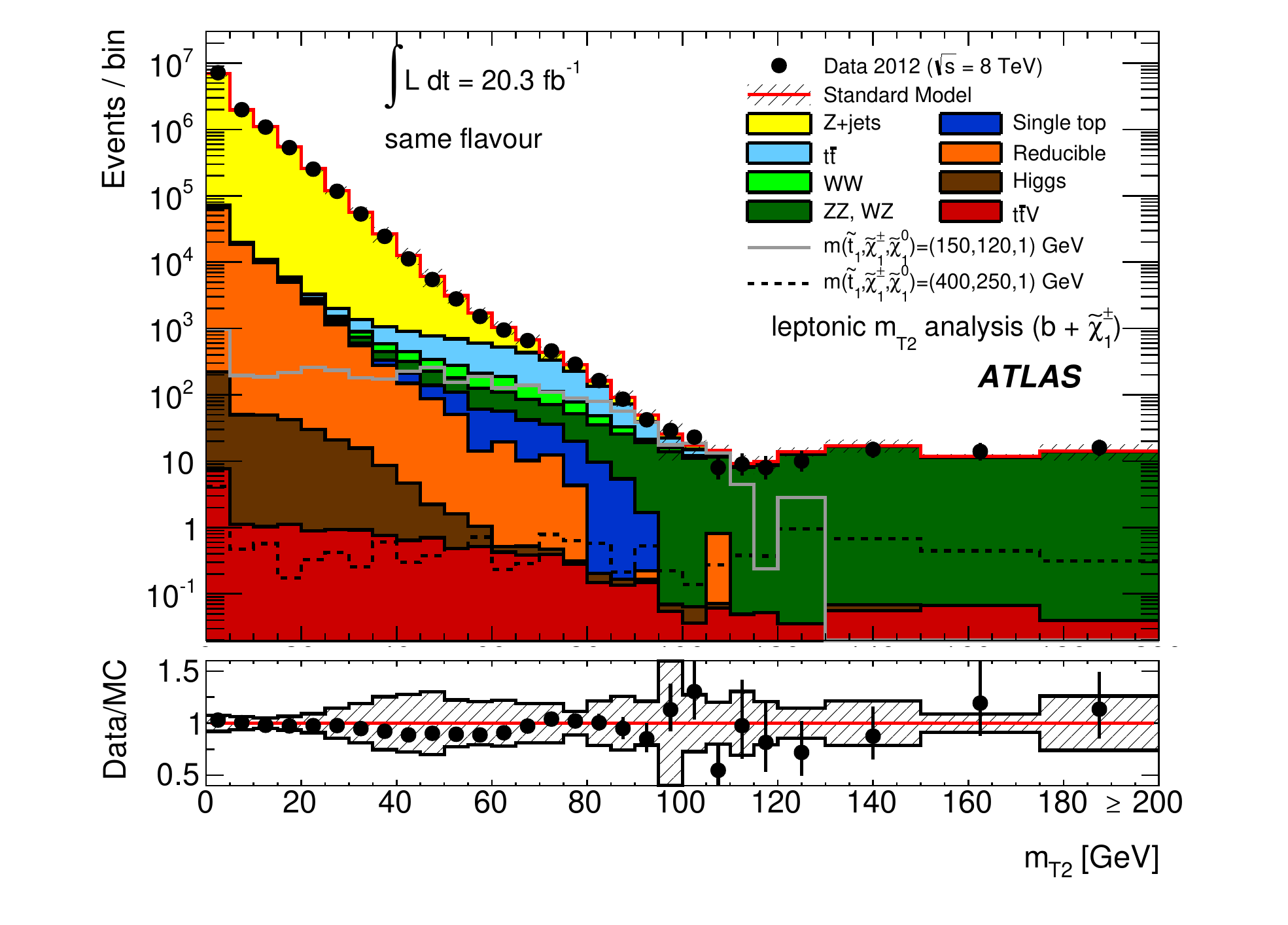}
\end{center}
\caption{Top: distribution of $p^{\ell\ell}_{\mathrm{Tb}}$ for  DF events with $40 < \mttwo < 80 \GeV$, $\Delta \phi_j > 1.0$ rad and $\Delta \phi_b < 1.5$~rad.
Bottom: Distribution of \mttwo for  SF events with a dilepton invariant mass in the 71--111~\GeV$\,$ range, $\Delta \phi > 1.0$~rad and $\Delta \phi_b < 1.5$~rad. 
The contributions from all SM backgrounds are shown as a histogram stack; the bands represent the total uncertainty. The components labelled ``Reducible'' correspond to the fake and non-prompt lepton backgrounds and are estimated from data as described in section~\ref{sec:fakes}; the other backgrounds are estimated from MC simulation. The expected distribution for two signal models is also shown.
The full line corresponds to a model with $m(\tone)=150$~GeV, $m(\chipm)=120$~GeV~ and $m(\neut)=1$~GeV; the dashed line to a model with $m(\tone)=400$~GeV, $m(\chipm)=250$~GeV and $m(\neut)=1$~GeV.} 

\label{fig:lep:CR}
\end{figure}
Figure~\ref{fig:lep:CR} (top) shows the $p^{\ell\ell}_{\mathrm{Tb}}$ distribution for DF events with $40<m_\textup{T2} < 80$~GeV, $\Delta \phi > 1.0$ and $\Delta \phi_b < 1.5$. The range   $p^{\ell\ell}_{\mathrm{Tb}}< 15$~GeV corresponds to the CRW$_\textup{L}$ while the events with  $p^{\ell\ell}_{\mathrm{Tb}} > 30$ GeV are those entering in CRT$_\textup{L}$.
Figure~\ref{fig:lep:CR} (bottom) shows the $m_\textup{T2}$ distribution for SF events with $\Delta \phi > 1.0$ and $\Delta \phi_b < 1.5$ and $m_{\ell\ell}$ within 20 GeV of the $Z$ boson mass. The events with $m_\textup{T2} > 90$~GeV in this figure are those entering CRZ$_\textup{L}$.

\clearpage

\subsection[Hadronic $m_\textup{T2}$ analysis]{Hadronic \boldmath$m_\textup{T2}$ analysis}
\label{sec:backgrounds-had}

Top-quark pair and single-top ($Wt$-Channel) production contribute significantly to the background event yields in the SR for this analysis. Simulation shows that 49\% of background events in the SR are from top-quark pair production and 37\% are from $Wt$.
The next most significant SM background contributions are those arising from fake or non-prompt leptons. The remainder of the background is composed of $Z/\gamma^*$+jets and $WW$ events. The contributions from other diboson ($WZ$ and $ZZ$), $t\bar{t}V$ and Higgs processes are negligible, and are estimated using the MC simulation.

The CRs are defined for the combined $t\bar{t}$ and $Wt$ process, and $Z/\gamma^*(\rightarrow ee,\mu\mu)+$jets backgrounds (the $Z/\gamma^*(\rightarrow \tau\tau)+$jets contribution is fixed at the MC expectation). The contribution from $Wt$ in the SR is dominated by its NLO contributions (which can be interpreted as top-pair production, followed by decay of one of the top-quarks).
These CRs are referred to as CRX$_\textup{H}$, where X$=$T,Z for the $(t\bar{t},Wt)$ and  $Z/\gamma^*(\rightarrow ee,\mu\mu)+$jet backgrounds respectively. The validity of the combined estimate of the $Wt$ and $t\bar{t}$ backgrounds is tested using a validation region for the top-quark background (VRT$_\textup{H}$). The definitions of these regions are given in table~\ref{tab:hadmT2:CVR}, and their composition before and after the likelihood fit described in section~\ref{sec:background:fit} is given in table~\ref{tab:hadMT2:CVRcomp}. Good agreement is found between data and SM prediction before and after the fit, leading to normalisations consistent with one: $0.93\pm0.32$ for the ($t\bar{t}$,$Wt$) and $1.5\pm0.5$ for the $Z/\gamma^*+$jets backgrounds.

The signal contamination in CRZ$_\textup{H}$ is negligible, whilst in CRT$_\textup{H}$ it is of order $10\%$ ($16\%$) for models with a 300~GeV top squark and a 150~GeV (100~GeV) chargino, for neutralino masses below 100~GeV, which the region where H160 is sensitive. The signal contamination in VRT$_\textup{H}$ is much higher ($\sim30\%$) in the same mass-space.

\begin{table}[!htp]

\caption{Definitions of the CRs and VR in the hadronic $m_\textup{T2}$ analysis: CRT$_\textup{H}$ (used to constrain $t\bar{t}$ and $Wt$), CRZ$_\textup{H}$ (used to constrain $Z/\gamma^*$+jets decays to $ee$ and $\mu\mu$) and VRT$_\textup{H}$ (validation region for $t\bar{t}$ and $Wt$). \label{tab:hadmT2:CVR}}
\begin{center}
\begin{tabular}{l|ccc}
\hline
Selection Variable & CRT$_\textup{H}$ & CRZ$_\textup{H}$ & VRT$_\textup{H}$ \\
 \hline
 \hline
 Flavour & any & SF & any \\
$b$-jets & $=1$ & $=2$ & $=2$ \\
leading lepton $p_\mathrm{T}$ [GeV]  & $<60$ & $>60$ & $>60$ \\
$m_{\ell\ell}$ (SF events only) [GeV] & - & $81--101$ & $<81$ or $>101$ \\
$m_\textup{T2}$ [GeV] & $<90$ & $<90$ & $<90$ \\
$m^{b-\textup{jet}}_\textup{T2}$ [GeV] & $>160$ & $>160$ & $>160$ \\
\hline
\end{tabular}
\end{center}
\end{table}

\begin{table}
\caption{Background fit results for the two CRs and VR region in the hadronic $m_\textup{T2}$ analysis. The nominal expectations from MC simulation
are given for comparison for those backgrounds ($t\bar{t}$, $Wt$ and $Z/\gamma^* (\rightarrow ee,\mumu) + $jets production) which are normalised to data. Combined statistical and systematic uncertainties are given. Events with fake or non-prompt leptons are estimated with the data-driven technique described in section~\ref{sec:fakes}. The observed events and the total (constrained) background are the same in the CRs by construction; this is not the case for the VR, where the consistency between these event yields is the test of the background model. Uncertainties on the predicted background event yields are quoted as symmetric except where the negative error reaches down to zero predicted events, in which case the negative error is truncated. }
\label{tab:hadMT2:CVRcomp}

\begin{center}
\setlength{\tabcolsep}{0.0pc}
{\small
\begin{tabular*}{\textwidth}{@{\extracolsep{\fill}}lrrr}
\noalign{\smallskip}\hline\noalign{\smallskip}
{\bf  Channel}           &  CRT$_\textup{H}$            &  CRZ$_\textup{H}$            &  VRT$_\textup{H}$              \\[-0.05cm]
\noalign{\smallskip}\hline\noalign{\smallskip}
Observed events          & $315$              & $156$              & $112$                    \\
\noalign{\smallskip}\hline\noalign{\smallskip}
 Total (constrained) bkg events         & $315 \pm 18$          & $156 \pm 13$          & $110 \pm 50$              \\
\noalign{\smallskip}\hline\noalign{\smallskip}
        Fit output, $t\bar{t},Wt$ events         & $256 \pm 27$          & $4 \pm 4$          & $70 \pm 40$              \\
        Fit output, $Z/\gamma^*\rightarrow ee,\mu\mu+$jets events         & $0.9_{-0.9}^{+1.1}$          & $147 \pm 13$          & $20 \pm 8$              \\
 \noalign{\smallskip}\hline\noalign{\smallskip}
 Total expected bkg events             & $335 \pm 90$          & $110 \pm 36$          & $110 \pm 60$              \\                                                           
\noalign{\smallskip}\hline\noalign{\smallskip}
         Fit input, expected $t\bar{t},Wt$ events         & $280 \pm 90$          & $5 \pm 5$          & $80 \pm 60$              \\
         Fit input, expected $Z/\gamma^*\rightarrow ee,\mu\mu+$jets events          & $0.6_{-0.6}^{+0.7}$          & $100 \pm 34$          & $13.8 \pm 2.4$              \\
         Expected $WW$ events         & $3_{-3}^{+4}$          & $0.07_{-0.07}^{+0.14}$          & $1_{-1}^{+3}$              \\
         Expected $t\bar{t}V$ events         & $2.3 \pm 0.8$          & $1.5 \pm 0.5$          & $2.3 \pm 0.7$              \\
         Expected $WZ$, $ZZ$ events         & $0.40 \pm 0.16$          & $0.06_{-0.06}^{+0.32}$          & $0.10_{-0.10}^{+0.15}$              \\
         Expected $Z/\gamma^*\rightarrow \tau\tau+$jets events         & $23 \pm 17$          & $0.14 \pm 0.09$          & $2.15 \pm 0.28$              \\
        Expected events with fake and non-prompt leptons         & $29.4 \pm 1.7$          & $0.36 \pm 0.24$          & $12.8 \pm 1.2$              \\
	Expected Higgs boson events          & $0.35 \pm 0.05$          & $2.06 \pm 0.30$          & $0.50 \pm 0.06$              \\
\noalign{\smallskip}\hline\noalign{\smallskip}
\end{tabular*}
}
\end{center}
\end{table}

Figure~\ref{fig:mt2bb:CRT1} shows the $m^{b-\textup{jet}}_\textup{T2}$ distribution for events with one $b$-jet (using the highest $p_\mathrm{T}$ jet which is not a $b$-jet with the single $b$-jet in the calculation of $m^{b-\textup{jet}}_\textup{T2}$), $m_\textup{T2}<90$~GeV and leading lepton $p_\mathrm{T}<60$~GeV. The events with $m^\textup{b-jet}_\textup{T2}>160$~GeV in the figure are those entering CRT$_\textup{H}$. The data are in agreement with the background expectation across the distribution.

\begin{figure}[htb!]
\begin{center}
\includegraphics[width=0.8\textwidth]{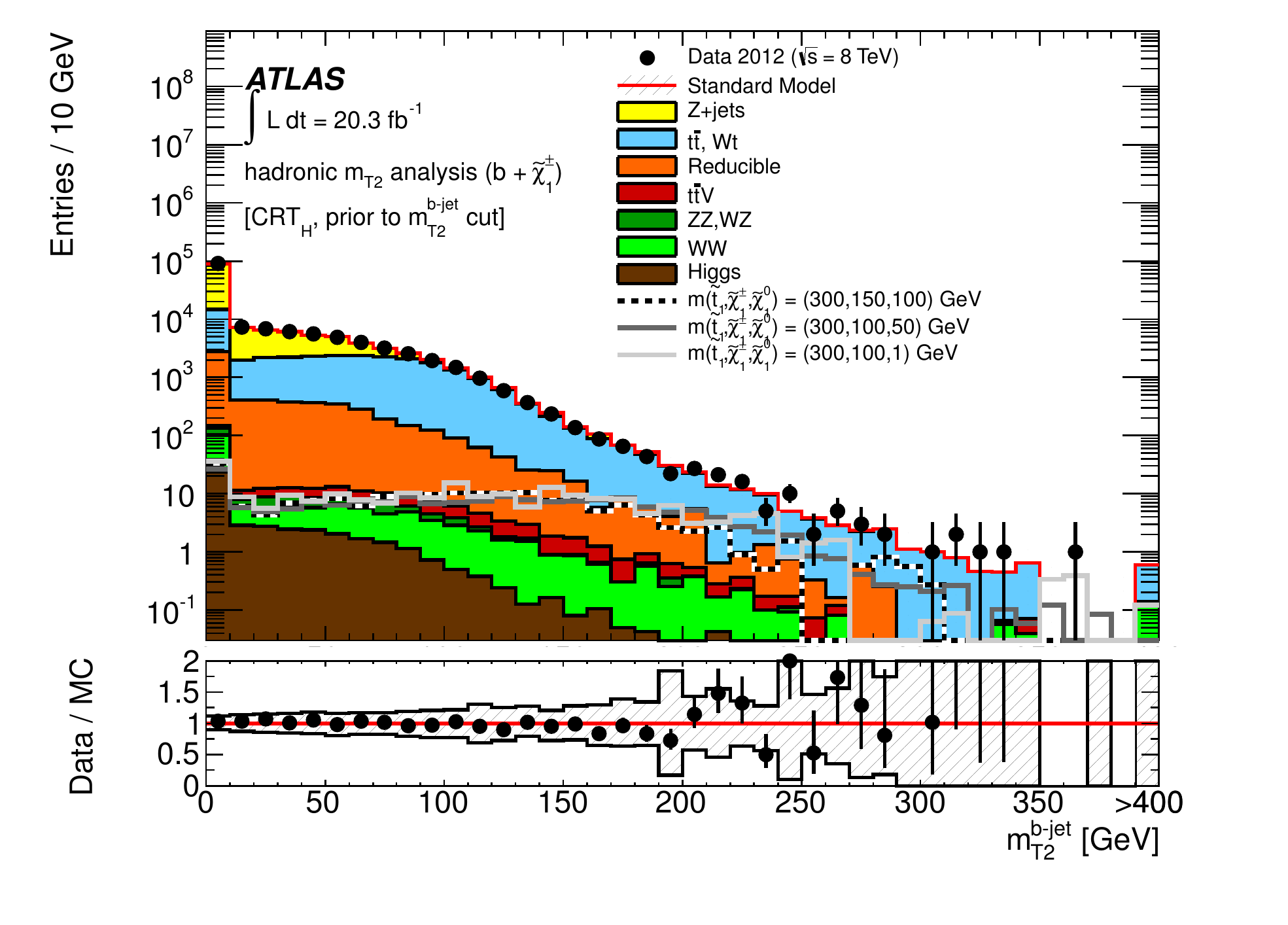}
\end{center}
\caption{Distribution of $m^{b-\textup{jet}}_\textup{T2}$ for events with 1 $b$-jet and all other CRT$_\textup{H}$ cuts, except that on $m^{b-\textup{jet}}_\textup{T2}$ itself.
The contributions from all SM backgrounds are 
shown as a histogram stack; the bands represent the total uncertainty. The component
labelled ``Reducible" corresponds to the fake and non-prompt lepton background and is estimated from data as described in section~\ref{sec:fakes}; the 
other backgrounds are estimated from MC samples normalised to the luminosity of the data and their respective cross-sections. The expected distribution 
for three signal models is also shown. The dotted line corresponds to a model
with $m(\tone)=300$~GeV, $m(\chipm)=150$~GeV and $m(\neut)=100$~GeV; the full line corresponds to a model
with $m(\tone)=300$~GeV, $m(\chipm)=100$~GeV and $m(\neut)=50$~GeV; the dashed
line to a model with $m(\tone)=300$~GeV, $m(\chipm)=100$~GeV and 
$m(\neut)=1$~GeV. The last bin includes the histogram overflow. }
\label{fig:mt2bb:CRT1}
\end{figure}

\subsection{Multivariate analysis}
\label{sec:backgrounds-mva}

In this analysis, the dominant SM background processes are top-quark pair production and diboson production. The $Z/\gamma^*+$jets contribution, relevant only for the SF channel, is strongly suppressed by the BDTG requirement. The CRs are defined for $t\bar{t}$ (table~\ref{tab:topreg}) in regions mutually exclusive to the SRs, using BDTG intervals much more populated with $t\bar{t}$ events, while all other SM background with two isolated leptons are small and evaluated using MC simulation. The fake and non-prompt lepton background is estimated using the method described in section~\ref{sec:fakes}. In addition to the application of all non-BDTG SR cuts, the following selections are applied in the CRs: $m_\textup{T2}>90$~GeV and, in SF events, $m_{\ell\ell}$ which must be less than 61~GeV or greater than 121~GeV.
The composition before and after the likelihood fit is given in tables \ref{tab:mva_CRDF} and \ref{tab:mva_CRSF} for the DF and SF CRs, respectively. The corresponding CR for the DF (SF) SR labelled N is denoted CRT$^\textup{DF(SF)}_\textup{MN}$. The normalisation factors derived in each CR for $t\bar{t}$ are consistent within one standard deviation ($1\sigma$) of the normalisation factor derived for $t\bar{t}$ in the leptonic-$m_\textup{T2}$ analysis.

\begin{table} [htb!]
\caption{Definitions of the CRs for the MVA analysis: the name of each CR is given in the first column and these have a one-to-one correspondence with the equivalently named SR. The middle column lists all selection cuts made, whilst the final column gives the BDTG range.}
\label{tab:topreg}
\begin{center}
\begin{tabular}{l|cc}
\hline
Control Region & Event Variable Selection [GeV] & BDTG range \\
\hline
$\textup{CRT}^\textup{DF}_\textup{M1}$ & C1, $m_\textup{T2}>90$    &  [$-1.00$, $-0.20$] \\
$\textup{CRT}^\textup{DF}_\textup{M2}$ & C1, $m_\textup{T2}>90$    &  [$-1.00$, $-0.30$] \\
$\textup{CRT}^\textup{DF}_\textup{M3}$    & C1, $m_\textup{T2}>90$    &[$-1.00$, 0.00] \\
$\textup{CRT}^\textup{DF}_\textup{M4}$ & C2, $m_\textup{T2}>90$    &  [$-1.00$, $-0.70$] \\
$\textup{CRT}^\textup{DF}_\textup{M5}$    & C4, $m_\textup{T2}>90$    & [$-1.00$, $-0.50$] \\
\hline
$\textup{CRT}^\textup{SF}_\textup{M1}$    & C1, $m_\textup{T2}>90$, $m_{\ell\ell}<61$ or $m_{\ell\ell}>121$   & [$-0.85$, $-0.75$] \\
$\textup{CRT}^\textup{SF}_\textup{M2}$    & C1, $m_\textup{T2}>90$, $m_{\ell\ell}<61$ or $m_{\ell\ell}>121$    &  [$-0.85$, $-0.20$] \\
$\textup{CRT}^\textup{SF}_\textup{M3}$ & C1, $m_\textup{T2}>90$, $m_{\ell\ell}<61$ or $m_{\ell\ell}>121$    & [$-0.95$, $-0.80$] \\
$\textup{CRT}^\textup{SF}_\textup{M4}$ & C3, $m_\textup{T2}>90$, $m_{\ell\ell}<61$ or $m_{\ell\ell}>121$    & [$-0.98$, $-0.78$] \\
\hline
\end{tabular} 
\end{center}
\end{table}

Figure \ref{fig:CRDF4} shows the BDTG distributions for data and MC simulation in $\textup{CRT}^\textup{DF}_\textup{M3}$ and $\textup{CRT}^\textup{SF}_\textup{M2}$.
The data are in agreement with the background expectations. The expected distribution for the signal point which was used to train the corresponding SR is also shown on each plot $m(\tilde{t}_1),m(\tilde{\chi}^0_1)=(300,50)$~GeV.

\begin{figure}[!hbp]
\begin{center}
\includegraphics[width=0.8\textwidth]{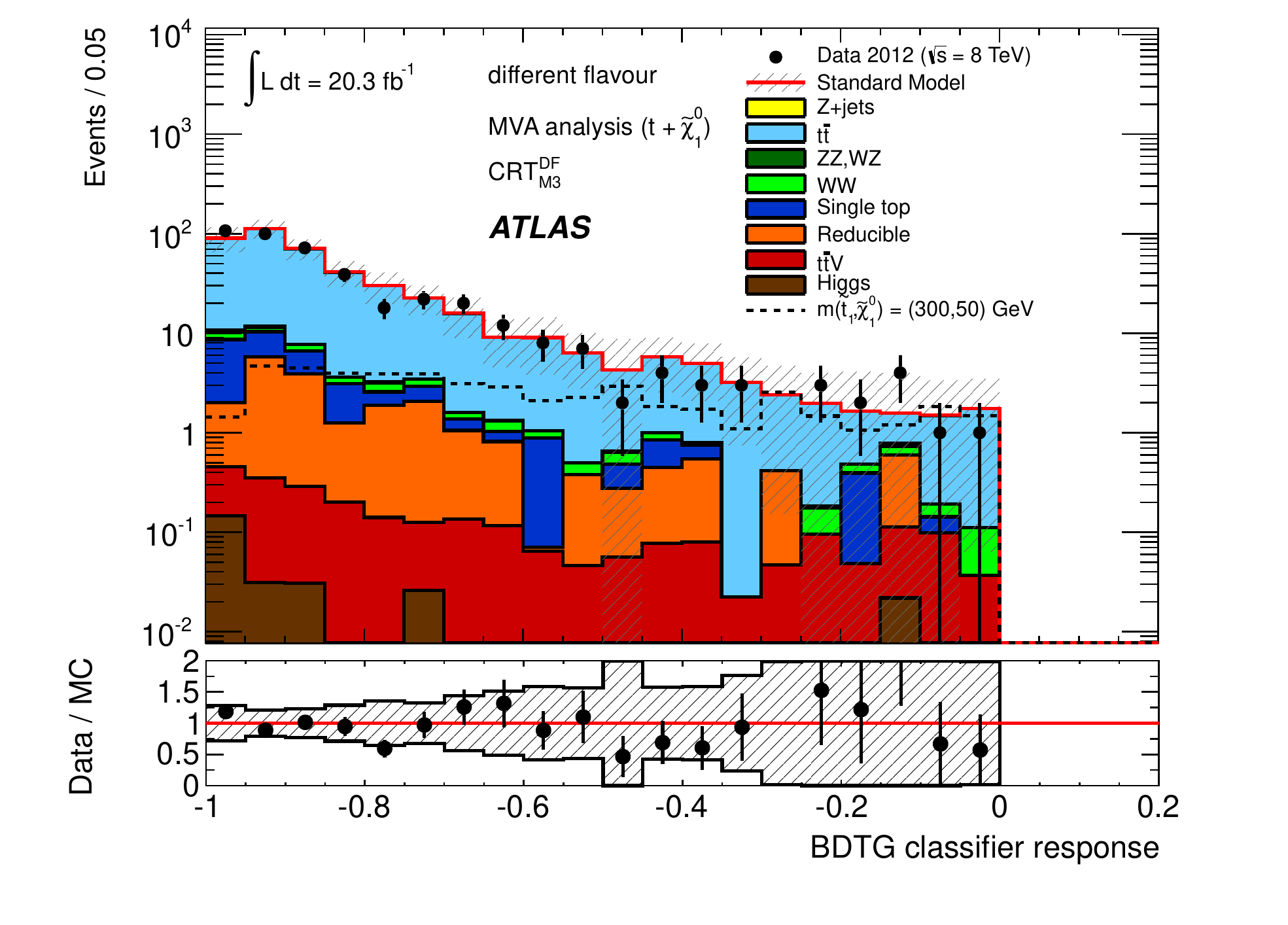}
\includegraphics[width=0.8\textwidth]{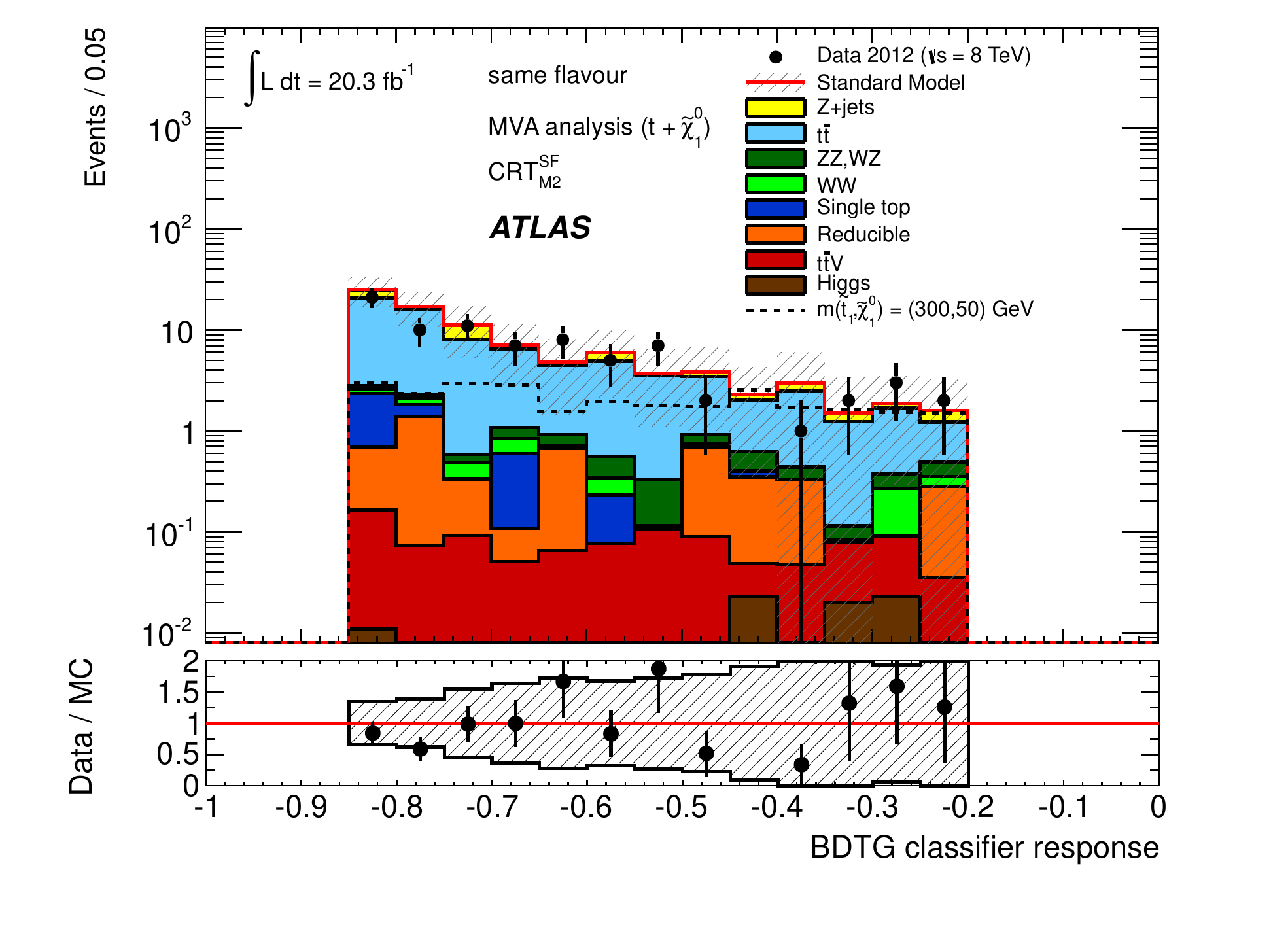}

\caption{{BDTG distributions of data and MC events in control regions $\textup{CRT}^\textup{DF}_\textup{M3}$  (top) and $\textup{CRT}^\textup{SF}_\textup{M2}$ (bottom). The contributions from all SM backgrounds are shown as a histogram stack. The bands represent the total uncertainty. The components labelled ``Reducible'' correspond to the fake and non-prompt lepton backgrounds and are estimated from data as described in section~\ref{sec:fakes}; the remaining backgrounds are estimated from MC samples normalised to the luminosity of the data. The expected distribution for the signal point which was used to train the corresponding SR is also shown on each plot (see text).}}
\label{fig:CRDF4}
\end{center}
\end{figure}

\begin{table}
\begin{center}
\caption{ \label{tab:mva_CRDF}
Background fit results for the DF CRs in the MVA analysis. The nominal expectations from MC simulation
are given for comparison for $t\bar{t}$,  which is normalised to data by the fit. Combined statistical and systematic uncertainties are given. Events with fake or non-prompt leptons are estimated with the data-driven technique described in section~\ref{sec:fakes}. The observed events and the total (constrained) background are the same in the CRs by construction. Uncertainties on the predicted background event yields are quoted as symmetric except where the negative error reaches down to zero predicted events, in which case the negative error is truncated. 
}
\setlength{\tabcolsep}{0.0pc}
{\scriptsize
\begin{tabular*}{\textwidth}{@{\extracolsep{\fill}}lccccc}
\noalign{\smallskip}\hline\noalign{\smallskip}

{\bf  Channel}          			 				& $\textup{CRT}^\textup{DF}_\textup{M1}$              	& $\textup{CRT}^\textup{DF}_\textup{M2}$              	& $\textup{CRT}^\textup{DF}_\textup{M3}$              	& $\textup{CRT}^\textup{DF}_\textup{M4}$             	& $\textup{CRT}^\textup{DF}_\textup{M5}$              	\\[-0.05cm]
\noalign{\smallskip}\hline\noalign{\smallskip}	
Observed events          								& $419$                 		& $410$                  		& $428$                    		& $368$                   			& $251$                    		\\
\noalign{\smallskip}\hline\noalign{\smallskip}
Total (constrained) bkg events         						 & $419 \pm 20$      		& $410 \pm 20$       		& $428 \pm 21$  		&$368 \pm 19$       			& $251 \pm 16$         \\
\noalign{\smallskip}\hline\noalign{\smallskip}
        Fit output, $t \bar t$ events         						& $369 \pm 23$       		& $363 \pm 23$     		& $379 \pm 24$       		& $325 \pm 22$      			& $214 \pm 19$          \\
 \noalign{\smallskip}\hline\noalign{\smallskip}
  Total expected bkg events             						&  $430 \pm 70$      		& $420 \pm 60$		& $440 \pm 70$ 		& $380 \pm 60$			&   $260 \pm 50$ \\                                               
  \noalign{\smallskip}\hline\noalign{\smallskip}
    
      Fit input, expected $t \bar t$         						& $380 \pm 60$       		& $375 \pm $  60    		& $390 \pm 70$       			& $340 \pm 50$     			& $220 \pm 40$          \\

        Expected $t \bar t V$ events   		 				& $2.7 \pm 0.8$            	& $2.2 \pm 0.7$             	& $2.4 \pm 0.7$            	& $2.7 \pm 0.8$             	 	 & $1.9 \pm 0.6$              \\
         Expected $Wt$ events     				   	 		& $20 \pm 5$           		& $19 \pm 5$           		& $20 \pm 5$            		& $16 \pm 5$         	 		& $15 \pm 4$           	\\
        Expected  $WW$  events          				    		& $8_{-8}^{+9}$ 		& $7_{-7}^{+8}$ 		& $7_{-7}^{+9}$ 		& $6_{-6}^{+8}$ 				& $6_{-6}^{+7}$  \\
        Expected $ZW, ZZ$  events         					& $1.0 \pm 1.0$            	& $0.9_{-0.9}^{+1.0}$ 	& $1.0 \pm 1.0$         		 &$0.5_{-0.5}^{+0.8}$ 	 	 & $1.0 \pm 0.8$            	 \\
       Expected $Z/\gamma^*\rightarrow \ell\ell+$jets events           						& $0.3_{-0.3}^{+0.4}$ 	 & $0.31_{-0.31}^{+0.35}$& $0.31_{-0.31}^{+0.35}$ 	& $0.3_{-0.3}^{+0.4}$ 		& $0.3_{-0.3}^{+0.4}$             	\\
        Expected Higgs boson events         					         & $0.26 \pm 0.10$            	 & $0.24 \pm 0.10$            	& $0.26 \pm 0.10$              & $0.12 \pm 0.05$            	       & $0.19 \pm 0.10$             	\\
        Expected events with fake and non-prompt leptons         				& $18 \pm 4$           		& $18 \pm 4$          		 & $19 \pm 4$         		 & $17 \pm 4$            			& $12.5 \pm 3.2$             \\
 \noalign{\smallskip}\hline\noalign{\smallskip}
\end{tabular*}
%%%%
}
\end{center}

\end{table}

\begin{table}
\caption{ \label{tab:mva_CRSF}
Background fit results for the SF CRs in the MVA analysis. The nominal expectations from MC simulation
are given for comparison for $t\bar{t}$,  which is normalised to data by the fit. Combined statistical and systematic uncertainties are given. Events with fake or non-prompt leptons are estimated with the data-driven technique described in section~\ref{sec:fakes}. The observed events and the total (constrained) background are the same in the CRs by construction. Uncertainties on the predicted background event yields are quoted as symmetric except where the negative error reaches down to zero predicted events, in which case the negative error is truncated.
}

\begin{center}
\setlength{\tabcolsep}{0.0pc}
{\small
\begin{tabular*}{\textwidth}{@{\extracolsep{\fill}}lcccc}
\noalign{\smallskip}\hline\noalign{\smallskip}

{\bf  Channel}           										& $\textup{CRT}^\textup{SF}_\textup{M1}$             		& $\textup{CRT}^\textup{SF}_\textup{M2}$               		 & $\textup{CRT}^\textup{SF}_\textup{M3}$              		& $\textup{CRT}^\textup{SF}_\textup{M4}$              		\\[-0.05cm]
\noalign{\smallskip}\hline\noalign{\smallskip}
Observed events          									&	 $99$                   					 & $79$                    			 			& $133$                    						& $27$                  		 	  \\
\noalign{\smallskip}\hline\noalign{\smallskip}
Total (constrained) bkg events     								& $99 \pm 10$              				& $79 \pm 9$          						& $133 \pm 12$        						& $27 \pm 5$              	  \\
\noalign{\smallskip}\hline\noalign{\smallskip}
        Fit output, $t \bar t$ events        								& $82 \pm 12$            					& $55 \pm 14$       						& $101 \pm 16$         					& $14 \pm 8$             	 \\
\noalign{\smallskip}\hline\noalign{\smallskip}
  Total expected bkg events             							&  $94 \pm 16$      						&$88\pm 16$  							&$129\pm 23$  
 & $32\pm 10$  \\                                               
\noalign{\smallskip}\hline\noalign{\smallskip}
         Fit input, expected $t \bar t$          						& $77 \pm 13$             	 				& $65 \pm 9$      						& $95 \pm 20$            					& $19 \pm 7$             	  \\

        Expected $t \bar t V$ events          						& $0.98 \pm 0.31$             					& $0.95 \pm 0.31$             					 & $1.4 \pm 0.4$              					& $0.70 \pm 0.23$             		 \\

        Expected $Wt$ events       								& $1.6 \pm 1.5$             					& $2.8 \pm 1.6$             					& $4 .0\pm 1.6$            					& $0.20_{-0.20}^{+0.33}$               \\
         Expected $WW$ events         						          & $1.3_{-1.3}^{+1.7}$              				& $1.4_{-1.4}^{+1.5}$					 &$1.7_{-1.7}^{+1.8}$   					& $0.7_{-0.7}^{+1.0}$             \\
        Expected $ZW, ZZ$ events      					          	& $1.3 \pm 0.8$             					& $2.1\pm 0.7$            		  			& $2.1 \pm 1.3$             					& $1.4 \pm 0.5$              	\\
        Expected $Z/\gamma^*\rightarrow \ell\ell+$jets events        								& $7 \pm 7$          					& $12\pm 11$ 		 					& $14\pm 9$            						& $7\pm 6$              \\
       Expected Higgs boson events         								& $0.06 \pm 0.06$          				& $0.08\pm 0.05$            		 			& $0.12 \pm 0.05$              				& $0.04\pm 0.04$              	 \\

         Expected events with fake and non-prompt leptons        					& $3.7 \pm 1.7$              					& $3.7 \pm 1.7$            					& $6.9 \pm 2.3$            					& $2.8 \pm 1.2$             		 \\
\noalign{\smallskip}\hline\noalign{\smallskip}
\end{tabular*}
%%%%
}
\end{center}
\end{table}

The validity of the background estimate is tested using a set of VRs. Analogously to the CR, the corresponding VR for the DF (SF) SR labelled N is referred to as VRT$^\textup{DF(SF)}_\textup{MN}$. The definitions of these regions are given in table~\ref{tab:topregVR}
and their composition before and after the likelihood fit is given in tables  \ref{tab:mva_VRDF} and \ref{tab:mva_VRSF} for the DF and SF VRs, respectively. 

The signal contamination in the CRs ranges from 1.5--$30\%$ (4.8--24$\%$) in the DF (SF) CRs, whilst the contamination in the DF (SF) VRs ranges from 0.4--20\% (0.9--13\%).
\begin{table} [htb!]
\caption{VRs for the MVA analysis. The name of each VR is given in the first column and these have a one-to-one correspondence with the equivalently named SR. The middle column lists all selection cuts made, whilst the final column gives the BDTG range.}
\label{tab:topregVR}
\begin{center}
\begin{tabular}{l|cc}
\hline
Validation Region & Event Variable Selection [GeV] & BDTG range \\
\hline
$\textup{VRT}^\textup{DF}_\textup{M1}$ & C1, $80< m_\textup{T2}< 90$    &  [$-0.75$, $-0.13$]\\
$\textup{VRT}^\textup{DF}_\textup{M2}$ & C1, $80< m_\textup{T2}< 90$    & [$-0.75$, $-0.18$]\\
$\textup{VRT}^\textup{DF}_\textup{M3}$    & C1, $80< m_\textup{T2}< 90$    &  [$-0.80$, 0.19]\\
$\textup{VRT}^\textup{DF}_\textup{M4}$ & C2, $80< m_\textup{T2}< 90$    &  [$-0.98$, $-0.65$]\\
$\textup{VRT}^\textup{DF}_\textup{M5}$    & C4, $80< m_\textup{T2}< 90$   & [$-0.998$, $-0.33$]\\
\hline
$\textup{VRT}^\textup{SF}_\textup{M1}$    & C1, $80< m_\textup{T2}< 90$, $m_{\ell\ell}<61$ or $m_{\ell\ell}>121$   &  [$-0.80$, $-0.66$]\\
$\textup{VRT}^\textup{SF}_\textup{M2}$    & C1, $80< m_\textup{T2}< 90$, $m_{\ell\ell}<61$ or $m_{\ell\ell}>121$    &  [$-0.85$, $-0.11$]\\
$\textup{VRT}^\textup{SF}_\textup{M3}$ & C1, $80< m_\textup{T2}< 90$, $m_{\ell\ell}<61$ or $m_{\ell\ell}>121$    &  [$-0.95$, $-0.77$]\\
$\textup{VRT}^\textup{SF}_\textup{M4}$ & C3, $80< m_\textup{T2}< 90$, $m_{\ell\ell}<61$ or $m_{\ell\ell}>121$    &  [$-0.995$, $-0.76$]\\
\hline
\end{tabular} 
\end{center}
\end{table}

\begin{table}
\begin{center}
\caption{ \label{tab:mva_VRDF}
Background fit results for the DF VRs in the MVA analysis. The nominal expectations from MC simulation
are given for comparison for $t\bar{t}$,  which is normalised to data. Combined statistical and systematic uncertainties are given. Events with fake or non-prompt leptons are estimated with the data-driven technique described in section~\ref{sec:fakes}. The observed events and the total (constrained) background are the same in the CRs by construction; this is not the case for the VRs, where the consistency between these event yields is the test of the background model. Entries marked - - indicate a negligible background contribution. Backgrounds which contribute negligibly to all VRs are not listed. Uncertainties on the predicted background event yields are quoted as symmetric except where the negative error reaches down to zero predicted events, in which case the negative error is truncated.}
\setlength{\tabcolsep}{0.0pc}
{\scriptsize
\begin{tabular*}{\textwidth}{@{\extracolsep{\fill}}lccccc}
\noalign{\smallskip}\hline\noalign{\smallskip}

{\bf  Channel}          			 				& $\textup{VRT}^\textup{DF}_\textup{M1}$              	& $\textup{VRT}^\textup{DF}_\textup{M2}$              	& $\textup{VRT}^\textup{DF}_\textup{M3}$              	& $\textup{VRT}^\textup{DF}_\textup{M4}$             	& $\textup{VRT}^\textup{DF}_\textup{M5}$               \\[-0.05cm]
\noalign{\smallskip}\hline\noalign{\smallskip}	
Observed events          								& $149$                 		& $57$                  		& $30$                    		& $40$                   		& $47$                    		\\
\noalign{\smallskip}\hline\noalign{\smallskip}
Total  bkg events         						 		& $144 \pm 24$      		& $59 \pm 8$       		& $30 \pm 6$  			&$43 \pm 9$       		& $41 \pm 10$         \\
\noalign{\smallskip}\hline\noalign{\smallskip}
        Fit output, $t \bar t$ events         						& $136 \pm 23$       		& $54 \pm 7$     		& $30 \pm 6$       		& $37 \pm 9$      		& $36 \pm 9$          \\
\noalign{\smallskip}\hline\noalign{\smallskip}
\noalign{\smallskip}\hline\noalign{\smallskip}
       Fit input, expected $t \bar t$         					& $141 \pm 20$       		& $56 \pm 10$      		& $30 \pm 8$       		& $39 \pm 10$     		& $37 \pm 7$          \\

        Expected $t \bar t V$ events   		 				& $0.64 \pm 0.21$            	& $0.34 \pm 0.13$             	& $0.32 \pm 0.14$            	& $0.50 \pm 0.17$             	 & $0.39 \pm 0.14$              \\
         Expected $Wt$ events     				   	 		& $4.4 \pm 2.2$           	& $2.4 \pm 1.6$           	& $0.4_{-0.4}^{+1.0} $       & $0.8_{-0.8}^{+1.2} $        & $2.6 \pm 1.5$           	\\
        Expected  $WW$  events          				    		& $1.0_{-1.0}^{+1.6}$ 	& $0.5_{-0.5}^{+1.0}$ 	& $0.4\pm 0.4$ 		& $0.9_{-0.9}^{+1.1}$ 	& $1.0_{-1.0}^{+1.2}$  \\
        Expected $ZW, ZZ$  events         					& $0.09 _{-0.09}^{+0.16}$	& $0.10_{-0.10}^{+0.16}$ 	& $0.08_{-0.08}^{+0.14} $ &$0.17_{-0.17}^{+0.21}$ 	 & $0.31 \pm 0.31$            	 \\
        Expected Higgs boson events         					         & $0.03 \pm 0.03$            	 & - -             			& $0.01 _{-0.01}^{+0.02}$ & $0.03 \pm 0.03$            	 & $0.02 \pm 0.02$             	\\
        Expected events with fake and non-prompt leptons         				& $1.7 \pm 1.7$           	 & $1.6 \pm 1.2$          	& $1.6 \pm 1.2$         		 & $3.0 \pm 1.5$            	& $0.3_{-0.3}^{+0.6}$             \\
\noalign{\smallskip}\hline\noalign{\smallskip}
\end{tabular*}
%%%%
}
\end{center}
\end{table}

\begin{table}
\caption{ \label{tab:mva_VRSF}
Background fit results for the SF VRs in the MVA analysis. The nominal expectations from MC simulation
are given for comparison for $t\bar{t}$,  which is normalised to data. Combined statistical and systematic uncertainties are given. Events with fake or non-prompt leptons are estimated with the data-driven technique described in section~\ref{sec:fakes}. The observed events and the total (constrained) background are the same in the CRs by construction; this is not the case for the VRs, where the consistency between these event yields is the test of the background models. Entries marked - - indicate a negligible background contribution. Uncertainties on the predicted background event yields are quoted as symmetric except where the negative error reaches down to zero predicted events, in which case the negative error is truncated.
}

\begin{center}
\setlength{\tabcolsep}{0.0pc}
{\small
\begin{tabular*}{\textwidth}{@{\extracolsep{\fill}}lcccc}
\noalign{\smallskip}\hline\noalign{\smallskip}

{\bf  Channel}           										& $\textup{VRT}^\textup{SF}_\textup{M1}$             		& $\textup{VRT}^\textup{SF}_\textup{M2}$               		 & $\textup{VRT}^\textup{SF}_\textup{M3}$              		& $\textup{VRT}^\textup{SF}_\textup{M4}$              		\\[-0.05cm]
\noalign{\smallskip}\hline\noalign{\smallskip}
Observed events          									&$65$                   					 	& $20$                    			 			& $140$                    						& $17$                  		 	  \\
\noalign{\smallskip}\hline\noalign{\smallskip}
Total  bkg events     										& $75 \pm 19$              					& $23 \pm 9$          						& $150 \pm 40$        						& $22 \pm 13$              	  \\
\noalign{\smallskip}\hline\noalign{\smallskip}
        Fit output, $t \bar t$ events        								& $69 \pm 19$            					& $19 \pm 10$       						& $130 \pm 40$         					& $17 \pm 13$             	 \\
\noalign{\smallskip}\hline\noalign{\smallskip}
\noalign{\smallskip}\hline\noalign{\smallskip}
  
           Fit input, expected $t \bar t$          						& $64 \pm 12$             	 				& $22 \pm 9$      						& $128 \pm 23$            					& $23 \pm 5$             	  \\
        Expected $t \bar t V$ events          						& $0.26 \pm 0.10$             					& $0.22 \pm 0.09$             					 & $0.6 \pm 0.2$              					& $0.20 \pm 0.09$             		 \\

        Expected $Wt$ events       								& $2.0 \pm 1.1$             					& $1.4 \pm 0.9$             					& $6.4 \pm 2.3$            					& $1.6\pm 1.0$               \\
         Expected $WW$ events         						          & $0.9 \pm 0.6$              					& $0.3_{-0.3}^{+0.5}$					 & $2.1\pm 1.7$  					& $0.4\pm 0.4$              \\
        Expected $ZW, ZZ$ events      					          	& $0.19 \pm 0.14$             					& $0.07_{-0.07}^{+0.18}$            		  	& $0.39 \pm 0.19$             					& $0.12 \pm 0.12$              	\\
        Expected $Z/\gamma^*\rightarrow \ell\ell+$jets events        								& $0.4_{-0.4}^{+0.6}$          				& $0.7_{-0.7}^{+0.9}$		 			&$0.9_{-0.9}^{+1.0}$             						& $0.3_{-0.3}^{+0.4}$              \\
       Expected Higgs boson events         							& - -           							& - -          		 					& $0.02 \pm 0.02$              				& - -              	 \\

         Expected events with fake and non-prompt leptons        					& $2.8 \pm 1.3$              					& $0.8 \pm 0.8$            					& $3.2 \pm 1.9$            					& $1.7 \pm 1.0$             		 \\
\noalign{\smallskip}\hline\noalign{\smallskip}
\end{tabular*}
%%%%
}
\end{center}
\end{table}

\clearpage

\section{Systematic uncertainties}
\label{sec:systematics}

Various systematic uncertainties affecting the predicted background rates in the signal regions are considered. Such uncertainties are either used directly in the evaluation of the predicted background in the SRs when this is taken directly from MC simulation, or to compute the uncertainty on the background fit.

The dominant detector-related systematic uncertainties considered in the analyses are:
\begin{itemize} 
\item[-] {\bf Jet energy scale and resolution.} The uncertainty on the jet energy scale (JES) was derived using a combination of MC simulations and data~\cite{Aad:2011he}, taking into account the dependence on $\pT$, $\eta$, jet flavour and number of primary vertices. The components of the JES uncertainty are varied by $\pm 1 \sigma$ in the MC simulations and propagated to the expected event yield. 
Uncertainties related to the jet energy resolution (JER) are obtained with in situ measurements of the jet response balance in dijet events~\cite{jer}. 
Their impact on the event yield is estimated by applying an additional smearing to the jet transverse momenta in the MC simulations. The JES and JER variations applied to jets are also propagated to the \EtMiss. 
\item[-] {\bf Clusters in the calorimeter energy scale, resolution and pile-up modelling.} The uncertainties related to the contribution to \EtMiss from the energy
scale and resolution of clusters in the calorimeter not associated to electrons, muons or jets (including low momentum ($7 < \pT < 20 \GeV$) jets), as well as the uncertainty due to the modelling of pile-up were evaluated. 
\item[-] {\bf \boldmath{\emph{b}}-tagging} (where applicable). The $b$-tagging uncertainty is evaluated by varying the $\pT$- and flavour-dependent correction factors applied to each jet in the simulation within a range that reflects the systematic uncertainty on the measured tagging efficiency and rejection rates. The relative impact of this uncertainty on the final event yield is dominated by the uncertainty on the $b$-tagging efficiency. 
\item[-] {\bf Fake and non-prompt lepton background uncertainties.}  The uncertainty on the fake and non-prompt lepton background arises from the limited size of the control samples used to measure the probabilities for loose leptons to pass the tight selections, the comparison of results obtained with probabilities computed with alternative control samples, and from the number of events in the loose and tight event samples.
\end{itemize}
The remaining detector-related systematic uncertainties, such as those on lepton reconstruction efficiency and on the modelling of the trigger, are of the order of a few percent. A 2.8\% uncertainty on the luminosity determination was measured using techniques similar to that of Ref.~\cite{lumi} from a calibration of the luminosity scale derived from beam-separation scans performed in November 2012, and it is included for all signal and background MC simulations.

Various theoretical uncertainties are considered in the MC modelling of the major SM backgrounds. 
In the case of top-quark contributions, the predictions of {\tt MC@NLO-4.06} are compared with {\tt POWHEG} interfaced to {\tt HERWIG} to estimate the uncertainty due to the choice of generator, while the difference in the yields obtained from {\tt POWHEG} interfaced to {\tt PYTHIA} and {\tt POWHEG} interfaced to {\tt HERWIG} is taken as the systematic uncertainty on parton showering, and the predictions of dedicated {\tt ACERMC-3.8} samples generated with different tuning parameters are compared to give the uncertainty related to the amount of ISR/FSR.

At next-to-leading order, contributions with an additional bottom quark in the final state lead to ambiguities in the distinction between the $Wt$ process ($gb\rightarrow Wt$) and top-quark pair production. In the hadronic $m_\textup{T2}$ analysis this becomes significant as the SR is a region of phase space where these ambiguities are important. All the $Wt$ samples, generated using {\tt MC@NLO-4.06} and {\tt POWHEG-1.0}, use the diagram removal~\cite{Frixione:2008yi} scheme. {\tt ACERMC-3.8} is used to generate a leading-order (LO) prediction of the $WWb$ and $WWb\bar{b}$ final state (which includes both $t\bar{t}$ and $Wt$ single-top processes); the predictions of these {\tt ACERMC-3.8} samples and {\tt MC@NLO-4.06} are then compared in order to assess the uncertainty on the background estimate from this interference.

The uncertainties on diboson production are evaluated by comparing the predictions of {\tt POWHEG-1.0} and {\tt SHERPA-1.4.1}, and the uncertainties on $Z/\gamma^*+$jets production are evaluated by comparing the predictions of {\tt SHERPA-1.4.1} and {\tt ALPGEN-2.14}. The former comparison includes the impact of choice of parton showering scheme.

The impact of the evaluated systematic uncertainties on the different SRs presented are shown in tables~\ref{tab:uncertainties-cutandcount}, \ref{tab:uncertainties-mvadf} and \ref{tab:uncertainties-mvasf}. These tables quote, for each SR, the percentage of the total systematic uncertainty on the background yield which is attributed to each source. Since these uncertainties are correlated, there is no requirement for these to sum in quadrature to 100\%. These correlations are particularly strong in H160, where there are strong cancellations between the $t\bar{t}$ and $Wt$ normalisation and the top-quark generator systematic uncertainties. The uncertainty on the $WZ/ZZ$ normalisation (where appropriate) has comparable statistical and systematic components, whilst the $t\bar{t}$ ($t\bar{t},Wt$) and $WW$ normalisation uncertainties are dominated by systematic effects.

 \begin{table} [htb!]
\caption{Summary of the systematic uncertainties on the background estimates for the two $m_\textup{T2}$-based analyses. The size of each uncertainty is quoted as a percent of the total uncertainty. Note that the individual uncertainties can be correlated, and thus do not necessarily sum in quadrature to 100\%. }
\label{tab:uncertainties-cutandcount}
\begin{center}
\begin{tabular}{lccccccc}
\hline
                                        &  L90 & L100 & L110 & L120&  H160\\
                                        \hline
                                        \hline
Background & $300\pm50$ & $5.2\pm2.2$ & $9.3\pm3.5$ & $19\pm9$ & $26\pm6$ \\
\hline
\hline
\multicolumn{6}{l}{Uncertainty Breakdown ($\%$):}\\
\hline
JES                                   & 2 &12 & 3 & 2 & 49 \\
JER 					& 46 & 47 & 1 & 9 & 67 \\
Cluster energy scale and resolution  & 44 & 30 & 11 & 4 & 4    \\
Pile-up                                    & 42 & 22 & 19 & 12    &   10        \\
$b$-tagging				& - &	- & - & - & 19 \\
Diboson generator              & 18 & 23 & 40 & 92 &   7   \\
Top-quark generator                      & 44 & 52 & 73& 4   &   19   \\
Top-quark decay: ISR/FSR                         & 19 & 27 & 1 & 8    &  16      \\
Top-quark decay: parton shower             & 17 & 20 & 21 & 5     & 33     \\
$t\bar{t}$, $Wt$ interference & - & - & - & - & 70\\
Simulation statistics          & 15 &31 &29 &15   & 40    \\
Fake and non-prompt leptons& 3 & 0 & 1 & 1  & 4      \\
$\ttbar$ normalisation      & 30 & 13 & 8 &1   & -     \\
$t\bar{t}$, $Wt$ normalisation & - & - & - & - & 125 \\
$WW$      normalisation    & 32 & 8 & 18 & 25  & -  \\
$WZ, ZZ$ normalisation    & 5 & 2 & 5 & 9   & - \\
$Z/\gamma^*\rightarrow ee,\mu\mu+$jets normalisation    & - & -  & - & -  & 1.5 \\
\hline
\end{tabular}
\end{center}
\end{table}

\begin{table} [htb!]
\caption{Summary of the systematic uncertainties on the background estimates for the MVA analysis DF signal regions. The size of each uncertainty is quoted as a percent of the total uncertainty. Note that the individual uncertainties can be correlated, and thus do not necessarily sum in quadrature to 100\%.}
\label{tab:uncertainties-mvadf}
\begin{center}
\begin{tabular}{lccccc}
\hline
                                        & $\textup{M1}^\textup{DF}$ & $\textup{M2}^\textup{DF}$ & $\textup{M3}^\textup{DF}$ & $\textup{M4}^\textup{DF}$ & $\textup{M5}^\textup{DF}$     \\
                                         \hline
                                        \hline
Background &  $ 5.8 \pm 1.9$               & $13 \pm 4$                  & $5.1 \pm 2.0$                   & $1.3 \pm 1.0$         & $1.0 \pm 0.5$        \\
\hline
\hline
\multicolumn{6}{l}{Uncertainty Breakdown ($\%$):}\\                                             
\hline
JES                                                   & 7         &28         &6         & 10         & 4                \\
JER                             & 12         &37         &29         & 14         & 25         \\
Cluster energy scale  & \multirow{2}{*}{31}         &\multirow{2}{*}{42}         & \multirow{2}{*}{33}         &\multirow{2}{*}{30}         & \multirow{2}{*}{11}            \\
and resolution & & & & &  \\
Pile-up                                            & 25     &35         & 14         & -            & 13                \\
Diboson generator                      & 26         &27         & 44        & 47         &  23        \\
Top-quark generator                                  & 100     &87         & 75        & 56       &  51            \\
Top-quark decay: ISR/FSR                                 & 27         &45         & 34         & 39        &  15          \\
Top-quark decay: parton shower                         & 35         & 1         & 33         & 5         & 15             \\
Simulation statistics                      & 40         &32         &39         & 30       & 44        \\
Fake and non-prompt leptons        & 15         & 8         & 15         & 27      & 66             \\
$\ttbar$ normalisation                  &47         & 48         & 30         &10       & 11         \\
\hline
\end{tabular}
\end{center}
\end{table}

\begin{table} [htb!]
\caption{Summary of the systematic uncertainties on the background estimates for the MVA analysis SF signal regions. The size of each uncertainty is quoted as a percent of the total uncertainty. Note that the individual uncertainties can be correlated, and thus do not necessarily sum in quadrature to 100\%.}
\label{tab:uncertainties-mvasf}
\begin{center}
\begin{tabular}{lcccc}
\hline
                                       & $\textup{M1}^\textup{SF}$ & $\textup{M2}^\textup{SF}$ & $\textup{M3}^\textup{SF}$ & $\textup{M4}^\textup{SF}$                        \\
                                         \hline
                                        \hline
Background         & $7.6 \pm 2.2$                    & $9.5 \pm 2.1$                 & $1.1 \pm 0.7$                    & $2.5 \pm 1.0$  \\
\hline
\hline
\multicolumn{5}{l}{Uncertainty Breakdown ($\%$):}\\
\hline
JES                                                           &12         &12         &21         & 13         \\
JER                                   & 48         &36         &53         & 26         \\
Cluster energy scale        & \multirow{2}{*}{21}         &\multirow{2}{*}{23}         &\multirow{2}{*}{23}         & \multirow{2}{*}{15}              \\
and resolution  & & & & \\
Pile-up                                            & 21         &32         &21         & 14              \\
Diboson generator                       & 6         &13         &5        & 2          \\
Top-quark generator                                  & 71         &50         &42        & 26          \\
Top-quark decay: ISR/FSR                                 & 25         &24         &12         & 17        \\
Top-quark decay: parton shower                           & 16         &14         &21         & 13          \\
Simulation statistics                       & 48         &38         &44         & 37       \\
Fake and non-prompt leptons         & 19         &38         &36         & 6           \\
$\ttbar$ normalisation                       &75         & 55         &27         &37        \\
\hline
\end{tabular}
\end{center}
\end{table}

Systematic uncertainties are also taken into account for expected signal yields.
The uncertainty on the signal cross-sections is calculated with an envelope of cross-section predictions which is defined using the 68\% CL ranges of the CTEQ~\cite{Nadolsky:2008zw} (including the $\alpha_s$ uncertainty) and MSTW~\cite{Martin:2009iq} PDF sets, together with  variations of the factorisation and renormalisation scales by factors of two or one half. The nominal cross-section value is taken to be the midpoint of the envelope and
the uncertainty assigned is half the full width of the envelope, using the procedure described in ref.~\cite{Kramer:2012kl}. The typical cross-section uncertainty is 15\% for the top-squark signal. Uncertainties on signal shape related to the generation of the SUSY samples are determined using additional samples with modified parameters. This includes uncertainties on the modelling of ISR and FSR, the choice of renormalisation/factorisation scales, and the parton-shower matching scale settings. 
These uncertainties are relevant only in the case of small $\Delta m(\tone, \chipm)$ for the $\tilde{t}_1\rightarrow b+\tilde{\chi}^\pm_1$ decay mode or when $m(\tone) \simeq m(t)+m(\tilde{\chi}^0_1)$ for the $\tilde{t}_1\rightarrow t+\tilde{\chi}^0_1$ decay mode. They have an impact of up to 10\% (20\%) on the acceptance in the $\tilde{t}_1\rightarrow b+\tilde{\chi}^\pm_1$ ($\tilde{t}_1\rightarrow b+\tilde{\chi}^0_1$) case depending on the SR, but yield negligible effects on the sensitivity.

\section{Results and interpretation}
\label{sec:results}

Tables~\ref{tab:lepSR} to~\ref{tab:mvaSF} report the background yields (before and after the background-only likelihood fit) and the observed numbers of events in the various SRs. In each, agreement is found between the SM prediction and the data, within uncertainties. In all tables the quoted uncertainty includes all the sources of statistical and systematic uncertainty considered (see section~\ref{sec:systematics}).

The agreement between the SM prediction and the data is tested separately for the SF and DF populations in L90 (the SR with the highest predicted background yield) as an additional check. Results of this check are consistent with the inclusive result in both the SF ($123$ observed and $136 \pm 19$ expected events) and DF ($151$ observed and $164 \pm 31$ expected events) samples, with the background composition being dominated by the flavour symmetric \ttbar\ and $WW$ backgrounds. Small differences in the background composition arise from the $WZ$ and $ZZ$ backgrounds, which account for 8\% of the total background SF events and $<1\%$ of the total background DF events. Other minor differences are a result of the fake and non-prompt lepton background which accounts for 6\% of the DF background but only 2\% of the SF background. $Z \gamma^{*} \rightarrow \ell \ell$ events contribute only to the SF channel, and are 2\% of the total background event yield.

\begin{table}[!htp]
\caption{Number of events and composition in the leptonic $m_\textup{T2}$ SRs for an integrated luminosity of 20.3~fb$^{-1}$. The nominal expectations from MC simulation are given for comparison for those backgrounds that are normalised to data. Combined statistical and systematic uncertainties are given. Events with fake or non-prompt leptons are estimated with the data-driven technique described in section~\ref{sec:fakes}.  Entries marked - - indicate a negligible background contribution. Uncertainties on the predicted background event yields are quoted as symmetric except where the negative error reaches down to zero predicted events, in which case the negative error is truncated. }
\label{tab:lepSR}
\begin{center}
\setlength{\tabcolsep}{0.0pc}
{\small
\begin{tabular*}{\textwidth}{@{\extracolsep{\fill}}lrrrr}
\noalign{\smallskip}\hline\noalign{\smallskip}
{\bf  Channel}           & L90            & L100            & L110            & L120              \\[-0.05cm]
\noalign{\smallskip}\hline\noalign{\smallskip}
Observed events          & $274$              & $3$              & $8$              & $18$                    \\
\noalign{\smallskip}\hline\noalign{\smallskip}
Total  bkg events         & $300 \pm 50$          & $5.2 \pm 2.2$          & $9.3 \pm 3.5$          & $19 \pm 9$              \\
\noalign{\smallskip}\hline\noalign{\smallskip}
        Fit output, $t\bar{t}$ events         & $172 \pm 33$          & $3.5 \pm 2.1$          & $3.4 \pm 2.9$          & $1.1 \pm 1.1$              \\
        Fit output, WW events         & $78 \pm 20$          & $1.0 \pm 0.5$          & $3.2 \pm 1.4$          & $12 \pm 7$              \\
        Fit output, $WZ$, $ZZ$ events         & $11.6 \pm 2.4$          & $0.22_{-0.22}^{+0.26}$          & $0.9 \pm 0.5$          & $4.1 \pm 2.1$            \\
\noalign{\smallskip}\hline\noalign{\smallskip}
\noalign{\smallskip}\hline\noalign{\smallskip}
        Fit input, expected \ttbar events       & $190 \pm 40$          & $3.9 \pm 2.4$          & $3.7 \pm 3.2$          & $1.2 \pm 1.2$              \\
        Fit input, expected $WW$ events         & $62 \pm 9$          & $0.75 \pm 0.38$          & $3 \pm 1$          & $9 \pm 5$              \\
        Fit input, expected $WZ$, $ZZ$ events         & $13.6 \pm 2.4$          & $0.26_{-0.26}^{+0.31}$          & $1.1 \pm 0.6$          & $4.8 \pm 2.5$             \\
        Expected $Z/ \gamma^{*} \rightarrow \ell \ell$ events         & $2.8 \pm 1.4$          & $0.14_{-0.14}^{+0.14}$          & $0.09_{-0.09}^{+0.14}$          & $0.07_{-0.07}^{+0.09}$              \\
        Expected $\ttbar V$ events         & $1.8 \pm 0.6$          & $0.35 \pm 0.14$          & $0.62 \pm 0.21$          & $0.51 \pm 0.18$              \\
        Expected $Wt$ events         & $21 \pm 7$          & $0.00_{-0.00}^{+0.19}$          & - -          & $0.35_{-0.35}^{+0.39}$              \\
        Expected Higgs boson events         & $0.65 \pm 0.22$          & $0.02_{-0.02}^{+0.02}$          & $0.03 \pm 0.03$          & $0.31 \pm 0.12$              \\
        Expected events with fake and non-prompt leptons      & $13.0 \pm 3.5$          &- -           & $1.0 \pm 0.6$          & $1.1 \pm 0.8$              \\
%%     \\
\noalign{\smallskip}\hline\noalign{\smallskip}
\end{tabular*}
%%%%
}
\end{center}
\end{table}

\begin{table}[!hbp]
\caption{Number of events and composition in SR H160 for an integrated luminosity of 20.3~fb$^{-1}$ in the hadronic $m_\textup{T2}$ analysis. The nominal expectations from MC simulation are given for comparison for those backgrounds ($t\bar{t}$, $Wt$ and $Z/\gamma^* (\rightarrow ee,\mumu) + $jets production) that are normalised to data. Combined statistical and systematic uncertainties are given. Events with fake or non-prompt leptons are estimated with the data-driven technique described in section~\ref{sec:fakes}.\label{tab:h160}. Uncertainties on the predicted background event yields are quoted as symmetric except where the negative error reaches down to zero predicted events, in which case the negative error is truncated. }

\begin{center}
\setlength{\tabcolsep}{0.0pc}
{\small
\begin{tabular*}{\textwidth}{@{\extracolsep{\fill}}lr}
\noalign{\smallskip}\hline\noalign{\smallskip}
{\bf  Channel}           & H160             \\[-0.05cm]
\noalign{\smallskip}\hline\noalign{\smallskip}
Observed events          & $33$                    \\
\noalign{\smallskip}\hline\noalign{\smallskip}
 Total  bkg events         & $26 \pm 6$              \\
\noalign{\smallskip}\hline\noalign{\smallskip}
        Fit output, $t\bar{t},Wt$ events         & $22 \pm 5$              \\
        Fit output, $Z/\gamma^*\rightarrow ee,\mu\mu+$jets events         & $0.2_{-0.2}^{+1.8}$              \\
\noalign{\smallskip}\hline\noalign{\smallskip}
\noalign{\smallskip}\hline\noalign{\smallskip}
         Fit input, expected $t\bar{t},Wt$ events         & $24 \pm 7$              \\
          Fit input, expected $Z/\gamma^*\rightarrow ee,\mu\mu+$jets events          & $0.2_{-0.2}^{+1.2}$        \\
         Expected $WW$ events         & $0.00_{-0.00}^{+0.35}$              \\
         Expected $t\bar{t}V$ events         & $0.47 \pm 0.16$              \\
         Expected $WZ$, $ZZ$ events         & $0.11 \pm 0.11$              \\
         Expected $Z/\gamma^*\rightarrow \tau\tau+$jets events         & $0.86 \pm 0.15$              \\
         Expected events with fake and non-prompt leptons         & $2.5 \pm 0.4$              \\
          Expected Higgs boson events          & $0.08 \pm 0.02$              \\
\noalign{\smallskip}\hline\noalign{\smallskip}
\end{tabular*}
%%%%                                                                                                                                                                                             
}
\end{center}
\end{table}

\begin{table}
\begin{center}

\caption{ \label{tab:mvaDF}
Number of events and composition of the DF signal regions for an integrated luminosity
of $20.3~\ifb$ in the MVA analysis. Nominal MC simulation expectation 
is given for comparison for the background ($t\bar{t}$) that is normalised to data.
Combined statistical and systematic uncertainties are given. Events with fake or non-prompt leptons are 
estimated with the data-driven technique described in section~\ref{sec:backgrounds-lep}. Entries marked - - indicate a negligible background contribution. Backgrounds which contribute negligibly to all SRs are not listed. Uncertainties on the predicted background event yields are quoted as symmetric except where the negative error reaches down to zero predicted events, in which case the negative error is truncated. 
 }

\setlength{\tabcolsep}{0.0pc}
{\scriptsize
\begin{tabular*}{\textwidth}{@{\extracolsep{\fill}}lccccc}
\noalign{\smallskip}\hline\noalign{\smallskip}

{\bf  Channel}           			& $\textup{M1}^\textup{DF}$ & $\textup{M2}^\textup{DF}$             & $\textup{M3}^\textup{DF}$           	& $\textup{M4}^\textup{DF}$           	& $\textup{M5}^\textup{DF}$           	        		\\[-0.05cm]               	           		           
\noalign{\smallskip}\hline\noalign{\smallskip}
Observed events          						& $9$              			 & $11$              				& $5$              				& $3$              			& $1$              			\\
\noalign{\smallskip}\hline\noalign{\smallskip}
Total  bkg events          						& $ 5.8 \pm 1.9$          	 & $13 \pm 4$      			& $5.1 \pm 2.0$       			& $1.3 \pm 1.0$ 		& $1.0 \pm 0.5$       	 \\
\noalign{\smallskip}\hline\noalign{\smallskip}
         Fit output, $t \bar t$ events    					& $5.0 \pm 1.9$           	 & $11 \pm 4$       			& $3.1 \pm 1.7$       			& $0.6_{-0.6}^{+0.8}$ 	& $0.29_{-0.29}^{+0.35}$	 \\
\noalign{\smallskip}\hline\noalign{\smallskip}
\noalign{\smallskip}\hline\noalign{\smallskip}
     Fit input, expected $t \bar t$   				& $5.2 \pm 2.6$                     & $11 \pm 5$       		          & $3.2 \pm 2.1$       			& $0.6_{-0.6}^{+0.8}$ 	 & $0.3_{-0.3}^{+0.4}$ 	 \\

         Expected $t \bar t V$ events                
        										& $0.43 \pm 0.15$           	& $0.83 \pm 0.27$       		& $0.73 \pm 0.24$       		& $0.38 \pm 0.13$       	& $0.23 \pm 0.09$       	 \\
        Expected $Wt$ events         				& $0.00_{-0.00}^{+0.09}$  & $0.9 \pm 0.7$ 				& $0.4 \pm 0.4$ 		& - -        			& - -        	    \\
        Expected $WW$ events          				& $0.3_{-0.3}^{+0.5}$	&$0.7_{-0.7}^{+1.1}$			& $0.8_{-0.8}^{+0.9}$ 	&$0.3_{-0.3}^{+0.5}$ 	& $0.49 \pm 0.19$       \\
         Expected $ZW, ZZ$ events      				& $0.05 _{-0.05}^{+0.06}$& $0.11 \pm 0.10$       		& $0.10_{-0.10}^{+0.12}$		& $0.05_{-0.05}^{+0.07}$ 	& $0.03\pm 0.03$	\\
          Expected events with fake and non-prompt leptons      		& $0.00_{-0.00}^{+0.29}$   & $0.00_{-0.00}^{+0.33}$ 	& $0.00_{-0.00}^{+0.30}$       	& $0.00_{-0.00}^{+0.27}$ &$0.00_{-0.00}^{+0.35}$           \\

\noalign{\smallskip}\hline\noalign{\smallskip}
\end{tabular*}
%%%%
}
\end{center}
\end{table}

\begin{table}
\begin{center}

\caption{ \label{tab:mvaSF}
Number of events and composition of the SF signal regions for an integrated luminosity
of $20.3~\ifb$ in the MVA analysis. Nominal MC simulation expectation 
is given for comparison for the background ($t\bar{t}$) that is normalised to data.
Combined statistical and systematic uncertainties are given. Events with fake or non-prompt leptons are 
estimated with the data-driven technique described in section~\ref{sec:backgrounds-lep}. Entries marked - - indicate a negligible background contribution. Backgrounds which contribute negligibly to all SRs are not listed. Uncertainties on the predicted background event yields are quoted as symmetric except where the negative error reaches down to zero predicted events, in which case the negative error is truncated. 
}

\setlength{\tabcolsep}{0.0pc}
{\small
\begin{tabular*}{\textwidth}{@{\extracolsep{\fill}}lcccc}
\noalign{\smallskip}\hline\noalign{\smallskip}

{\bf  Channel}           	&		 $\textup{M1}^\textup{SF}$   		 & $\textup{M2}^\textup{SF}$ 				& $\textup{M3}^\textup{SF}$  			& $\textup{M4}^\textup{SF}$     					\\[-0.05cm]               	           		           
\noalign{\smallskip}\hline\noalign{\smallskip}

Observed events          										& $6$              		 		& $9$              			& $0$              				& $5$              		  	\\
\noalign{\smallskip}\hline\noalign{\smallskip}
Total  bkg events         										& $7.6 \pm 2.2$        			& $9.5 \pm 2.1$        		 & $1.1 \pm 0.7$        			& $2.5 \pm 1.0$            \\
\noalign{\smallskip}\hline\noalign{\smallskip}
         Fit output, $t \bar t$ events     									& $7.1 \pm 2.2$        			& $3.8\pm 1.6$  		& $0.7 \pm 0.7$ 			& $0.6\pm 0.5$     \\
\noalign{\smallskip}\hline\noalign{\smallskip}
\noalign{\smallskip}\hline\noalign{\smallskip}
     Fit input, expected $t \bar t$   								& $6.6 \pm 2.2$        			& $4.4 \pm 1.8$         		& $0.7\pm 0.7$ 			& $0.7\pm 0.6$     \\
         Expected $t \bar t V$ events               
        														& $0.07 \pm 0.03$   			& $0.50 \pm 0.17$      	& $0.06 \pm 0.04$        		& $0.17 \pm 0.10$            \\
        Expected $Wt$ events                     							&$0.02_{-0.02}^{+0.08}$		& $0.02_{-0.02}^{+0.20}$  & - -        			 	& - -             \\
         Expected $WW$ events          								& $0.08_{-0.08}^{+0.14}$ 		& $0.18_{-0.18}^{+0.30}$  & $0.00_{-0.00}^{+0.04}$         & $0.06_{-0.06}^{+0.07}$     \\
         Expected $ZW, ZZ$ events           							& $0.03_{-0.03}^{+0.05}$ 		& $2.3 \pm 0.5$         		  & $0.08_{-0.08}^{+0.15}$	 & $1.2 \pm 0.9$            \\
         Expected $Z/\gamma^*\rightarrow \ell\ell+$jets events         									&$0.02_{-0.02}^{+0.03}$        	& $1.4_{-1.4}^{+1.6}$ 	 & - -  					&  $0.5_{-0.5}^{+0.6}$	\\
        Expected events with fake and non-prompt leptons       						& $0.3_{-0.3}^{+0.4}$       	      	& $1.1 \pm 0.8$         		& $0.25_{-0.25}^{+0.26}$		& $0.00_{-0.00}^{+0.06}$            \\
\noalign{\smallskip}\hline\noalign{\smallskip}
\end{tabular*}
%%%%
}
\end{center}
\end{table}

Figures~\ref{fig:mt2ll:SRa} to \ref{fig:mt2ll:SRc} illustrate the distribution of $m_\textup{T2}$ in the different SRs of the leptonic $m_\textup{T2}$ analysis, prior to any cut on $m_\textup{T2}$, after the background fit. In this figure, the events are separated into DF and SF lepton pairs, illustrating the similarity of the background composition between the two populations (and the negligible size of $Z/\gamma^*+$jets in the SRs themselves).
Figure~\ref{fig:mt2bb:SR} illustrates the distribution of $m^{b-\textup{jet}}_\textup{T2}$ in SR H160, prior to any cut on $m^{b-\textup{jet}}_\textup{T2}$, after the background fit. Figure~\ref{fig:mva:BDTG_SR} illustrates the BDTG distribution, prior to any cut on BDTG and after the background fit, for the DF and SF channels of the MVA analysis as obtained from the trainings which used the point $(m(\tilde{t}),m(\tilde{\chi}^0_1))=(300,50)$ GeV and $(m(\tilde{t})$,$m(\tilde{\chi}^0_1))=(300,100)$~GeV, respectively. 

\begin{figure}[htb!]
\begin{center}
\includegraphics[width=0.79\textwidth]{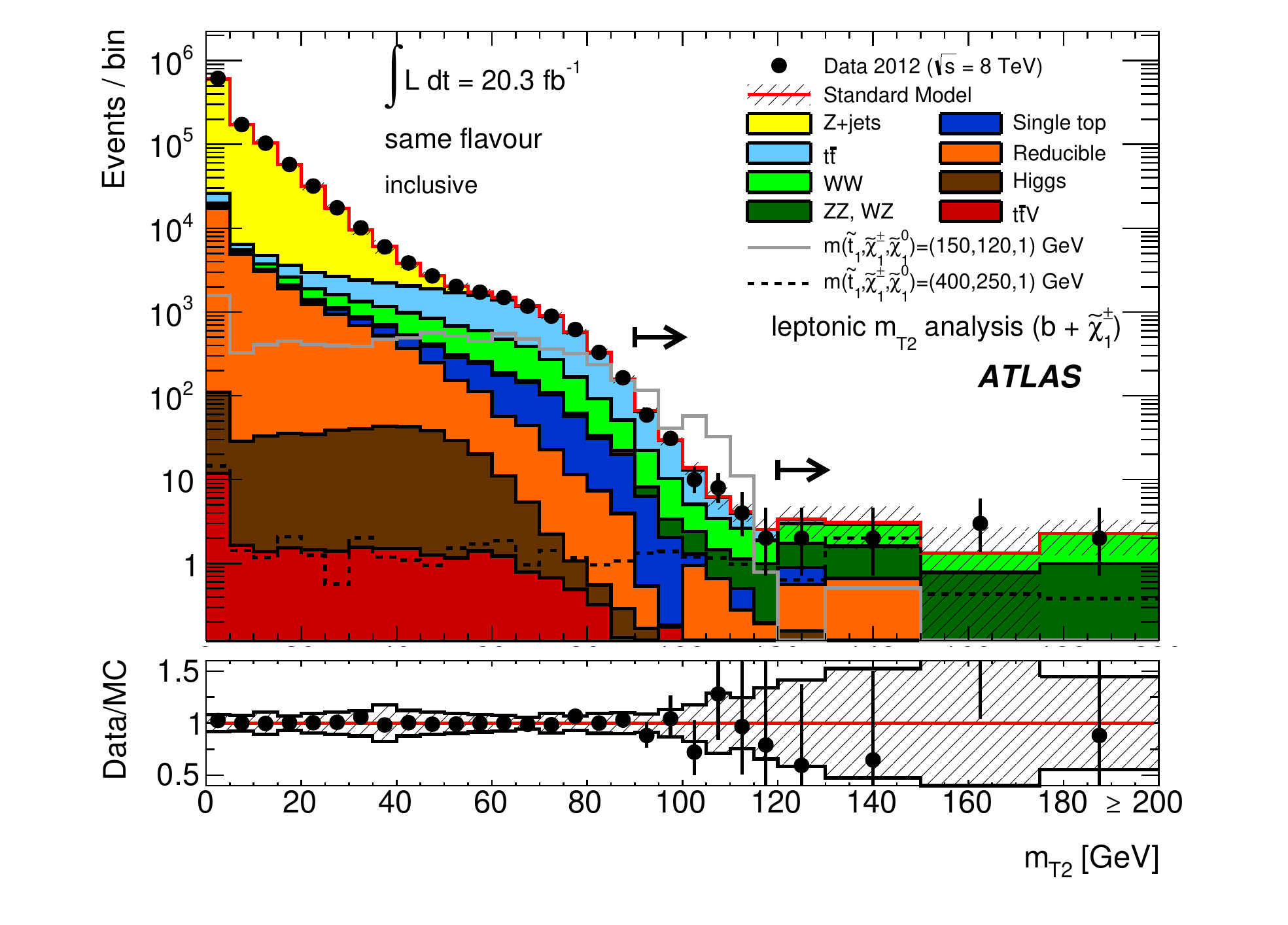}
\includegraphics[width=0.79\textwidth]{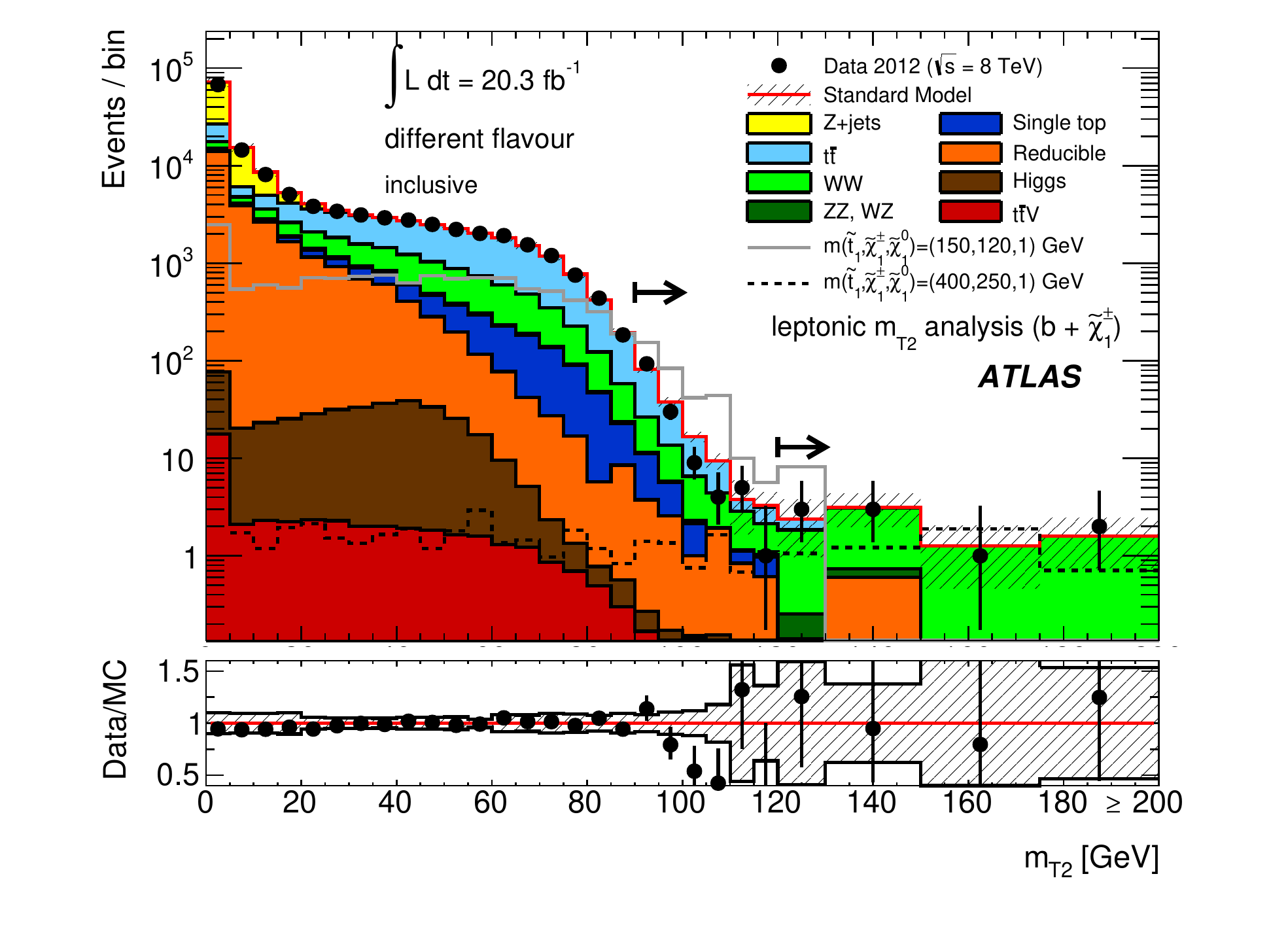}
\end{center}
\caption{Distribution of $m_\textup{T2}$ for events passing all the signal candidate selection requirements, except that on $m_\textup{T2}$ of the L90 and L120 selections, for SF (top) and DF (bottom) events.
The contributions from all SM backgrounds are shown as a histogram stack; the bands represent the total uncertainty. The components labelled ``Reducible'' correspond to the fake and non-prompt lepton backgrounds and are estimated from data as described in section~\ref{sec:fakes}; the other backgrounds are estimated from MC simulation with normalisations measured in control regions described in section~\ref{sec:backgrounds-lep} for \ttbar and 
diboson backgrounds. The expected distribution for two signal models is also shown. The full line corresponds to a model with $m(\tone)=150$~GeV, $m(\chipm)=120$~GeV and $m(\neut)=1$~GeV; the dashed line to a model with $m(\tone)=400$~GeV, $m(\chipm)=250$~GeV and $m(\neut)=1$~GeV. The arrows mark the cut values used to define the SRs. } 
\label{fig:mt2ll:SRa}
\end{figure}

\begin{figure}[htb!]
\begin{center}
\includegraphics[width=0.79\textwidth]{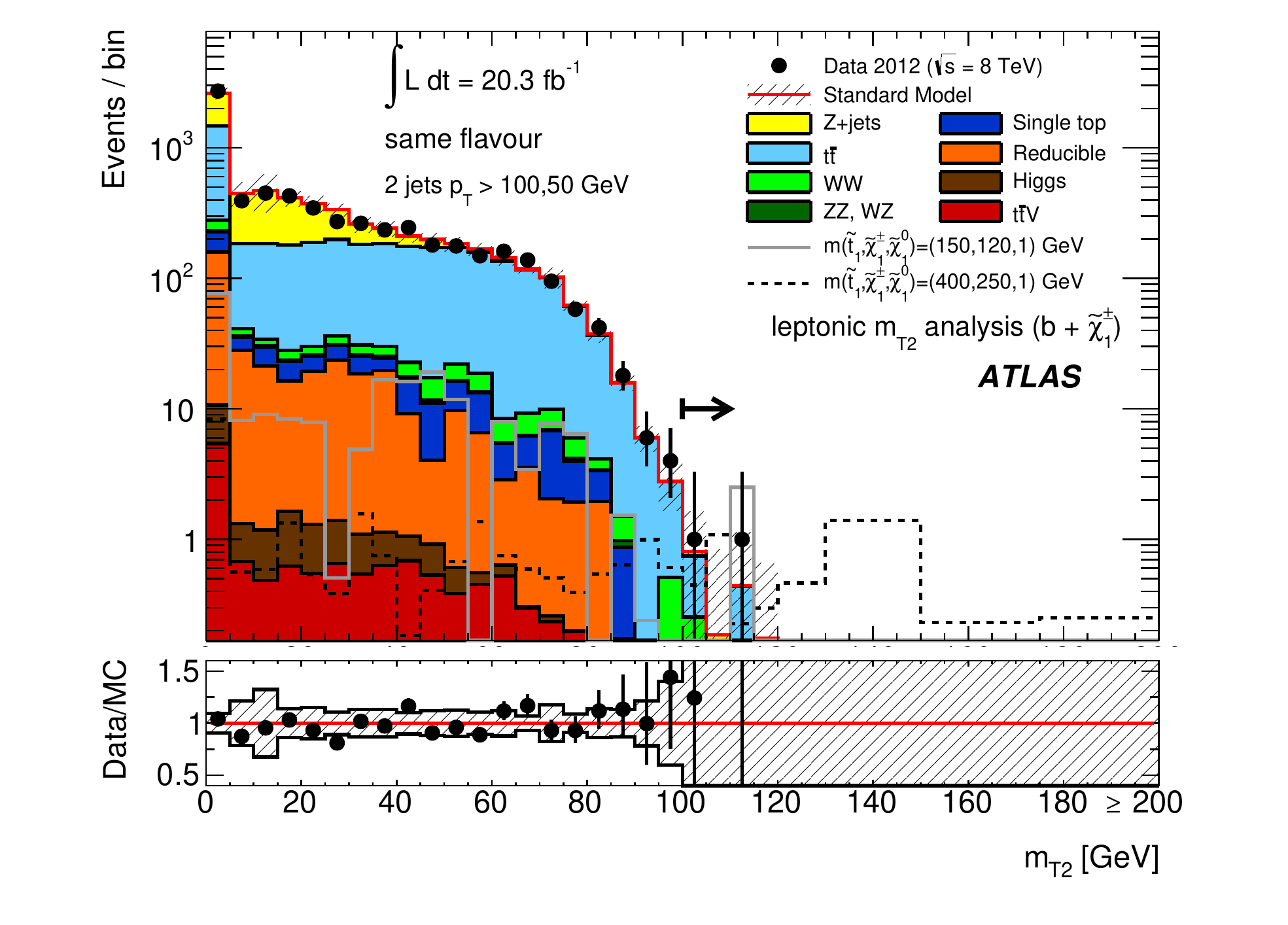}
\includegraphics[width=0.79\textwidth]{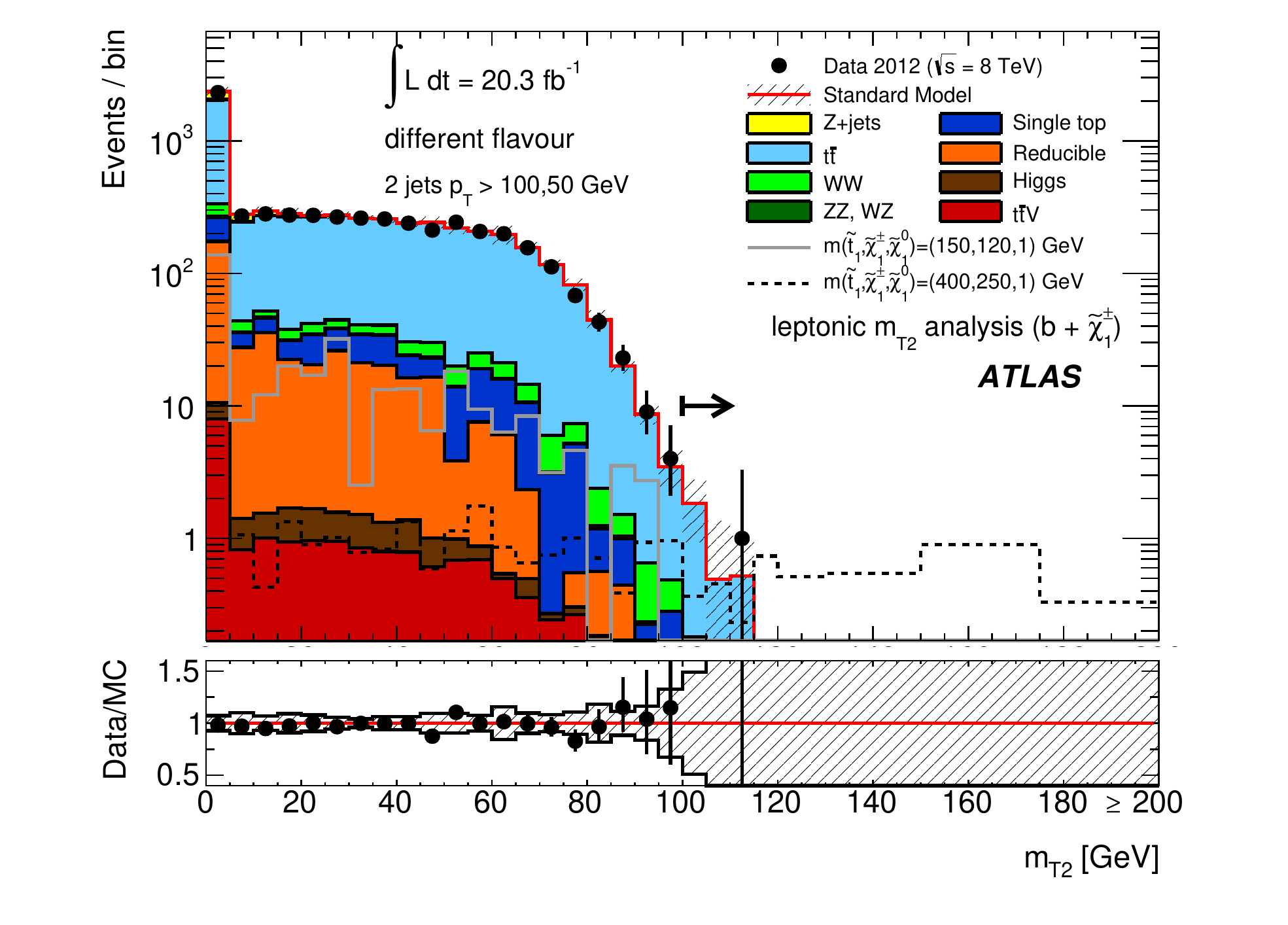}
\end{center}
\caption{Distribution of $m_\textup{T2}$ for events passing all the signal candidate selection requirements, except that on $m_\textup{T2}$ of the L100 selection, for SF (top) and DF (bottom) events.
The contributions from all SM backgrounds are shown as a histogram stack; the bands represent the total uncertainty. The components labelled ``Reducible'' correspond to the fake and non-prompt lepton backgrounds and are estimated from data as described in section~\ref{sec:fakes}; the other backgrounds are estimated from MC simulation with normalisations measured in control regions described in section~\ref{sec:backgrounds-lep} for \ttbar and 
diboson backgrounds. The expected distribution for two signal models is also shown. The full line corresponds to a model with $m(\tone)=150$~GeV, $m(\chipm)=120$~GeV and $m(\neut)=1$~GeV; the dashed line to a model with $m(\tone)=400$~GeV, $m(\chipm)=250$~GeV and $m(\neut)=1$~GeV. The arrows mark the cut values used to define the SRs. } 
\label{fig:mt2ll:SRb}
\end{figure}

\begin{figure}[htb!]
\begin{center}
\includegraphics[width=0.79\textwidth]{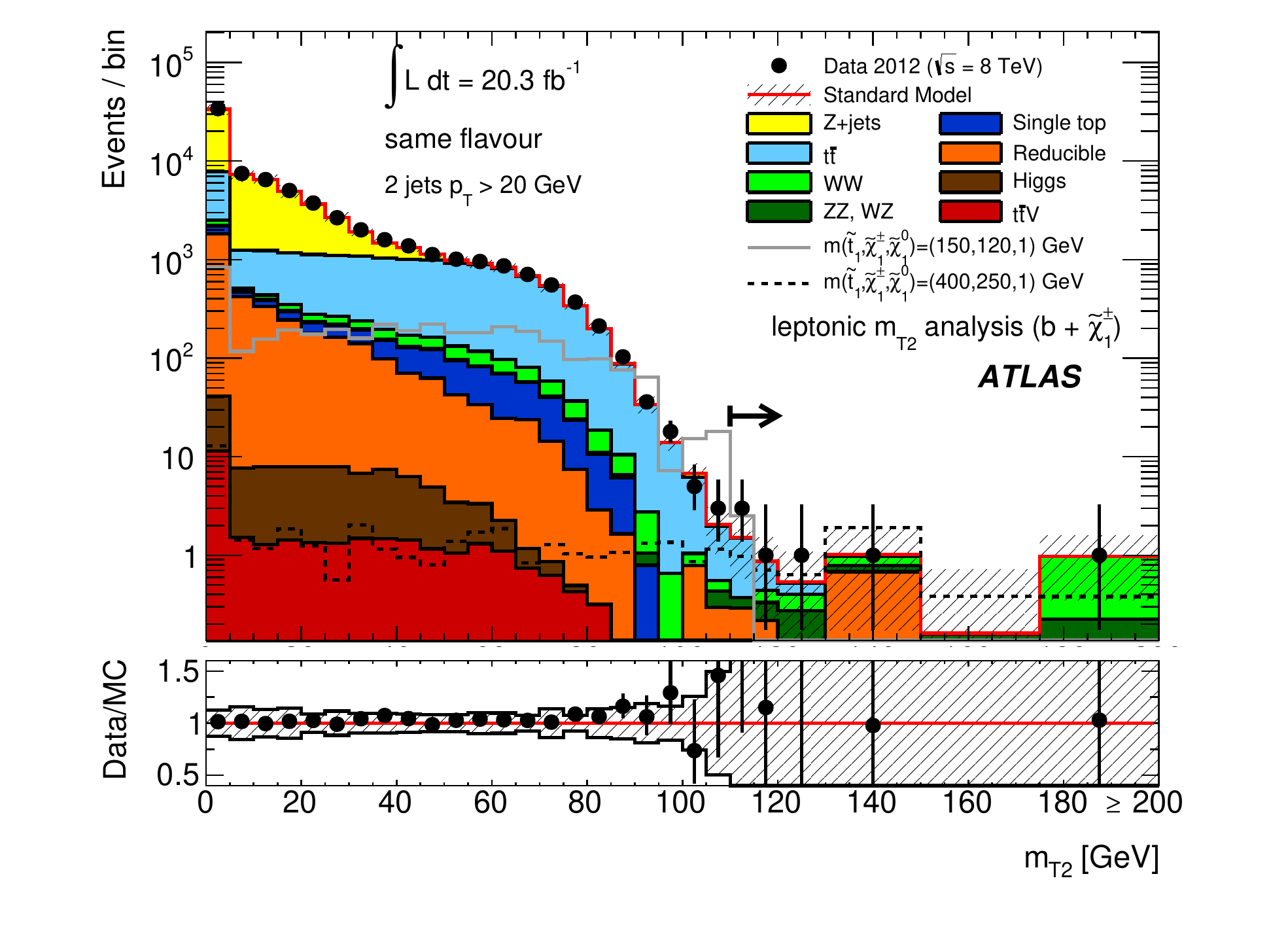}
\includegraphics[width=0.79\textwidth]{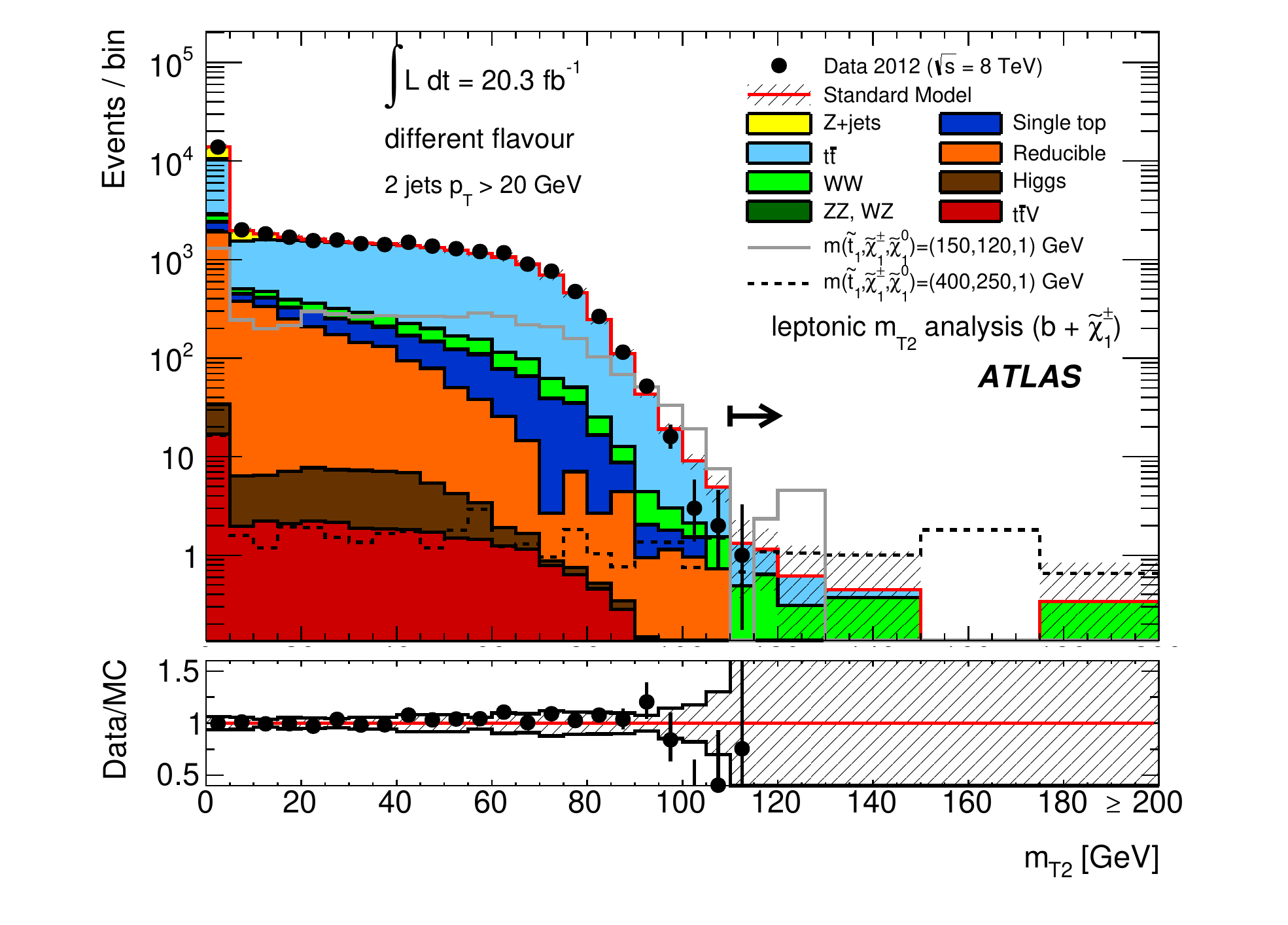}
\end{center}
\caption{Distribution of $m_\textup{T2}$ for events passing all the signal candidate selection requirements, except that on $m_\textup{T2}$ of the L110 selection, for SF (top) and DF (bottom) events.
The contributions from all SM backgrounds are shown as a histogram stack; the bands represent the total uncertainty. The components labelled ``Reducible'' correspond to the fake and non-prompt lepton backgrounds and are estimated from data as described in section~\ref{sec:fakes}; the other backgrounds are estimated from MC simulation with normalisations measured in control regions described in section~\ref{sec:backgrounds-lep} for \ttbar and 
diboson backgrounds. The expected distribution for two signal models is also shown. The full line corresponds to a model with $m(\tone)=150$~GeV, $m(\chipm)=120$~GeV and $m(\neut)=1$~GeV; the dashed line to a model with $m(\tone)=400$~GeV, $m(\chipm)=250$~GeV and $m(\neut)=1$~GeV. The arrows mark the cut values used to define the SRs. } 
\label{fig:mt2ll:SRc}
\end{figure}

\begin{figure}[htb!]
\begin{center}
\includegraphics[width=0.8\textwidth]{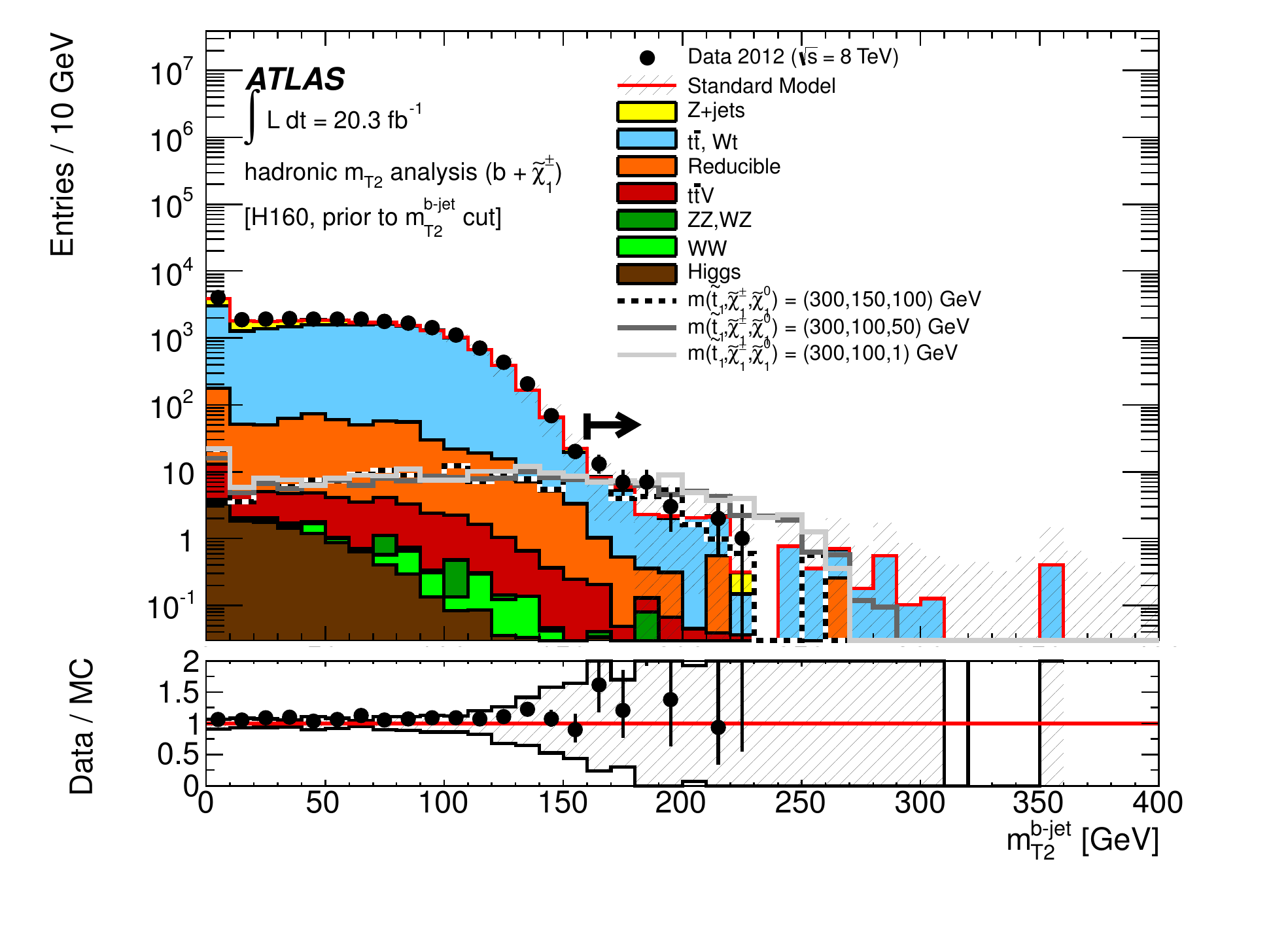}
\end{center}
\caption{Distribution of $m^{b-\textup{jet}}_\textup{T2}$ for events with two $b$-jets and all other H160 cuts, minus that on $m^{b-\textup{jet}}_\textup{T2}$ itself.
The contributions from all SM backgrounds are 
shown as a histogram stack; the bands represent the total uncertainty. The component
labelled ``Reducible" represents the fake and non-prompt lepton background and is estimated from data as described in section~\ref{sec:fakes} and the combined $t\bar{t}$ and $Wt$ component is shown renormalised after the background fit; the 
other backgrounds are estimated from MC samples normalised to the luminosity of the data and their respective cross-sections. The expected distribution 
for three signal models is also shown. The dotted line corresponds to a model
with $m(\tone)=300$~GeV, $m(\chipm)=150$~GeV and $m(\neut)=100$~GeV; the full line corresponds to a model
with $m(\tone)=300$~GeV, $m(\chipm)=100$~GeV and $m(\neut)=50$~GeV; the dashed
line to a model with $m(\tone)=300$~GeV, $m(\chipm)=100$~GeV and 
 $m(\neut)=1$~GeV. The arrow marks the cut value used to define the SR.}
\label{fig:mt2bb:SR}
\end{figure}

\begin{figure}[!hbp]
\begin{center}
\includegraphics[width=0.8\textwidth]{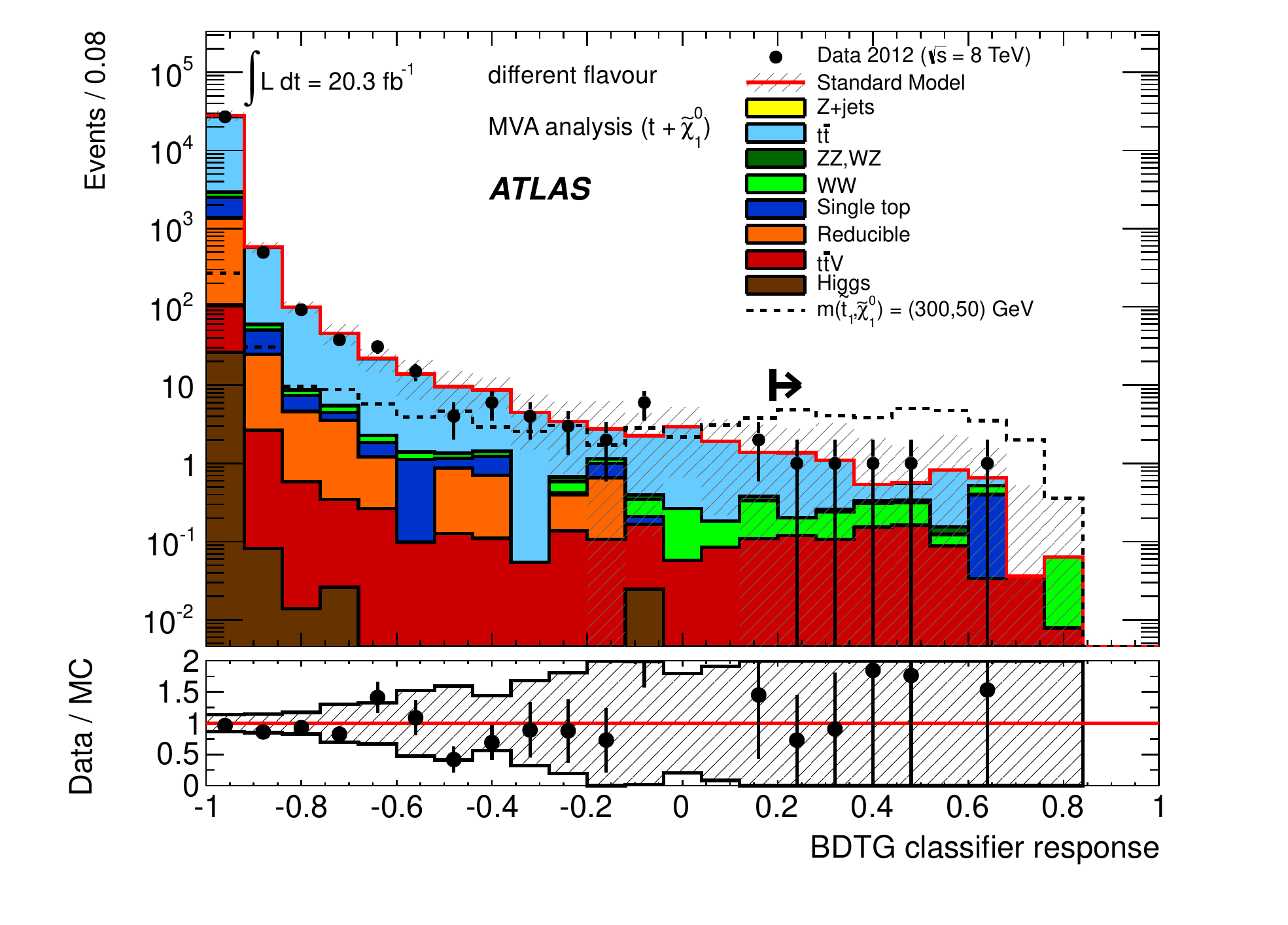}
\includegraphics[width=0.8\textwidth]{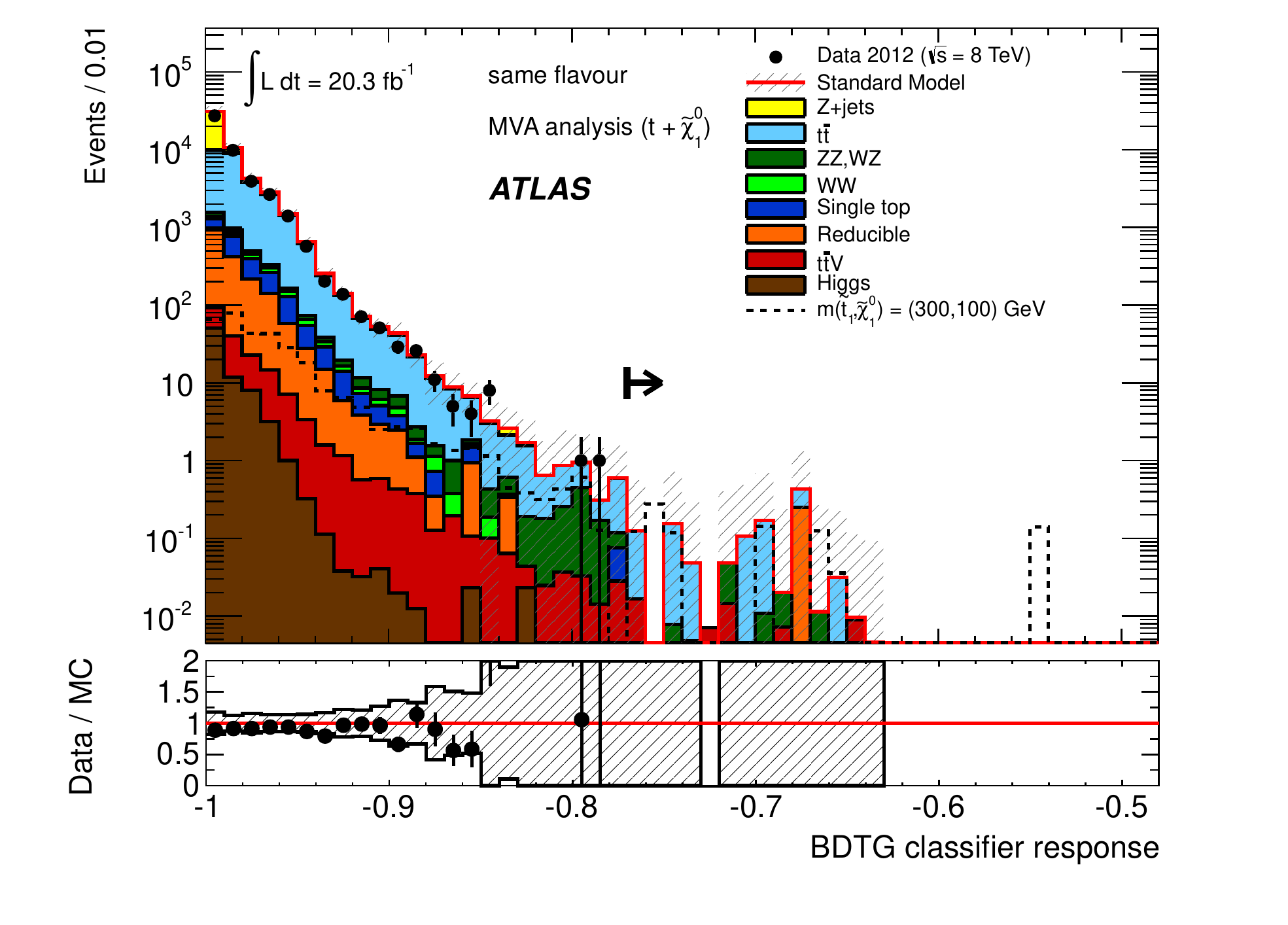}
\end{center}
\caption{BDTG distribution after all selection requirements, except the cut on the BDTG itself, after the background fit and for the DF (top) and SF (bottom) channels, as obtained from the trainings which used the point $(m(\tilde{t}_1),m(\tilde{\chi}^0_1))=(300,50)$~GeV and $(m(\tilde{t}_1),m(\tilde{\chi}^0_1))=(300,100)$~GeV, respectively. The contributions from all SM backgrounds are shown as a histogram stack.  The components labelled ``Reducible'' correspond to the fake and non-prompt lepton backgrounds and are estimated from data as described in section~\ref{sec:fakes}; the remaining backgrounds are estimated from MC samples normalised to the luminosity of the data. The reference signal points used in the training of each channel are also shown. The bands represent the total uncertainty. The arrows mark the cut values used to define the SRs: $\textup{M3}^\textup{DF}$ (top) and $\textup{M3}^\textup{SF}$ (bottom).
\label{fig:mva:BDTG_SR}}
\end{figure}
\clearpage

Upper limits at 95\% CL on the number of beyond-the-SM (BSM) events for each SR are derived using the CL$_s$ likelihood ratio prescription as described in ref.~\cite{Read:2002hq} and neglecting any possible contamination in the control regions. Normalising these by the integrated luminosity of the data sample, they can be interpreted as upper limits on the visible BSM cross-section, $\sigma_{\mathrm{vis}} = \sigma \times \epsilon \times \mathcal{A}$, where $\sigma$ is the production cross-section for the BSM signal, $\mathcal{A}$ is the acceptance defined by the fraction of events passing the geometric and kinematic selections at particle level, and $\epsilon$ is the detector reconstruction, identification and trigger efficiency (see appendix~\ref{app:truth}). Table~\ref{tab:upperlimits} summarises, for each SR, the estimated SM background yield, the observed numbers of events, and the expected and observed upper limits on event yields from a BSM signal and on $\sigma_\mathrm{vis}$.

\begin{table}[!hbp]
\caption{\label{tab:upperlimits}
Left to right: Expected background, observed events, and 95\% CL expected (observed) upper limits on the number of
BSM events ($S_{\rm exp. (obs.)}^{95}$) and the visible cross-section ($\langle\mathcal{A}\epsilon\sigma\rangle_{\rm exp. (obs.)}^{95}$). For each SR the numbers are calculated using toy MC pseudo-experiments. The equivalent limits on the visible cross-section calculated using an asymptotic method~\cite{2011EPJC...71.1554C} are given inside the square brackets. }
\begin{center}
\begin{tabular}{c|cccc}
\hline
{Signal Region}  & Background & Observation & $S_{\rm exp. (obs.)}^{95}$  &  $\sigma_\textup{vis}$ [fb] \\
\hline
\hline
L90 & $300\pm50$ & 274 & 85 (74) & 4.2 (3.6) [4.3 (3.7)] \\
L100 & $5.2 \pm 2.2$ & 3 & 6.4 (5.6)& 0.32 (0.28) [0.30 (0.24)]  \\
L110 & $9.3 \pm 3.5$ & 8 & 9.4 (9.0) & 0.46 (0.44) [0.45 (0.42)]  \\
L120 & $19 \pm 9$ & 18 & 17 (17) & 0.89 (0.86)  [0.85 (0.82)] \\
\hline
H160  & $26\pm6$ & 33 & 17 (22) & 0.85 (1.1) [0.83 (1.1)] \\
\hline
$\textup{M1}^\textup{DF}$ 		&		 	$5.8 \pm 1.9$	&9		& 7.7 (9.7)	 	& 0.38 (0.48)   [0.37 (0.44)]\\ 	
$\textup{M2}^\textup{DF}$ 		&			$13 \pm 4$	&11		&10.5 (9.4)  	& 0.52 (0.46)   [0.51 (0.45)] \\ 	
$\textup{M3}^\textup{DF}$ 		& 			$5.1 \pm 2.0$ 	& 5		& 7.1 (7.1)	  	& 0.35 (0.35)  [0.33 (0.33)]\\ 	
$\textup{M4}^\textup{DF}$ 		& 			$1.3\pm 1.0$	& 3		& 4.5 (6.5)	  	& 0.22 (0.32)  [0.22 (0.31)]\\ 	
$\textup{M5}^\textup{DF}$ 		& 			$1.0\pm 0.5$	& 1		& 3.7 (3.7)	  	& 0.18 (0.18) [0.18 (0.17)] \\ 	
$\textup{M1}^\textup{SF}$ 		& 			$7.6\pm 2.2$	& 6		& 7.6 (6.7)	  	& 0.37 (0.33)  [0.37 (0.32)]\\ 	
$\textup{M2}^\textup{SF}$ 		& 			$9.5\pm 2.1$	& 9		& 8.4 (8.2)	  	& 0.41 (0.40)  [0.41 (0.39)]\\ 	
$\textup{M3}^\textup{SF}$ 		& 			$1.1\pm 0.7$	& 0		& 3.1 (3.1)	  	& 0.15 (0.15)  [0.15 (0.11)]\\ 	
$\textup{M4}^\textup{SF}$  		&			$2.5\pm 1.0$	&5		& 5.2 (8.0) 	& 0.26 (0.39)  [0.26 (0.38)]\\ 	
\hline
\end{tabular}
\end{center}
\end{table}

The results obtained are used to derive limits on the mass of a pair-produced top squark $\tilde{t}_1$ decaying with 100\% BR into the lightest chargino and a $b$-quark (for the leptonic and hadronic $m_\textup{T2}$ analyses), an off-shell $t$-quark and the lightest neutralino (for the leptonic $m_\textup{T2}$ analyses) or an on-shell top quark and the lightest neutralino (for the MVA). 

The inclusive SRs in the leptonic $m_\textup{T2}$ analysis were designed to maximise the discovery potential of the analysis. In the absence of any excess, a set of statistically exclusive SR can be defined in order to maximise the exclusion power of the search.
Thus, in order to allow a statistical combination of the leptonic $m_\textup{T2}$ SRs and maximise this potential, a set of seven statistically independent SRs is defined in the (jet selections, $m_\textup{T2}$) plane, as shown in figure \ref{fig:excl:srdef}. These SRs are labelled \textit{Sn}, with $n$ ranging from one to seven. Table~\ref{table.results.yields.fit.S1_S2_S3_S4_S5_S6_S7} reports the background yields (after the likelihood fit) and upper limits on the visible cross-sections for each of these SRs. In each, agreement is found between the SM prediction and the data.

\begin{figure}[!hbp]
\begin{center}
\includegraphics[width=0.8\textwidth]{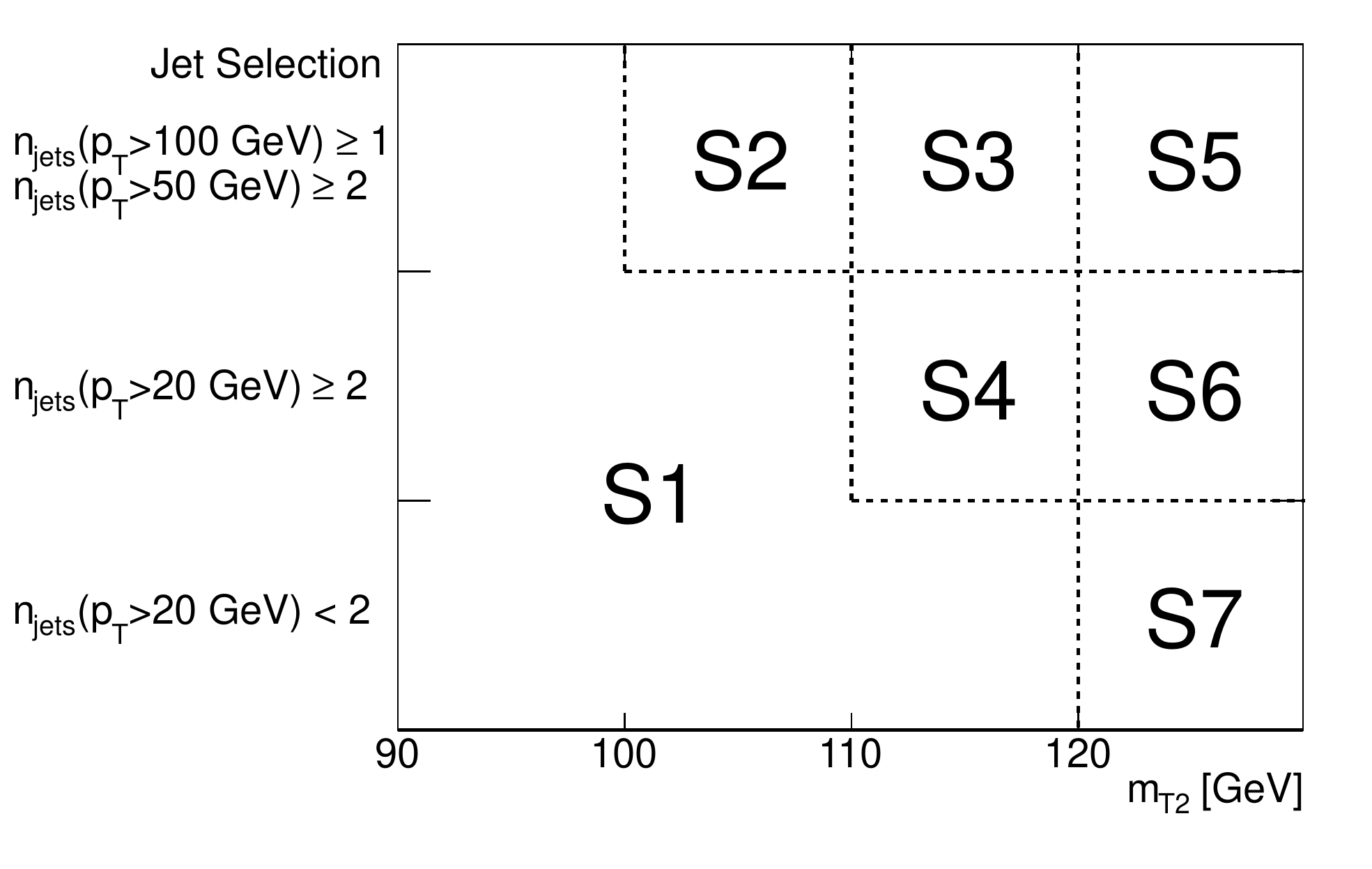}
\end{center}
\caption{\label{fig:excl:srdef} Definition of the ``leptonic $m_\textup{T2}$''  SRs used in the exclusion. The (jet selections, $m_\textup{T2}$) plane is divided into 7 non-overlapping SRs.}
\end{figure}

\begin{table}
\caption{Number of events in the leptonic $m_\textup{T2}$ SRs used in the exclusion interpretation for an integrated luminosity of 20.3~fb$^{-1}$. Combined statistical and systematic uncertainties are given. Upper limits on the visible cross-section ($\langle\mathcal{A}\epsilon\sigma\rangle_{\rm exp. (obs.)}^{95}$) are also reported for each SR using toy MC pseudo-experiments. The equivalent limits on the visible cross-section calculated using an asymptotic method~\cite{2011EPJC...71.1554C} are given inside the square brackets. }
\label{table.results.yields.fit.S1_S2_S3_S4_S5_S6_S7}
\setlength{\tabcolsep}{0.3pc}
{\small
\begin{tabular*}{1.\textwidth}{@{\extracolsep{\fill}}lrrrr}
\noalign{\smallskip}\hline\noalign{\smallskip}
{\bf  Channel}           & S1            & S2            & S3            & S4          \\[-0.05cm]
\noalign{\smallskip}\hline\noalign{\smallskip}
Observed events          & $250$              & $1$              & $2$              & $3$                           \\
Total bkg events         & $270 \pm 40$          & $3.4 \pm 1.8$          & $1.3 \pm 0.6$          & $3.7 \pm 2.7$           \\
\noalign{\smallskip}\hline\noalign{\smallskip}
$\langle\mathcal{A}\epsilon{\rm \sigma}\rangle_{\rm exp. (obs.)}^{95}$ [fb] & 3.79 (3.76) & 0.22 (0.18) & 0.20 (0.23) & 0.32 (0.32) \\
  & [3.85 (3.79)] & [0.23 (0.17)] & [0.19 (0.23)] & [0.13 (0.11)] \\
 \noalign{\smallskip}\hline\noalign{\smallskip} 
 \noalign{\bigskip}\hline\noalign{\smallskip}
{\bf  channel}           &      S5            & S6            & S7          &    \\[-0.05cm]
\noalign{\smallskip}\hline\noalign{\smallskip}
Observed events          &  $0$              & $3$              & $15$              &      \\
Total bkg events             & $0.5 \pm 0.4$          & $3.8 \pm 1.6$          & $15 \pm 7$        &      \\
\noalign{\smallskip}\hline\noalign{\smallskip}
$\langle\mathcal{A}\epsilon{\rm \sigma}\rangle_{\rm exp. (obs.)}^{95}$ [fb] & 0.15 (0.15) & 0.28 (0.28) & 0.46 (0.48) & \\
 & [0.13 (0.11)] & [0.28 (0.25)] & [0.48 (0.48)] & \\
 \noalign{\smallskip}\hline\noalign{\smallskip}
\end{tabular*}
%%%%
}
\end{table}
A fit similar to that described in section~\ref{sec:background:fit} is used to evaluate exclusion contours in various two-dimensional mass parameter planes. In this fit, the CRs and SR(s) are fit simultaneously taking into account the experimental and theoretical systematic uncertainties as nuisance parameters. The signal contamination of the CRs is taken into account in the fit. The fit thus differs from the ``background-only'' fit described in section~\ref{sec:background:fit} as follows:
\begin{enumerate}
\item An extra free parameter for a possible BSM signal strength which is constrained to be non-negative is added.
\item The number of events observed in the signal region is now also considered as an input to the fit.
\item The expected contamination of the control regions by the signal is included in the fit.
\end{enumerate}
Systematic uncertainties on the signal expectations stemming from detector effects are included in the fit in the same way as for the backgrounds. Systematic uncertainties on the signal cross-section due to the choice of renormalisation and factorisation scale and PDF uncertainties are calculated as described earlier but not included directly in the fit.
In all resulting exclusion contours the dashed (black) and solid (red) lines show the 95\% CL expected and observed limits, respectively, including all
uncertainties except for the theoretical signal cross-section uncertainty (PDF and scale). The (yellow) bands around the expected limits show the $\pm1\sigma$ expectations. The dotted $\pm1\sigma$ (red) lines around the observed limit represent the results obtained when moving the nominal signal cross-section up or down by its theoretical uncertainty. Quoted numerical limits on the particle masses are taken from these $-1\sigma$ ``theory lines''.

For the leptonic and hadronic $m_\textup{T2}$ analyses, various two-dimensional slices in the three-dimensional mass parameter space $m(\tilde{t}_1,\tilde{\chi}^\pm_1,\tilde{\chi}^0_1)$ are used to quantify the exclusion contours on these parameters in the $\tilde{t}_1\rightarrow b+\tilde{\chi}^\pm_1$ mode: in the $(\tone,\tilde{\chi}^\pm_1)$ mass plane for a neutralino with a mass of 1~GeV (figure~\ref{fig:excl:top squarkchargino1}); in the $(\tone,\tilde{\chi}^0_1)$ mass plane for a fixed value of $m(\tilde{t}_1)-m(\tilde{\chi}^\pm_1)=10$~GeV (figure~\ref{fig:excl:top squarkneutralino}); in the $(\tilde{\chi}^\pm_1,\tilde{\chi}^0_1)$ mass plane for a fixed 300 GeV top squark (figure~\ref{fig:excl:top squark300}); and in the ($\tone,\tilde{\chi}^0_1$) mass plane for $m(\tilde{\chi}^\pm_1)=2m(\tilde{\chi}^0_1)$ (figure~\ref{fig:excl:chi2neu}).
For the above limits, in each case all the exclusive SRs of the leptonic $m_\textup{T2}$ analysis are combined when setting the exclusions. The hadronic $m_\textup{T2}$ SR, H160, is added into the combination in the plane with fixed 300~GeV top-squark mass, a projection in which the $m^{b-\textup{jet}}_\textup{T2}$ variable is expected to increase sensitivity, and for points in the 1~GeV neutralino and the $m(\tilde{\chi}^\pm_1)=2m(\tilde{\chi}^0_1)$ planes with $m(\tilde{t}_1)=300$~GeV. In particular, in this last plane (figure~\ref{fig:excl:chi2neu}), the contribution from the hadronic $m_\textup{T2}$ SR is the narrow corridor at $m(\tilde{t}_1)=300$~GeV and low $m(\tilde{\chi}^0_1)$: this is the result of the sensitivity being limited on the higher $m(\tilde{t}_1)$ side by the decreasing $\tilde{t}_1$ production cross-section and at lower masses by the $m_\textup{T2}^{b\textup{-jet}}$ cut acceptance. The optimal choice of $m_\textup{T2}^{b\textup{-jet}}$ cut-value is heavily dictated by the shape and expected sharp end-point of $m_\textup{T2}^{b\textup{-jet}}$ for the $t\bar{t}$ background, rather than the end-points expected for signal events.

For the MVA analysis, the exclusion contours for an on-shell top-quark in a $\tilde{t}_1\rightarrow t+\tilde{\chi}^0_1$ decay are quantified in the $m(\tilde{t}_1)-m(\tilde{\chi}^0_1)$ plane (figure~\ref{fig:excl:MVA}), taking the best expected DF and SF SRs (defined as the regions with the lowest value of the expected CL$_{s}$), for each point, and combining them statistically. 

The results of the leptonic $m_\textup{T2}$ analysis are used to derive limits on the mass of a top squark decaying with 100\% BR into $bW\tilde{\chi}^0_1$ (figure~\ref{fig:excl:3body}) and the results of the hadronic $m_\textup{T2}$ analysis are also used to derive limits on $\tilde{t}_1\rightarrow b+\tilde{\chi}^\pm_1$ for fixed 106 GeV chargino mass (figure~\ref{fig:excl:fixchi106}), a grid introduced by CDF in ref.~\cite{Aaltonen:2009sf}.

\begin{figure}[!hbp]
\begin{center}
\includegraphics[width=0.8\textwidth]{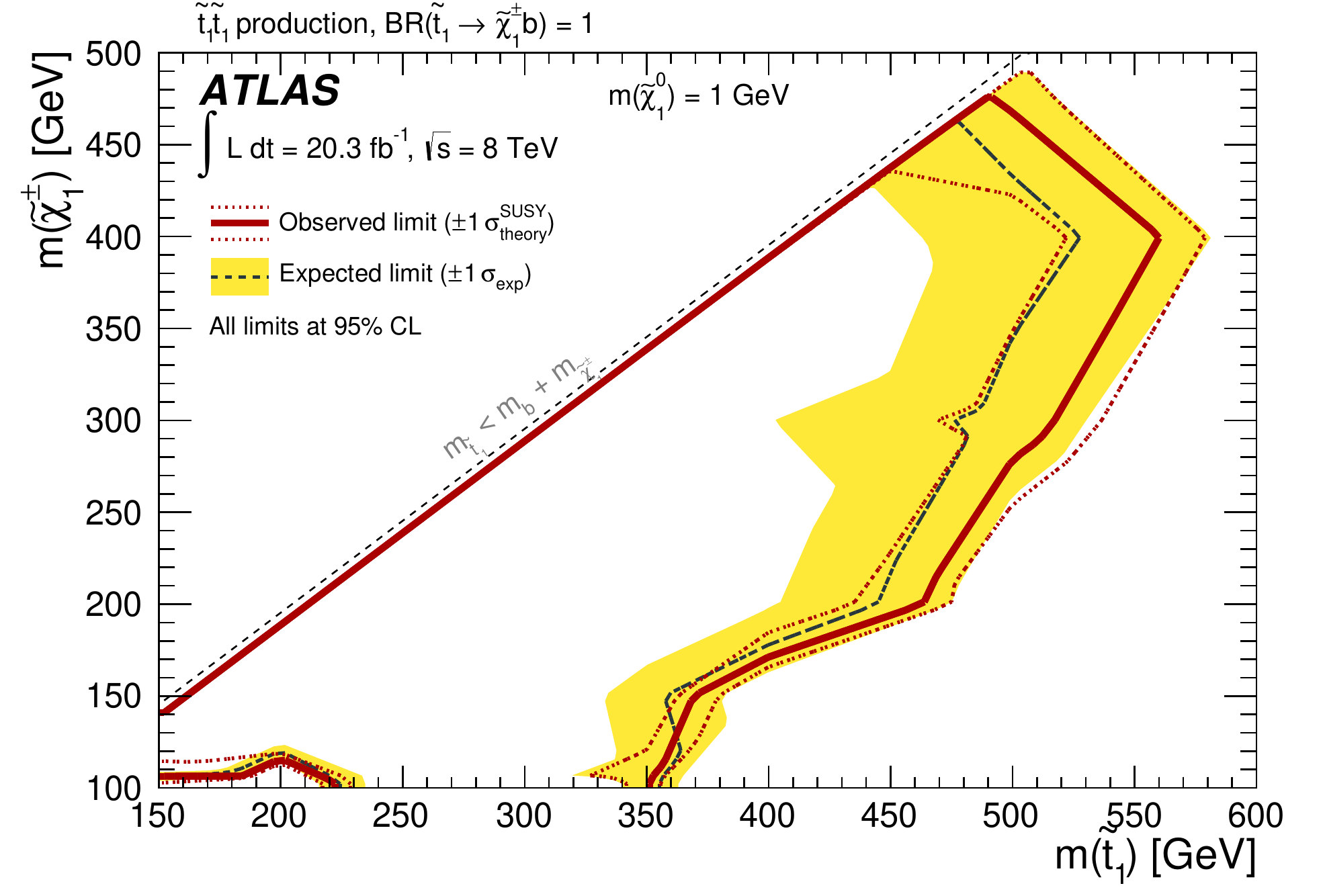}
\end{center}
\caption{\label{fig:excl:top squarkchargino1}Observed and expected exclusion contours at 95\% CL in the ($\tilde{t}_1,\tilde{\chi}^\pm_1$) mass plane for a fixed value of $m(\tilde{\chi}^0_1)=1$~GeV. The dashed and solid 
lines show the 95\% CL expected and observed limits, respectively, including all
uncertainties except for the theoretical signal cross-section uncertainty (PDF and scale). The band around the expected limit shows the $\pm1\sigma$ expectation. The dotted $\pm1\sigma$ lines around the observed limit represent the results obtained when moving the nominal signal cross-section up or down by the theoretical uncertainty.}
\end{figure}

\begin{figure}[!hbp]
\begin{center}
\includegraphics[width=0.8\textwidth]{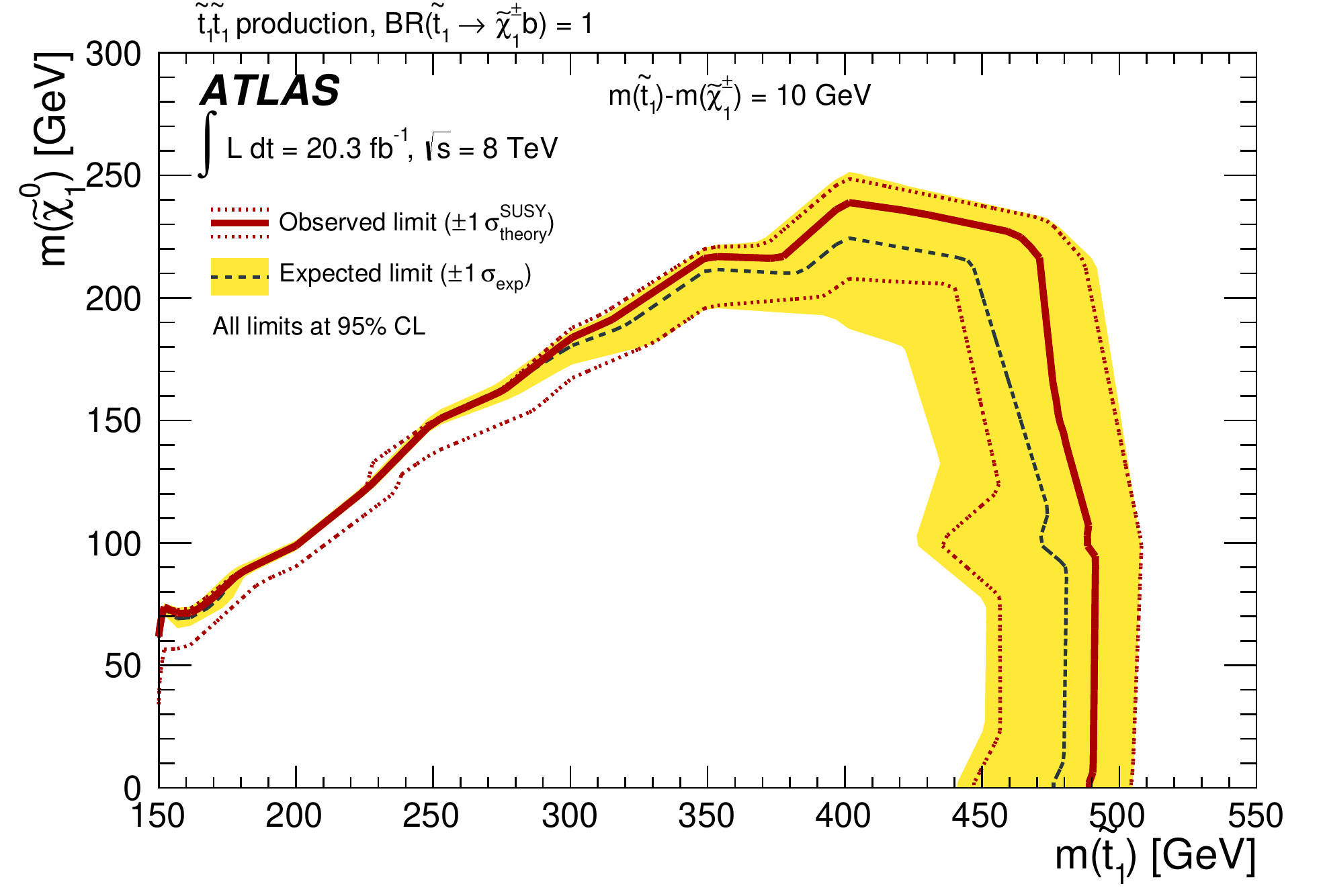}
\end{center}
\caption{\label{fig:excl:top squarkneutralino}Observed and expected exclusion contours at 95\% CL in the ($\tilde{t}_1,\tilde{\chi}^0_1$) mass plane for a fixed value of $m(\tilde{t}_1)-m(\tilde{\chi}^\pm_1)=10$~GeV. The dashed and solid 
lines show the 95\% CL expected and observed limits, respectively, including all
uncertainties except for the theoretical signal cross-section uncertainty (PDF and scale). The band around the expected limit shows the $\pm1\sigma$ expectation. The dotted $\pm1\sigma$ lines around the observed limit represent the results obtained when moving the nominal signal cross-section up or down by the theoretical uncertainty.}
\end{figure}

\begin{figure}[!hbp]
\begin{center}
\includegraphics[width=0.8\textwidth]{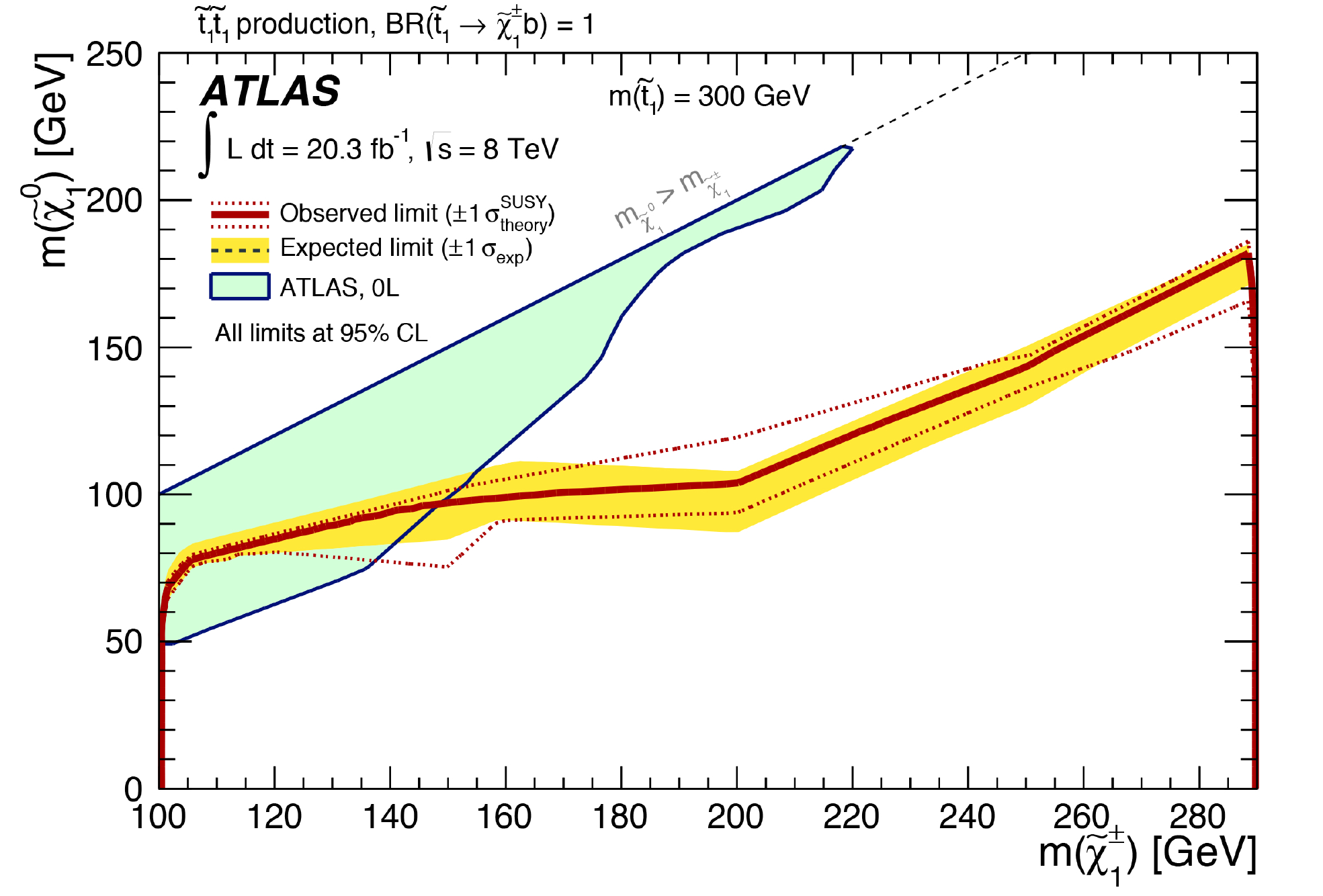}
\end{center}
\caption{\label{fig:excl:top squark300}Observed and expected exclusion contours at 95\% CL in the ($\tilde{\chi}^\pm_1,\tilde{\chi}^0_1$) mass plane for a fixed value of $m(\tilde{t}_1)=300$~GeV. The dashed and solid 
lines show the 95\% CL expected and observed limits, respectively, including all
uncertainties except for the theoretical signal cross-section uncertainty (PDF and scale). The band around the expected limit shows the $\pm1\sigma$ expectation. The dotted $\pm1\sigma$ lines around the observed limit represent the results obtained when moving the nominal signal cross-section up or down by the theoretical uncertainty. The solid light azure area labelled 0L is the exclusion contour from the ATLAS zero lepton direct top squark analysis~\cite{Aad:2013ija}.}
\end{figure}

\begin{figure}[!hbp]
\begin{center}
\includegraphics[width=0.8\textwidth]{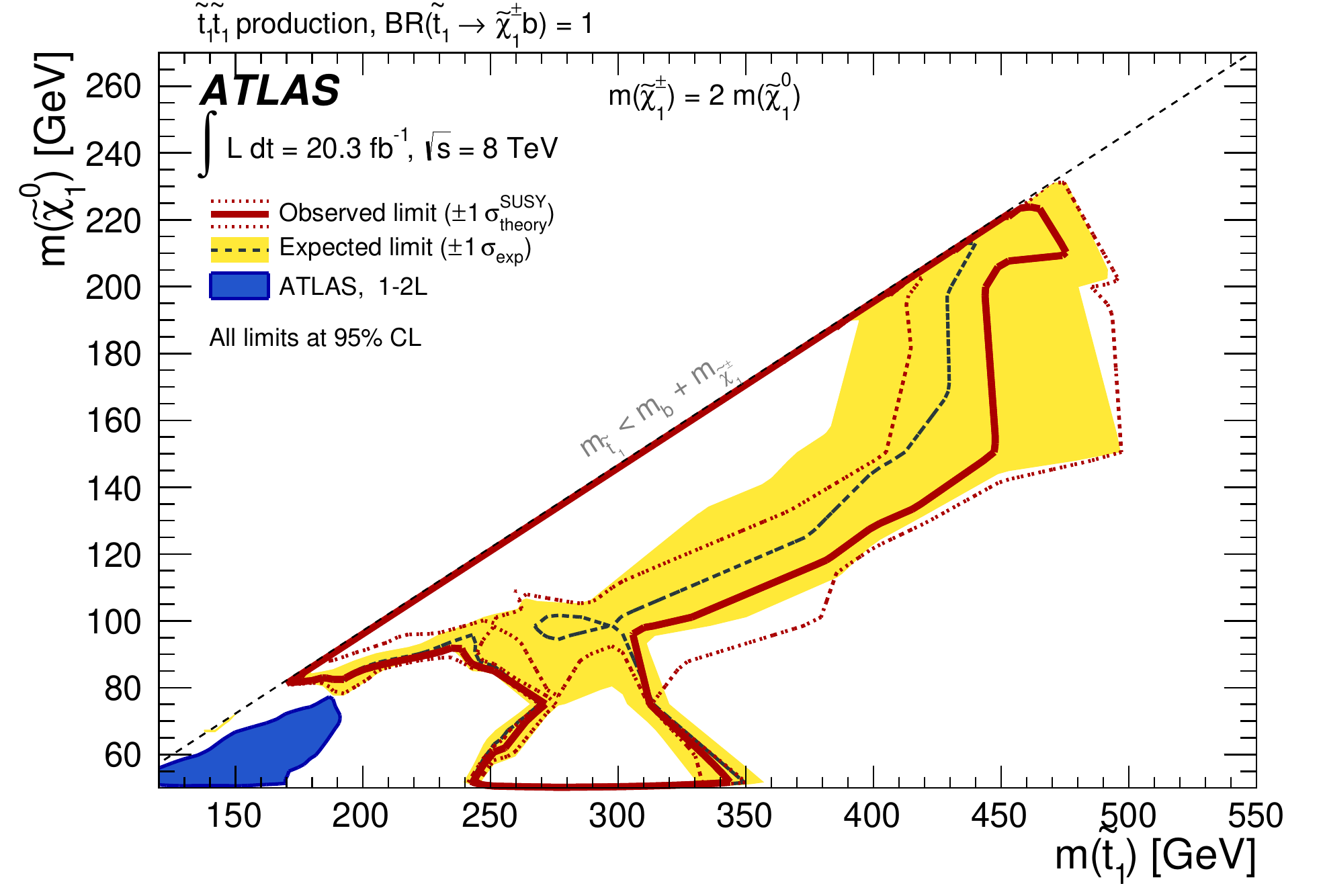}
\end{center}
\caption{\label{fig:excl:chi2neu}Observed and expected exclusion contours at 95\% CL in the ($\tilde{t}_1,\tilde{\chi}^0_1$) mass plane for $m(\tilde{\chi}^\pm_1)=2m(\tilde{\chi}^0_1)$. The dashed and solid 
lines show the 95\% CL expected and observed limits, respectively, including all uncertainties except for the theoretical signal cross-section uncertainty (PDF and scale). The band around the expected limit shows the $\pm1\sigma$ expectation. The dotted $\pm1\sigma$ lines around the observed limit represent the results obtained when moving the nominal signal cross-section up or down by the theoretical uncertainty. The solid blue area labelled 1-2L is the exclusion contour from an ATLAS search for direct top squark production in events with one or two leptons~\cite{Aad:2012yr}.
}
\end{figure}

\begin{figure}[!hbp]
\begin{center}
\includegraphics[width=0.8\textwidth]{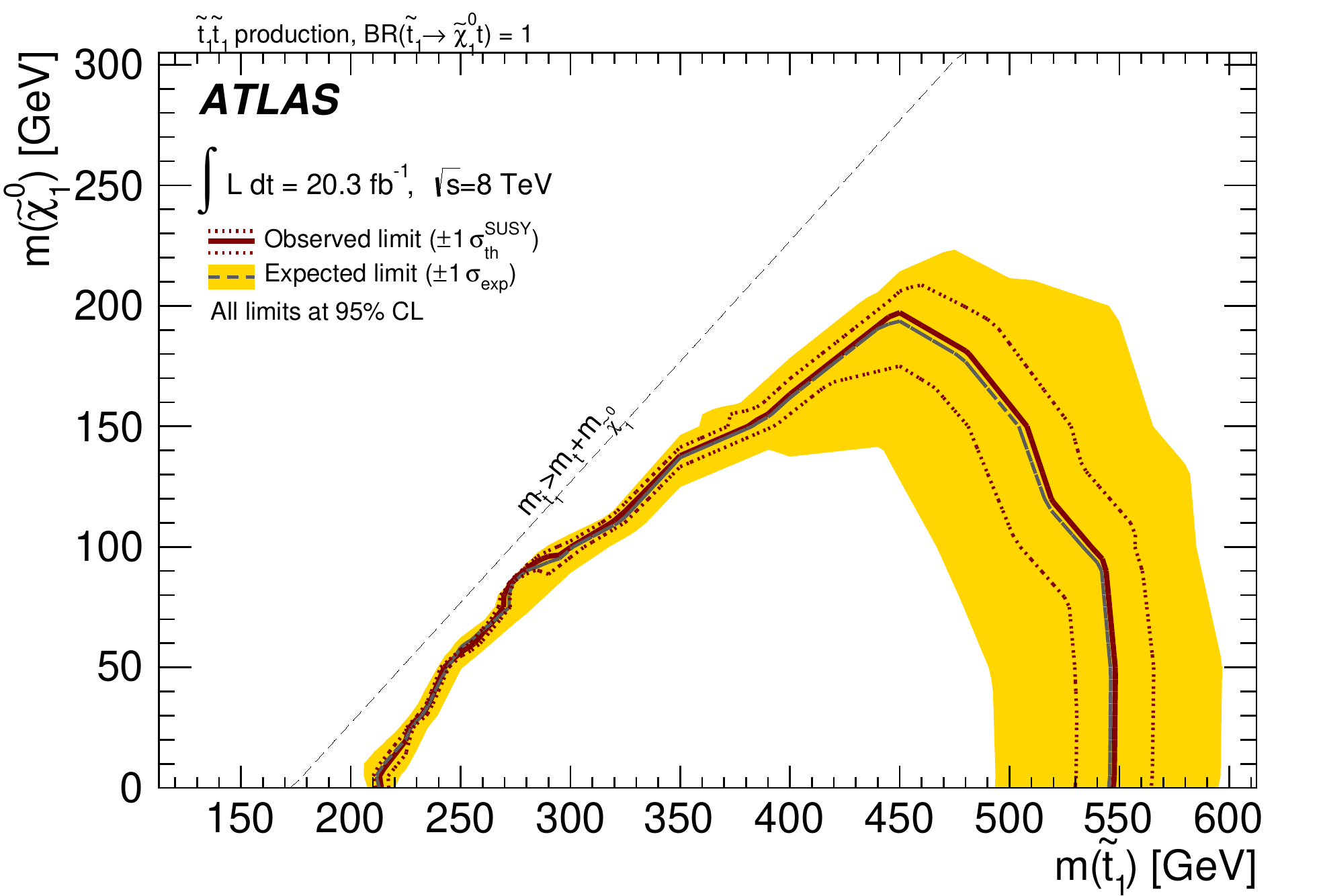}	
\end{center}
\caption{\label{fig:excl:MVA}Observed and expected exclusion contours at 95\% CL in the ($\tilde{t}_1,\tilde{\chi}^0_1$)  mass plane assuming $\tilde{t}_1\rightarrow t+\tilde{\chi}^0_1$. The dashed and solid 
lines show the 95\% CL expected and observed limits, respectively, including all
uncertainties except for the theoretical signal cross-section uncertainty (PDF and scale). The band around the expected limit shows the $\pm1\sigma$ expectation. The dotted $\pm1\sigma$ lines around the observed limit represent the results obtained when moving the nominal signal cross-section up or down by the theoretical uncertainty.}
\end{figure}

\begin{figure}[!hbp]
\begin{center}
\includegraphics[width=0.8\textwidth]{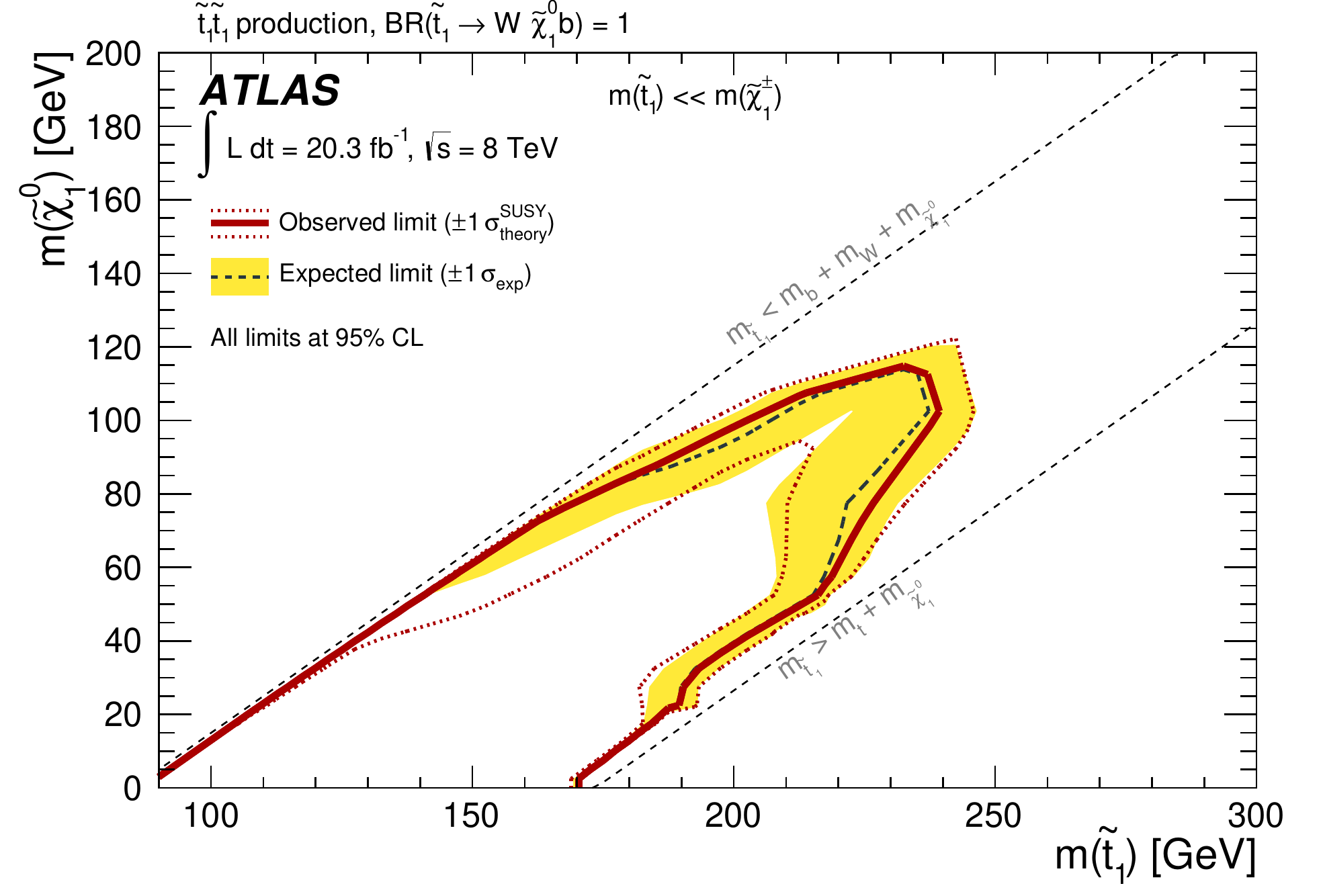}
\end{center}
\caption{\label{fig:excl:3body}Observed and expected exclusion contours at 95\% CL in the ($\tilde{t}_1,\tilde{\chi}^0_1$) mass plane assuming $\tilde{t}_1\rightarrow bW\tilde{\chi}^0_1$ with 100\% BR. The dashed and solid 
lines show the 95\% CL expected and observed limits, respectively, including all
uncertainties except for the theoretical signal cross-section uncertainty (PDF and scale). The band around the expected limit shows the $\pm1\sigma$ expectation. The dotted $\pm1\sigma$ lines around the observed limit represent the results obtained when moving the nominal signal cross-section up or down by the theoretical uncertainty.}
\end{figure}

\begin{figure}[!hbp]
\begin{center}
\includegraphics[width=0.8\textwidth]{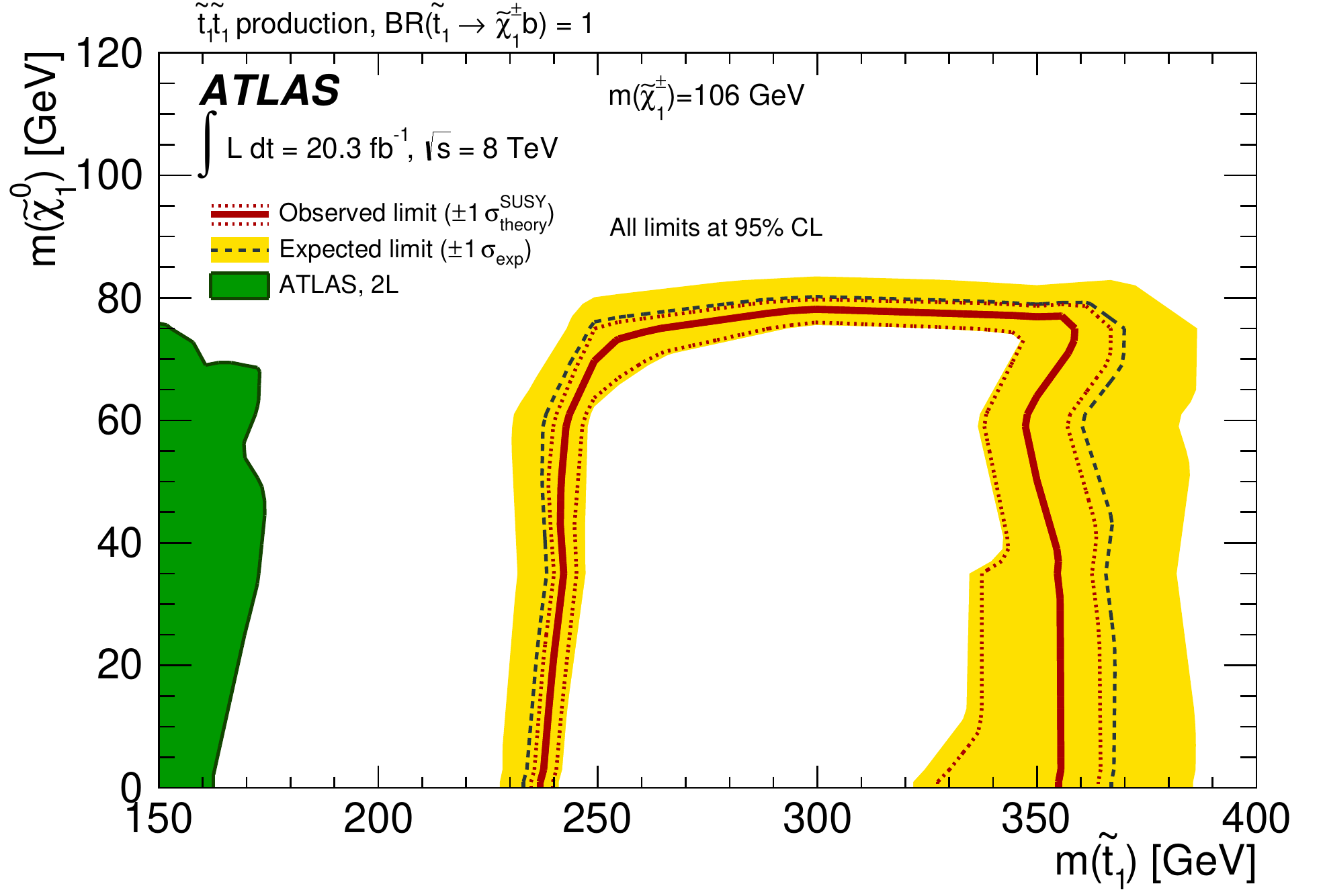}
\end{center}
\caption{\label{fig:excl:fixchi106}Observed and expected exclusion contours at 95\% CL in the ($\tilde{t}_1,\tilde{\chi}^0_1$) mass plane for a fixed value of $m(\tilde{\chi}^\pm_1)=106$~GeV. The dashed and solid 
lines show the 95\% CL expected and observed limits, respectively, including all
uncertainties except for the theoretical signal cross-section uncertainty (PDF and scale). The band around the expected limit shows the $\pm1\sigma$ expectation. The dotted $\pm1\sigma$ lines around the observed limit represent the results obtained when moving the nominal signal cross-section up or down by the theoretical uncertainty. The solid green area shows the excluded region from a previous ATLAS two-lepton analysis~\cite{Aad:2012yr}.}
\end{figure}

\section{Conclusions}
\label{sec:conclusions}

The results of a search for the production of the lightest top squark $\tilde{t}_1$ in a 20.3~fb$^{-1}$ dataset of LHC $pp$ collisions at $\sqrt{s}=8$~TeV recorded by ATLAS are reported. Events with two oppositely charged leptons (electrons or muons) were analysed and data compared to SM predictions in a variety of SRs. 
Results are in agreement with SM predictions across all SRs.
The observations in the various SRs are used to produce 95\% CL upper limits on $\tilde{t}_1$ pair production assuming either the decay $\tilde{t}_1\rightarrow  b+\tilde{\chi}^\pm_1$ or the decay $\tilde{t}_1\rightarrow t+\tilde{\chi}^0_1$ (each with 100\% BR) for different assumptions on the mass hierarchy of the top squark, chargino and lightest neutralino. In the $\tilde{t}_1\rightarrow t+\tilde{\chi}^0_1$ case, and for an on-shell $t$-quark, the SRs considered utilised an MVA technique.

For the case of a  1~GeV neutralino, a top-squark $\tone$ with a mass between 150~GeV and 445~GeV decaying to a $b$-quark and a chargino is excluded at 95\% CL for a chargino approximately degenerate with the top squark. For a 300~GeV top squark decaying to a $b$-quark and a chargino, chargino masses between 100~GeV and 290~GeV are excluded for a lightest neutralino with mass below 70~GeV.
Top squarks of masses between 215~GeV and 530~GeV decaying to an on-shell $t$-quark and a neutralino of mass 1~GeV are excluded at 95\% CL.
Limits are also set on the direct three-body decay mode, $\tilde{t}_1\rightarrow t+\tilde{\chi}^0_1$ with an off-shell $t$-quark ($\tilde{t}_1\rightarrow W\tilde{\chi}^0_1 b$), excluding a top squark between 90~GeV and 170~GeV, under the assumption of a 1~GeV neutralino.

\clearpage

\include{Acknowledgements}

\bibliographystyle{JHEP}
\bibliography{2LepStopPaper2013}

\clearpage
\appendix

\clearpage

\section{Generator-level object and event selection}

\label{app:truth}

    The generator-level MC information is used to determine the acceptance and the efficiency
    for simulated signal events in this analysis. The acceptance is defined as the fraction of signal events which pass the analysis 
selection performed on generator-level objects, therefore emulating an ideal detector with perfect particle identification
and no measurement resolution effects. The efficiency is the ratio between the expected signal rate calculated 
with simulated data passing all the reconstruction level cuts applied to reconstructed objects, and the signal rate for the ideal detector. In this section, the details of the generator-level object and event selection information are given.

      The input to the object selection algorithm is the particles from the generated primary proton-proton collision after parton 
shower and final-state radiation, and after the decay of unstable supersymmetric particles, hadrons and $\tau$ leptons. 
Muons and hadrons with a lifetime comparable to or larger than the time of flight through the detector are not decayed.
            
     Jets are reconstructed using the anti-$k_t$ jet clustering algorithm with radius parameter of 0.4, as for the 
simulated and observed data, but the particle input to the algorithm is restricted to MC particles other than muons, neutrinos, and neutralinos. All jets which have a $b$-quark with $\pT > 5 \GeV$ within a $\Delta R < 0.4$ of the jet axis are considered as $b$-jet. 

     Electrons or muons are considered if they are produced by the decay of a $W$,$Z$, or Higgs boson, a supersymmetric 
particle, or if they are produced by the decay of a $\tau$~lepton which was produced by the decay of these particles. 
The same selections on $p_\mathrm{T}$ and $\eta$ applied to reconstructed electrons, muons and jets, as well as the 
$\Delta R$ selections between them, described in section~4, are applied also at generator-level.

    The truth \EtMiss is taken as the sum of momenta of weakly interacting particles (neutrinos and neutralinos). 

     The event selection described in section~\ref{sec:selection} is then performed on the selected electrons, muons, jets, and \EtMiss.
 
\onecolumn
\clearpage 
\input{atlas_authlist} 
%%%%%%%%%%%%%%%%%%%%%%%%%%%%%%%%%%%%%%%%%%%%%%%%%%%%%%%%%%%%%%%%%%%%%%%%%%%%%%

%% file: Acknowledgements.tex
%\documentclass[11pt,a4paper,dvips]{article}

%\begin{document}

% Acknowledgements for papers with collision data
% Version 19-Feb-2014

\section{Acknowledgements}

% Standard acknowledgements start here
%----------------------------------------------
We thank CERN for the very successful operation of the LHC, as well as the
support staff from our institutions without whom ATLAS could not be
operated efficiently.

We acknowledge the support of ANPCyT, Argentina; YerPhI, Armenia; ARC,
Australia; BMWF and FWF, Austria; ANAS, Azerbaijan; SSTC, Belarus; CNPq and FAPESP,
Brazil; NSERC, NRC and CFI, Canada; CERN; CONICYT, Chile; CAS, MOST and NSFC,
China; COLCIENCIAS, Colombia; MSMT CR, MPO CR and VSC CR, Czech Republic;
DNRF, DNSRC and Lundbeck Foundation, Denmark; EPLANET, ERC and NSRF, European Union;
IN2P3-CNRS, CEA-DSM/IRFU, France; GNSF, Georgia; BMBF, DFG, HGF, MPG and AvH
Foundation, Germany; GSRT and NSRF, Greece; ISF, MINERVA, GIF, I-CORE and Benoziyo Center,
Israel; INFN, Italy; MEXT and JSPS, Japan; CNRST, Morocco; FOM and NWO,
Netherlands; BRF and RCN, Norway; MNiSW and NCN, Poland; GRICES and FCT, Portugal; MNE/IFA, Romania; MES of Russia and ROSATOM, Russian Federation; JINR; MSTD,
Serbia; MSSR, Slovakia; ARRS and MIZ\v{S}, Slovenia; DST/NRF, South Africa;
MINECO, Spain; SRC and Wallenberg Foundation, Sweden; SER, SNSF and Cantons of
Bern and Geneva, Switzerland; NSC, Taiwan; TAEK, Turkey; STFC, the Royal
Society and Leverhulme Trust, United Kingdom; DOE and NSF, United States of
America.

The crucial computing support from all WLCG partners is acknowledged
gratefully, in particular from CERN and the ATLAS Tier-1 facilities at
TRIUMF (Canada), NDGF (Denmark, Norway, Sweden), CC-IN2P3 (France),
KIT/GridKA (Germany), INFN-CNAF (Italy), NL-T1 (Netherlands), PIC (Spain),
ASGC (Taiwan), RAL (UK) and BNL (USA) and in the Tier-2 facilities
worldwide.
%----------------------------------------------
%\end{document}

%% file: atlas_authlist.tex
% ATLAS Collaboration author list
% Data extracted on 06-Mar-2014 for paper reference SUSY-2013-19
%\documentclass[11pt]{article}
%\usepackage{a4wide}\begin{document}
\begin{flushleft}
{\Large The ATLAS Collaboration}

\bigskip

G.~Aad$^{\rm 84}$,
T.~Abajyan$^{\rm 21}$,
B.~Abbott$^{\rm 112}$,
J.~Abdallah$^{\rm 152}$,
S.~Abdel~Khalek$^{\rm 116}$,
O.~Abdinov$^{\rm 11}$,
R.~Aben$^{\rm 106}$,
B.~Abi$^{\rm 113}$,
M.~Abolins$^{\rm 89}$,
O.S.~AbouZeid$^{\rm 159}$
% ...
\bigskip
\\
% ...
$^{11}$ Institute of Physics, Azerbaijan Academy of Sciences, Baku, Azerbaijan\\
% ...
$^{21}$ Physikalisches Institut, University of Bonn, Bonn, Germany\\
% ...
$^{84}$ CPPM, Aix-Marseille Universit{\'e} and CNRS/IN2P3, Marseille, France\\
% ...
$^{89}$ Department of Physics and Astronomy, Michigan State University, East Lansing MI, United States of America\\
% ...
$^{106}$ Nikhef National Institute for Subatomic Physics and University of Amsterdam, Amsterdam, Netherlands\\
% ...
$^{112}$ Homer L. Dodge Department of Physics and Astronomy, University of Oklahoma, Norman OK, United States of America\\
$^{113}$ Department of Physics, Oklahoma State University, Stillwater OK, United States of America\\
% ...
$^{116}$ LAL, Universit{\'e} Paris-Sud and CNRS/IN2P3, Orsay, France\\
% ...
$^{152}$ Institute of Physics, Academia Sinica, Taipei, Taiwan\\
% ...
$^{159}$ Department of Physics, University of Toronto, Toronto ON, Canada\\
% ...
\end{flushleft}

\end{document}
% Created with ./xml2latex.py